\documentclass{stile/aa}

\usepackage{txfonts}

\usepackage[english]{babel}
\usepackage{amsmath,amssymb}
\usepackage{graphicx, subfig}
\usepackage[normalem]{ulem}

\usepackage{verbatim}

\usepackage{multirow}
\usepackage{mathtools}
\usepackage{epstopdf}

\usepackage{url}
\usepackage{xcolor,color}
\usepackage{orcidlink}

\usepackage{natbib,twoopt}
\usepackage[hyphenbreaks]{breakurl}

\bibpunct{(}{)}{;}{a}{}{,}             
\definecolor{cobalt}{rgb}{0.06, 0.2, 0.65}
\hypersetup{
  colorlinks,
  citecolor=cobalt,
  linkcolor=[rgb]{0.8, 0.2, 1.0},
  urlcolor=cobalt,
}
\makeatletter
  \newcommandtwoopt{\citeads}[3][][]{\href{http://adsabs.harvard.edu/abs/#3}%
    {\def\hyper@linkstart##1##2{}%
     \let\hyper@linkend\@empty\citealp[#1][#2]{#3}}}
  \newcommandtwoopt{\citepads}[3][][]{\href{http://adsabs.harvard.edu/abs/#3}%
    {\def\hyper@linkstart##1##2{}%
     \let\hyper@linkend\@empty\citep[#1][#2]{#3}}}
  \newcommandtwoopt{\citetads}[3][][]{\href{http://adsabs.harvard.edu/abs/#3}%
    {\def\hyper@linkstart##1##2{}%
     \let\hyper@linkend\@empty\citet[#1][#2]{#3}}}
  \newcommandtwoopt{\citeyearads}[3][][]%
    {\href{http://adsabs.harvard.edu/abs/#3}
    {\def\hyper@linkstart##1##2{}%
     \let\hyper@linkend\@empty\citeyear[#1][#2]{#3}}}
\makeatother

\def\angstrom{\textrm{A\kern -1.3ex\raisebox{0.6ex}{$^\circ$}}}

\def\HII{\hbox{H~$\scriptstyle\rm II $~}}

\def\HeII{\hbox{He~$\scriptstyle\rm II $~}}
\def\HeIII{\hbox{He~$\scriptstyle\rm III $~}}

\def\be{\begin{equation}} 
\def\ee{\end{equation}} 
\def\ba{\begin{eqnarray}} 
\def\ea{\end{eqnarray}} 
\def\gtsima{$\; \buildrel > \over \sim \;$}
\def\ltsima{$\; \buildrel < \over \sim \;$}
\def\gsim{\lower.5ex\hbox{\gtsima}} 
\def\lsim{\lower.5ex\hbox{\ltsima}}
\def\prosima{$\; \buildrel \propto \over \sim \;$} 
\def\simgt{\lower.5ex\hbox{\gtsima}} 
\def\simlt{\lower.5ex\hbox{\ltsima}} 
\def\simpr{\lower.5ex\hbox{\prosima}}

\definecolor{mkcolor}{HTML}{01abdf}
\definecolor{apcolor}{HTML}{b3003b}
\definecolor{afcolor}{HTML}{01bdff}
\definecolor{lvcolor}{HTML}{ff9933}
\definecolor{colorecommenta}{HTML}{9a24d1}

\usepackage{mathrsfs}  
\usepackage{graphicx}
\usepackage{txfonts}
\begin{document}

\title{Explaining JWST star formation history at $z \sim 17$ by modifying $\Lambda$CDM}
\titlerunning{Star formation beyond $\Lambda$CDM up to $z\sim 17$}
\author{
	O.\ Sokoliuk\inst{1,2,3}\thanks{Corresponding author: \url{oleksii.sokoliuk@mao.kiev.ua}}\orcidlink{0000-0003-4503-7272}
	}
\authorrunning{O. Sokoliuk}
\institute{
	\tiny Main Astronomical Observatory of the National Academy of Sciences of Ukraine, 27 Akademik Zabolotny St., Kyiv, 03143, Ukraine \label{1}
	\and
	\tiny Astronomical Observatory, Taras Shevchenko National University of Kyiv, 3 Observatorna St., 04053 Kyiv, Ukraine \label{2} 
    \and 
    \tiny Department of Physics, University of Aberdeen, Aberdeen AB24 3UE, UK \label{3}\\
	}
 \date{Received: \today}

 
\abstract{
Recent cosmological observations indicate a $5\sigma$ discrepancy between the values of the Hubble constant $H_0$ derived from late and early universe probes. A further possible tension at the $\sim 3\sigma$ level arises from different measurements of $\sigma_8$. These measurements suggest the existence of new physics. Here, we explore several theories of modified gravity that may help to resolve these cosmological tensions. These include a family of phenomenological modified theories, where only Newton's gravitational constant and the Einstein-Boltzmann equations are affected. We consider one particular class of these theories: cosmologies with varying growth index $\gamma$ and varying dark energy Equation of State (EoS) $w_\Lambda$. We also consider the normal branch of the Dvali-Gabadadze-Porrati (nDGP) model as well as $k$-mouflage gravity, which involves a non-trivially coupled scalar field. Our main aim is to narrow down the modified gravity landscape by constraining each model using high-redshift JWST data. Several probes are considered in this work: Stellar Mass Function (SMF), Stellar Mass Density (SMD), Star Formation Rate Density (SFRD) and Ultra-Violet Luminosity Function (UVLF) along with the Epoch of Reionization (EoR). The theories we have considered are parameterized using $r_c$ (the scale length on which 5-dimensional gravity transitions to 4-dimensions, for nDGP), $\beta$ (associated with the Weyl re-scaling of the metric tensor) and $K_0$ (quantifying aspects of the scalar field, for $k$-mouflage). The analysis carried out in this paper provides new constraints on these parameters. We find that generally, the choice of $r_c\gtrsim 10^{3.5}$ Mpc is preferred for nDGP, while $\beta\sim0.1$, $K_0\gtrsim 0.9$ is favored for $k$-mouflage. Moreover, in the context of phenomenological gravity, phantom-like dark energy EoS $w_\Lambda\lesssim -1$ is preferred over the quintessence.
}
\maketitle
%

\section{Introduction}\label{sec:1}
Following the discovery of the late-time accelerated expansion of the universe through the distance moduli of supernovae \citep{1998AJ....116.1009R}, the simple $\Lambda$ Cold Dark Matter ($\Lambda$CDM) theory was established as a standard cosmological model, as it both accurately explained the expansion of the universe and matched most of the observations at the time. With recent technological advances, uncertainties in the data have been largely reduced, making it possible to obtain a constraint on cosmological parameters with a percent accuracy. Such a level of precision revealed several discrepancies within the $\Lambda \rm CDM$ model. For more details on that matter, see \cite{2022NewAR..9501659P, 2022JHEAp..34...49A, 2021JHEAp..32...28A, 2016IJMPD..2530007B, 2021CQGra..38o3001D} and references therein. There exist several cosmological tensions. Of particular concern are the values of Hubble parameter $H_0$, and cosmic variance $\sigma_8$. These tensions indicate a significant deviation for measured parameters from late and early universe probes, such as supernova distance moduli, Cosmic Microwave Background (CMB), and Baryon Acoustic Oscillations (BAO). Specifically, a recent study has found that \textit{Planck} and \textit{S$H_0$ES} measurements of $H_0$ deviate at $>5.9\sigma$ significance \citep{2021MNRAS.502.2065D}, suggesting the presence of new physics. 

Another important tension arises from the context of Big-Bang Nucleosynthesis (BBN). $\Lambda$CDM predicts significantly lower lithium abundance than expected from BBN models \citep{2011ARNPS..61...47F}, with a tension of up to $4\sigma$. Yet another diffuculty for $\Lambda$CDM was found in the collision velocity of two galaxy clusters, jointly named as an "El Gordo". The measured colliding velocity for the cluster appears to exceed the upper bound provided by the standard cosmological model \citep{2021MNRAS.500.5249A}. Many other problems within the fiducial cosmology exhibit smaller tensions of order $\sim3\sigma$ \citep{2022NewAR..9501659P}. 

There are other issues apart from cosmological tensions that challenge $\Lambda$CDM. For instance, horizon and flatness problems require an extreme level of fine-tuning, such that $\rho/\rho_c\sim 10^{-53}$ at present, which can be provided by the inclusion of an inflationary phase around $t\sim 10^{-35}$s \citep{1994PhRvD..49.3830H}. Inflation is usually considered to be a part of the standard cosmological model, but it has several shortcomings. Firstly, it includes an extremely large set of solutions that fit the observational data, creating an "inflationary landscape" \citep{2013JCAP...11..040M}. Secondly, inflationary spacetimes have incomplete past geodesics, which hints on the existence of the initial singularity \citep{2003PhRvL..90o1301B}. Some theories beyond $\Lambda$CDM can address the fine-tuning without resorting to inflation or by solving the singularity problem (see \cite{2017PhR...692....1N} for review). Other propositions to solve the flatness problem include the varying speed of light \citep{1999CQGra..16.1435B} and additional dimensions \citep{2002PhRvD..66b3519G}. Hence, at minimum, one needs to modify the early universe physics within $\Lambda $CDM to match the observational data.

From the quantum perspective, Lagrangian density for $\Lambda$CDM is linear in the Ricci scalar, $R$, preventing it from being canonically quantizable \citep{1974AIHPA..20...69T}. Specifically, a Feynman diagram for a gravitational field interacting with a scalar field shows divergencies at a one loop level. As a result, $\Lambda$CDM lacks the required properties to be a viable theory of quantum gravity, which can pose a problem as it is expected that quantum gravitational effects dominate at $t<t_{\rm Pl}$. Additionally, predictions for the value of the cosmological constant, derived from the Quantum Field Theory (QFT) and cosmological observations differ by up to 122 orders of magnitude, commonly referred to as the cosmological constant problem (see \cite{1989RvMP...61....1W, 2012CRPhy..13..566M} for further discussion).

\begin{figure*}
    \centering
    \includegraphics[width=0.75\linewidth]{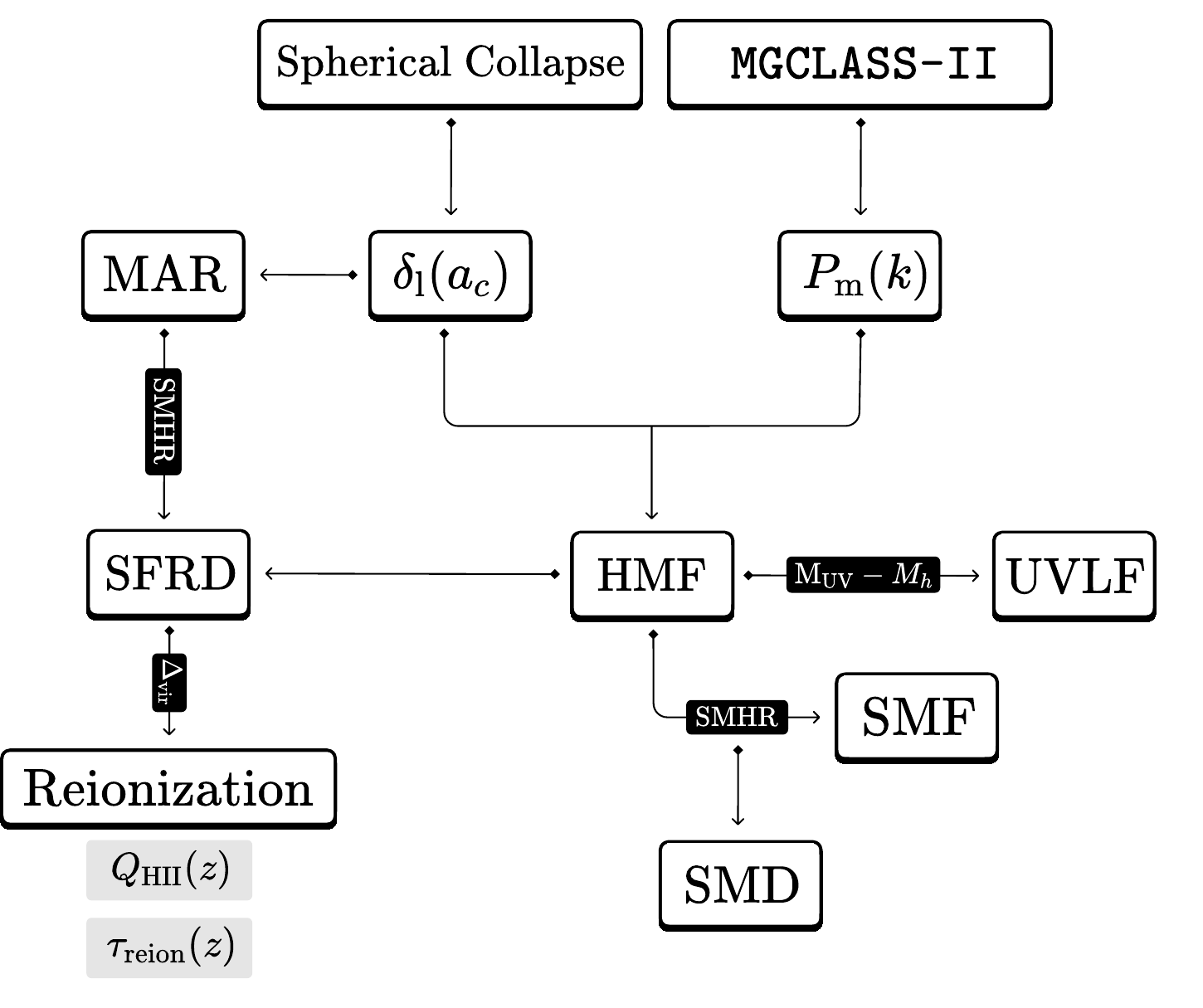}
    \caption{Structure of this paper}
    \label{fig:intro}
\end{figure*}

As shown in \cite{1977PhRvD..16..953S}, the renormalization issue can be resolved by adding higher-order terms, such as the contraction of a Riemann tensor, $R_{\mu\nu\alpha\beta}R^{\mu\nu\alpha\beta}$, to the gravitational sector, making the theory renormalizable at the one-loop level. However, this approach suffers from the presence of Weyl ghosts. A model that avoids this issue is the $R^2$ model, a promising candidate for a fiducial cosmology. It also serves as a viable model of cosmological inflation due to the presence of massive scalaron \citep{1980PhLB...91...99S}, but it remains UV-incomplete. 

Interestingly, modified gravity can also partially address most of the problems in classical cosmology. For instance, \cite{2021arXiv210512582S} found that numerous models of modified gravity can alleviate both Hubble and $\sigma_8$ tensions to an acceptable significance of $\lesssim2\sigma$. Moreover, theories like the well-known $f(R)$ one have successfully unified both the inflationary and late time accelerated expansion phases \citep{2011PhR...505...59N,2003PhRvD..68l3512N}. On the other hand, some special cases of the Gauss-Bonnet theory passed the Solar System and other tests on larger scales, addressed the hierarchy problem \citep{2006PhRvD..73h4007C} and the UV-completion at the quantum level. Another interesting approach to go beyond $\Lambda$CDM is to construct theories based on various affine connections. For example, the choice of a Weitzenb\"ock connection leads to the presence of the so-called torsion and the emergence of teleparallel gravitation \citep{2009PhRvD..79l4019B}, which can mimic the behavior of dark energy without actually introducing any other fields apart from the Standard Model (SM) Lagrangian. As noted in \cite{2022arXiv221006810T}, some attempts have also been made to approach the cosmological constant problem with the help of modified gravity. However, we still lack the understanding of the nature of $\Lambda$ in both the classical and quantum sense and thus we cannot properly assess the reason behind the cosmological constant problem. These examples demonstrate that the use of modified gravity may be very effective in solving long-standing problems in modern cosmology and high-energy physics. Next, we will discuss the specific models that have been chosen in the present work and the motivation behind our choice.

\textit{Phenomenological modifications to gravity:} these represent the simplest way to minimally modify $\Lambda$CDM model. Such modifications avoid adding any additional fields, instead making only slight alterations, e.g. setting a time-dependent cosmological constant or changing its EoS, and adjusting the lensing equations. Interestingly, phenomenological theories can significantly impact the non-linear evolution of matter, and consequently it's power spectrum, which is a key focus of this work. However, these remain minor adjustments to $\Lambda$CDM, allowing us to preserve many successful features of the fiducial model, such as mass conservation and the preservation of the Cosmological Principle at scales greater than $\sim100$ Mpc. These features may be absent in theories that depart from $\Lambda$CDM significantly (e.g., those with different affine connections, new matter fields, or non-minimal interactions between sectors). The exact formalism for phenomenological modified gravity was first introduced in the pioneering works of \cite{2008JCAP...04..013A,2007PhRvL..99n1302Z} and then rigorously tested with the CMB data by the \cite{2016A&A...594A..14P}. Being phenomenological, these models lack a strong theoretical motivation. However, the functions representing the effective gravitational constant and gravitational slip were derived taking into account cosmological observations. Hence, as phenomenological gravity modifies solely the dark energy component, all of the affected quantities will revert to $\Lambda$CDM values as $z\to \infty$, similarly to other effective dark energy models. This will ensure the fact that beneficial features of the $\Lambda$CDM in the early universe will be preserved.

\textit{$w\boldsymbol{\gamma}$CDM model:} this is another, rather interesting minor modification of $\Lambda$CDM. It affects both the dark energy equation of state, allowing us to explore the behavior of phantom and quintessence cosmologies by varying $w_\Lambda$, and also the evolution of matter overdensity via the growth index $\boldsymbol{\gamma}$. This theory is also largely phenomenological. However, unlike the $\mu-\Sigma$ parameterization discussed in the Section (\ref{sec:2}), $w\boldsymbol{\gamma}$CDM generally does not respect the law of baryon conservation. The methodology on the implementation of varying $\boldsymbol{\gamma}$ into cosmological background was first introduced in \cite{2010PhRvD..81j4023P} and later extended to $w_\Lambda\neq -1$ cases in \cite{2022JCAP...05..030S}.

\textit{nDGP theory}: the normal branch of the Dvali-Gabadadze-Porrati (DGP) theory is a widely-used modified gravity model. It introduces an additional scalar field for a screening of gravity. This way, one can retrieve General Relativity (GR) at small scales and high overdensities. Key properties of nDGP gravity and effects of the braneworld formalism on different observable quantities in modern cosmology were discussed in the review paper \cite{2010LRR....13....5M}. It is one of the few theories for which the full non-linear power spectrum was derived using the high-resolution $N$-body simulations (for instance, see \cite{2016RPPh...79d6902K, 2009PhRvD..80l3003S}). Baryonic effects on the power spectrum have also been quantified through the series of complex simulations of galaxy formation and evolution \citep{2021MNRAS.503.3867H}. Studies show that some particular cases of nDGP pass CMB, SN Ia and BAO, Integrated Sachs-Wolfe (ISW), $f\sigma_8$ tests \citep{2009PhRvD..80f3536L, 2016MNRAS.457.2377Z, 2019arXiv190803430B, 2016PhRvD..94h4022B}. However, it is worth mentioning that this model still requires the presence of small amount of dark energy to be viable, which may be considered as a major flaw.

\textit{$k$-mouflage theories:} the final model of modified gravity that we consider is the class of $k$-mouflage theories. It is a generalisation of the $k$-essence cosmologies, and in a physical sense it is close to Jordan-Brans-Dicke (JDB) gravity, as both models use a non-canonical scalar field to drive the accelerated expansion of the universe. Unlike JBD gravity, which is generally unscreened, $k$-mouflage always has a screening mechanism. However, this topic is still actively discussed, as JBD can be taken as a special limit of Horndeski scalar-tensor theory that is prone to screening \citep{2014PhRvL.113a1101A}. In a way, this mechanism is similar to nDGP, but the scales at which screening occurs are vastly different \citep{Brax:2014wla}. $k$-mouflage was first introduced in \cite{2009IJMPD..18.2147B}, and its cosmological properties at both linear and non-linear scales were explored in a series of works \citep{Brax:2014wla, Brax:2015pka, 2014PhRvD..90b3508B, 2015PhRvD..92d3519B}. These studies showed that in some cases, $k$-mouflage can rival $\Lambda$CDM given a constrained set of parameters and an appropriate form of the non-canonical scalar field potential $K(\chi)$, where $\chi$ is the scalar field. However, at smaller scales, such as within the Solar System, $k$-mouflage becomes highly non-linear and unstable due to the presence of higher-order modes in the field equation. This makes it challenging to decide on the validity of a theory on such scales \citep{2014PhRvD..90l3521B}.

Aside from the problems already mentioned, another major obstacle is the modification of gravity itself. The CANTATA network \citep{2021arXiv210512582S} reports that there are tens, if not hundreds of modified gravity theories that can provide viable cosmological evolution and potentially solve Hubble tension. However, as mentioned in \cite{2016A&A...594A..14P}, some of those "viable" theories may arise due to statistical bias in the data. The issue is that there are too many models, and we have yet to find an effective way to test those models and distinguish the only few that are actually viable. One of the key aims of this work is to use JWST data and determine the viability of the models that we consider, hopefully narrowing the so-called "modified gravity landscape".

Our manuscript is structured as follows: Section (\ref{sec:1}) provides a clear motivation for this study, a brief overview of $\Lambda$CDM model and it's modifications, as well as a literature survey of past studies on this topic. Section (\ref{sec:2}) introduces the basic formalism of modified gravity, including the background field equations, and first-order linear perturbations of energy density, with modified values of effective gravitational constant and Bardeen potentials. We then introduce each modified gravity theory in a greater detail within the same section and show the exact form of functions $\mu(a,k)$ and $\gamma(a,k)$, which are implemented into the altered version of \texttt{CLASS}\footnote{\href{https://lesgourg.github.io/class_public/class.html}{\texttt{CLASS} homepage}} code. Section (\ref{sec:3}) presents derivation of a Halo Mass Function (HMF) and the critical overdensity threshold $\delta_{c}$ for each theory. In Section (\ref{sec:4}), we present a similar formalism for the Stellar Mass Function (SMF), Stellar Mass Density (SMD). Section (\ref{sec:5}) introduces the Ultra-Violet Luminosity Function (UVLF), while Section (\ref{sec:6}) the Epoch of Reionization (EoR). Section (\ref{sec:7}) shows the observational data and theoretical predictions along with a detailed discussion on the constrained parameter space. Section (\ref{sec:8}) provides some concluding remarks. Figure (\ref{fig:intro}) 
presents a step-by-step flow chart for each quantity of interest derived in this work.

\section{A class of MG Theories}\label{sec:2}
In the current section, we introduce models of modified gravitation used throughout this paper and various cosmological observables that are related to the respective models. We adopt the \textit{Planck2018} TT,TE,EE+lowE+lensing+BAO cosmological parameters \citep{2020A&A...641A...6P}
\begin{equation*}
    \Omega_b h^2 = 0.02242 \pm  0.00014,  \;  \Omega_c h^2 = 0.11933 \pm 0.00091
\end{equation*}
\begin{equation*}
    \ln(10^{10}A_s) = 3.047 \pm 0.014, \;  n_s = 0.9665 \pm 0.0038
\end{equation*}
\begin{equation*}
    H_0  = 67.66 \pm 0.42\, \mathrm{km/s/Mpc}, \; Y_{\rm He} = 0.243 \pm 0.023
\end{equation*}
\begin{equation*}
    N_{\rm eff} = 3.046, \; \tau =  0.0561 \pm 0.0071, \;  \sigma_8 =  0.8102 \pm 0.0060
\end{equation*}
Ideally, a more detailed analysis would involve the application of each modified gravity model we consider to the Planck (or other) cosmological datasets, to derive new parameters in each case. Other studies \citep{2018MNRAS.476.3263L,2022MNRAS.512.1967R} have considered the implications of this assumption, noting that there should be no significant alterations to the main conclusions. 

We do not use the natural system of units $G_N=c=\hbar=1$, but rather work in the units of solar mass $M_\odot$, which is required to derive observable quantities mentioned in the previous section. Also, the metric signature is taken to be mostly positive, i.e. $\mathrm{sig}g=(-,+,+,+)$. Such a choice is purely a matter of preference, but can also be motivated while working in both relativistic and non-relativistic QFT.
\subsection{Phenomenological Modified Gravity}
\label{sec:2.1}
The first model we consider is a phenomenological Modified Gravity (MG). Throughout this paper, we will not depart from the Friedmann–Lemaître–Robertson–Walker (FLRW) line element. We add a small scalar perturbation in a Newtonian conformal gauge:
\begin{equation}
    ds^2 = a^2(\tau)\bigg[-(1+2\Psi)d\tau^2+\sum_{i,j}(1-2\Phi)\delta_{ij}dx^idx^j\bigg],
\end{equation}
where $a(\tau)$ is the scale factor. $\Psi$ and $\Phi$ are Bardeen potentials, that depend both on conformal time $\tau$ and conformal coordinates $\mathbf{x}=(x_1,x_2,x_3)$. To describe the matter content of the universe, we adopt a perfect fluid energy-momentum tensor of form:
\begin{equation}
    T^{\mu\nu} = (\rho+p)U^\mu U^\nu + pg^{\mu\nu}.
\end{equation}
Let $\mu,\nu\in\{0,...,3\}$ be free indices, $\rho$, $p$ represent the energy density and isotropic pressure of the perfect fluid respectively, and $\mathbf{U}=(1,0,0,0)$ denote the timelike 4-vector. Now we can perturb the perfect fluid energy-momentum tensor \citep{2019JCAP...05..001Z}:
\begin{equation}
\begin{gathered}
    T^0_0 + \delta T_0^0 = - \rho(1+\delta), \quad\quad\quad\;\;\;\;  \rm (time-time) \hfill\\
    T^0_i + \delta T_i^0 = - (\rho+p)v_i ,\quad\quad\quad\;\;\;  \rm (time-spatial) \hfill \\
    T^i_j + \delta T_j^i = (p+\delta p)\delta^i_{j}+\pi^i_j.\quad\quad\,  \rm (spatial-spatial) \hfill \hfill
    \end{gathered}
    \label{eq:conservation}
\end{equation}
Here, $v_i$ denotes the velocity field, while $\delta T$, $\delta p$ are the perturbed parts of energy-momentum tensor and pressure respectively. Besides, $\pi^i_j$ is the spatial part of anisotropic energy-momentum tensor, that is also traceless, such that $\mathrm{Tr}\,\mbox{\boldmath$\pi$}=0$. Additionally, $\delta=\overline{\rho}/\rho$ represents the overdensity of the perfect fluid field. Since the energy-momentum conservation, $\nabla_\mu T^\mu_\nu$ is the key feature of most cosmologies, one can derive a set of field equations based on it in the Fourier space \citep{2006ApJ...648..797B}:
\begin{equation}
    \dot{\delta} = -(1+w)(\theta -3 \dot{\Phi})-3\mathcal{H}\left(\frac{\delta p}{\delta \rho}-w\right)\delta,
    \label{eq:6}
\end{equation}
\begin{equation}
    \dot{\theta}=-\mathcal{H}(1-3w)\theta -\frac{\dot{w}}{1+w}\theta +\frac{\delta p/\delta \rho}{1+w}k^2\delta-k^2\sigma+k^2\Psi.
    \label{eq:7}
\end{equation}
Following the usual notation, we define $\theta=\mathrm{div}\,\mathbf{v}$. Parameter $w$ is the matter Equation of State (EoS), that is taken to be a constant. Moreover, $k$ is the wavenumber, typically measured in the units of $\rm Mpc^{-1}$. Finally,
\begin{equation}
    \mathcal{H}=\frac{1}{a}\frac{da}{d\tau}
\end{equation}
is the conformal Hubble parameter. Equations (\ref{eq:6}) and (\ref{eq:7}) can be used to study any kind of decoupled fluid, such as Cold Dark Matter (CDM), photon-baryons, or massive neutrinos. However, aforementioned equations still contain several unknowns, requiring two additional relations for potentials $\Phi$ and $\Psi$:
\begin{equation}
    k^2\Phi = -4\pi G a^2 \rho \Delta.
    \label{eq:9}
\end{equation}
This equation is derived by combining the time-time and the divergence of the time-spatial components of the conservation equation (\ref{eq:conservation}). An additional unknown is introduced while deriving the exact form of the potential $\Phi$ in Eq. (\ref{eq:9}). It is exactly the the gauge invariant density contrast:
\begin{equation}
    \rho\Delta = \rho\delta +\frac{3\mathcal{H}}{k^2}(\rho+p)\theta.
\end{equation}
From the traceless part of the spatial-spatial component of the conservation equation, we derive the following equality:
\begin{equation}
    k^2(\Phi-\Psi)=12\pi G a^2 (\rho+p)\sigma.
    \label{eq:11}
\end{equation}
Combining (\ref{eq:11}) and (\ref{eq:9}), we also obtain the parameterization of $\Psi$ in terms of  $\Delta$:
\begin{equation}
    k^2\Psi = -4\pi G a^2(\rho\Delta + 3(\rho+p)\sigma).
    \label{eq:12}
\end{equation}
Equations (\ref{eq:6}), (\ref{eq:7}), and (\ref{eq:11}), (\ref{eq:12}) provide a closed system, governing the evolution of density contrast, velocity divergence, and both Bardeen potentials. These equations can be used as an input for Einstein-Boltzmann solvers \texttt{CAMB} \citep{2000ApJ...538..473L}, and \texttt{CLASS} \citep{2011arXiv1104.2932L}, their numerous modifications. Here, we are primarily interested in the evolution of those quantities within the phenomenological modified gravity. So far, we have only used $\Lambda$CDM background cosmology. In certain modified gravity theories (for instance, family of scalar-tensor models), the relation between $\Psi$ and $\Phi$ (\ref{eq:11}) is altered. This is effectively represented by an additional, "fifth" force between particles, which accelerates (or decelerates) the development of the large scale structure of the universe. These changes can be introduced by adding two functions of time $\tau$ and scale $k$ into the evolution equations:
\begin{equation}
\begin{gathered}
    k^2\Psi = - 4\pi G\mu(a,k) a^2[\rho\Delta + 3(\rho+p)\sigma],\hfill \\
        k^2(\Phi-\gamma(a,k)\Psi)=12\pi G\mu(a,k) a^2 (\rho+p)\sigma.\hfill
\end{gathered}
\end{equation}
The function $\mu(a,k)$ either reduces or enhances the gravitational force between particles (i.e., scales the gravitational constant $G_{\rm N}$). In a similar manner, $\gamma(a,k)$ scales Bardeen potential $\Psi$. In the literature, $\gamma(a,k)$ is often referred to as a gravitational slip function, and it is not directly related to any cosmological observable \citep{2008JCAP...04..013A,2013PhRvD..87b3501A}. This makes it challenging to constrain such a function using present-day observational datasets, such as CMB, BAO, OHD or Pantheon. The issue can be addressed by adopting a slightly different notion of phenomenological modified gravity \citep{2019JCAP...05..001Z}:
\begin{equation}
        k^2(\Phi+\Psi)=-4\pi G\Sigma(a,k) a^2[2\rho\Delta +3(\rho+p)\sigma].
\end{equation}
The function $\Sigma(a,k)$ relates the sum of Bardeen potentials to the gauge-invariant density contrast. In an absence of anisotropic stress, it is given by an equality $\Sigma=\mu(1+\gamma)/2$. To derive the matter power spectrum and transfer functions, from the beginning of the epoch of reionization ("cosmic dawn") to the present time, we will use the \texttt{MG-CLASS} code \citep{2022JCAP...05..030S}, which adopts $\mu-\Sigma-\gamma$ parameterization. Modified gravity background is turned on at $z=99$, beyond that point, it is assumed that the background is represented by the $\Lambda$CDM. At this stage, all we know about $\mu$ and $\Sigma$ functions is that they are time and scale dependent. To perform numerical analysis, their exact forms must be properly specified. \texttt{MG-CLASS} implemented a couple of parameterizations. The first one that we are going to use is the Planck parameterization \citep{2009PhRvD..79h3513Z}:
\begin{equation}
\begin{gathered}
    \mu(k,a) = 1+f_1(a)\frac{1+c_1\mathcal{H}/k^2}{1+\mathcal{H}/k^2},\hfill \\
    \gamma(k,a) = 1+f_2(a)\frac{1+c_2\mathcal{H}/k^2}{1+\mathcal{H}/k^2}.\hfill
\end{gathered}
\label{eq:13}
\end{equation}
Here, $i\in\{1,2\}$, and $f_i$ represents the set of functions controlling the amplitude of scale deviations, and depends on the dimensionless dark energy cosmological parameter and free parameter $E_{ii}$:
\begin{equation}
    f_i(a) = E_{ii}\Omega_\Lambda(a).
\end{equation}
\begin{figure*}[!htbp]
    \centering
    \includegraphics[width=0.95\linewidth]{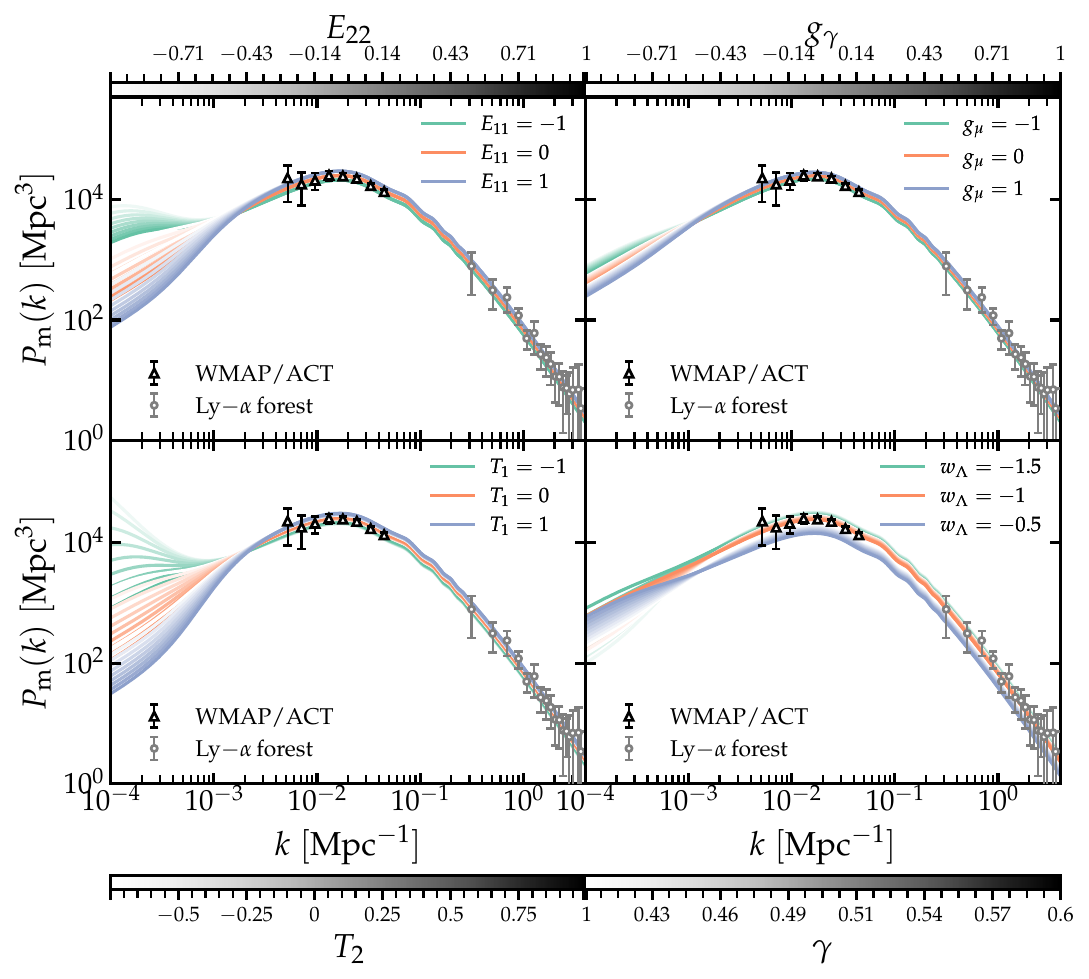}
    \caption{Matter power spectrum predictions for phenomenological modified gravity versus observational data with varying values of $E_{11}$, $E_{22}$ (first subplot), $g_\mu$ and $g_\gamma$, constant $n=1$ (second subplot), for varying values of $T_{1}=T_3$ and $T_2=T_4$, constant $n=1$ (third subplot), running growth index $\boldsymbol{\gamma}$ as well as dark energy EoS $w_\Lambda$ (fourth subplot)}
    \label{fig:1}
\end{figure*}
\begin{figure*}[!htbp]
    \centering
    \includegraphics[width=0.95\linewidth]{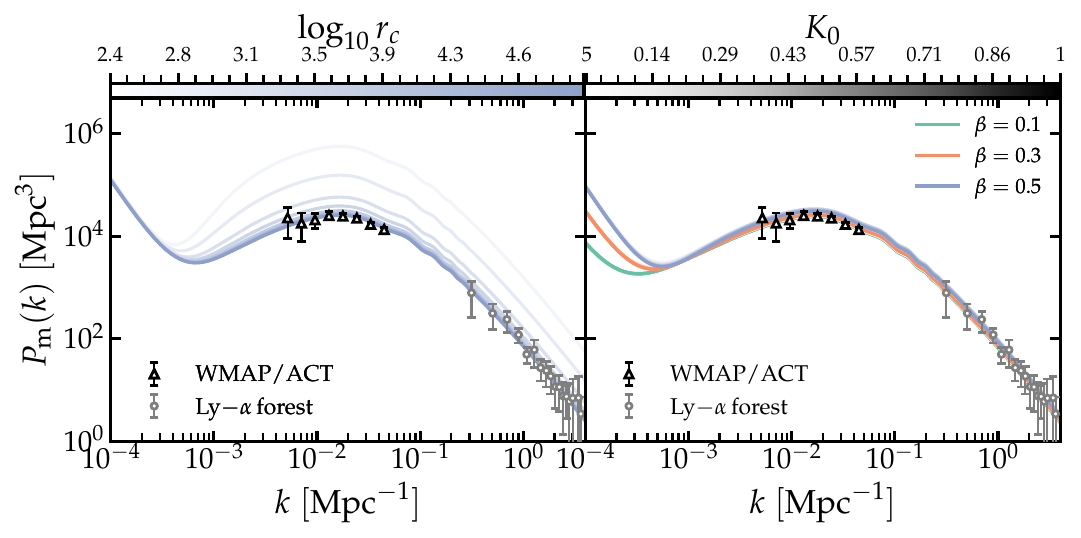}
    \caption{Figure (\ref{fig:1}) continued, but for the nDGP gravity with varying $r_c$ (first subplot) and $k$-mouflage gravity with varying $\beta$, $K_0$ (second subplot)}
    \label{fig:2}
\end{figure*}
\noindent As $f_i$ vanish at early times, given that $\lim_{a\to0}\Omega_\Lambda(a)=0$, $\Lambda$CDM is recovered at sufficiently high redshifts. Following the pioneering work of \cite{2022JCAP...05..030S}, we assume that functions $\mu$ and $\gamma$ are independent of wavenumber, setting $c_i=1$. We consider all possible permutations of two sets, $\{1,-1,0\}\in E_{11}$, $\{-1,...,1\}\in E_{22}$ with $|E_{22}|=10$, resulting in 30 possible realizations of Planck parameterization.

The second parameterization, often referred to as "z\_flex", is independent of cosmological observables and always recovers GR at early times \citep{2017PhRvD..96b3542N,2022JCAP...05..030S}:
\begin{equation}
\begin{gathered}
    \mu(a) = 1+g_\mu(1-a)^n-g_\mu(1-a)^{2n},\hfill\\
   \gamma(a) = 1+g_\gamma(1-a)^n-g_\gamma(1-a)^{2n},\hfill
\end{gathered}
\label{eq:15}
\end{equation}
$g_{\mu}$, $g_{\gamma}$, and $n$ are free parameters representing additional degrees of freedom to be constrained. To reduce the parameter space, we impose the constraint $n=1$.

The last kind of phenomenological modified gravity theory does not have any particular motivations from high energy physics or cosmology, but is still an interesting choice. This model relies on the Taylor expansion of $\mu(a)$ and is usually referred to as a Dark Energy Survey (DES) parameterization \citep{2022JCAP...05..030S}:
\begin{equation}
\begin{gathered}
    \mu(a) = 1+T_1\Omega_\Lambda^n+T_2\Omega_\Lambda^{2n},\hfill \\
    \gamma(a) = 1+T_3\Omega_\Lambda^n+T_4\Omega_\Lambda^{2n}.\hfill
    \end{gathered}
    \label{eq:16}
\end{equation}
Here, $T_i,\,i=\{1,2,3,4\}$ and $n$ represent the set of free parameters to be constrained in the later sections. With five free parameters, the calculations become computationally-expensive. To reduce the parameter space, we take $n=1$, similarly to the z\_flex case, and adopt $T_1=T_3$, $T_2=T_4$. In this way, we are left with only two free parameters. The parameter spaces for both z\_flex and DES models are sampled similarly to the Planck parameterization. Using the \texttt{MG-CLASS} code, we plot the matter power spectrum $P_{\rm m}(k)$ for each phenomenological modified gravity theory under our consideration. For comparison, we also show \textit{WMAP/ACT} measurements \citep{2002PhRvD..66j3508T} as well as observational data from Ly-$\alpha$ forest spectroscopy of distant quasars \citep{2002ApJ...581...20C,2002MNRAS.334..107G}. The resulting plots are located on the Figure (\ref{fig:1}).

Having introduced all of the necessary quantities needed for the case of phenomenological modified gravity, we can proceed to review other models.
\subsection{Varying $\boldsymbol{\gamma}$ and $w_{\Lambda}$}
In this case, we depart from the $\Lambda$CDM model by varying both the growth index of linear perturbations $\boldsymbol{\gamma}$ and Dark Energy Equation of State $w_{\Lambda}$. Please note that this $\boldsymbol{\gamma}$ is the scalar and has nothing to do with function $\gamma(k,a)$ from Eqs. (\ref{eq:13}), (\ref{eq:15}) and (\ref{eq:16}). The growth index can be defined in terms of a growth rate as follows:
\begin{equation}
    f = \frac{d\ln D}{d\ln a} \approx \Omega_m^{\boldsymbol{\gamma}} (a).
\end{equation}
Here $D(a)=\delta(a)/\delta(1)$ represents the normalized growth of perturbations, obeying a second-order partial differential equation of the form:
\begin{equation}
    \delta''(a) + \bigg[2+\frac{H'(a)}{H(a)}\bigg]\delta'(a)=-\frac{3}{2}\frac{\Omega_m(a)}{E(a)}\mu(a)\delta(a).
    \label{eq:24}
\end{equation}
According to \cite{2022JCAP...05..030S}, for this model, the evolution of both Bardeen potentials is governed by a system of equations:
\begin{equation}
\begin{gathered}
    -k^2 \Psi(a) = \frac{4\pi G}{c^4}a^2 \overline{\rho}(a)\delta(a)\mu(a)\\
    \Phi(a) = \gamma(a)\Psi(a)\hfill.
\end{gathered}
\end{equation}
By combining the system of equations above with the evolution equation for $\delta(a)$ from (\ref{eq:24}), the exact form of $\mu(a)$ can be derived \citep{2022JCAP...05..030S}:
\begin{equation}
\begin{gathered}
    \mu(a) = \frac{2}{3}\Omega^{\boldsymbol{\gamma}-1}_m(a)\bigg[\Omega^{\boldsymbol{\gamma}}_m(a)+2-3\boldsymbol{\gamma}+3\bigg(\boldsymbol{\gamma}-\frac{1}{2}\bigg)\\
    \times(\Omega_m(a)+(1+w_\Lambda)\Omega_\Lambda)\bigg].
\end{gathered}
\end{equation}
Plugging this function into \texttt{MG-CLASS} provides us with the accurate predictions for the matter power spectrum and CMB temperature angular power spectrum. In the $\Lambda$CDM limit, $\boldsymbol{\gamma}\approx0.545$ \citep{1990ApJS...74..831L,1991MNRAS.251..128L}. We will take values of $\boldsymbol{\gamma}$ that deviate slightly around the $\Lambda$CDM prediction. For the dark energy equation of state parameter, we are going to probe both the phantom dark energy $w_\Lambda<-1$ and quintessence $-1<w_\Lambda<0$ cases. This specific model has been chosen due to the fact that as it appears, the most recent cosmological data have a preference for $\gamma$ beyond $\Lambda$CDM \citep{2024PhRvD.109d3528S}.
\subsection{nDGP model}
The widely known Dvali-Gabadadze-Porrati (DGP) model belongs to the large class of braneworld theories, where one assumes the existence of an $n$-dimensional brane embedded into a $d$ dimensional bulk with $d>n$. Most braneworld models, such as Kaluza-Klein \citep{1921SPAW.......966K,1926ZPhy...37..895K} or Randall-Sundrum \citep{1999PhRvL..83.4690R} take the simplest case, $n=4$ and $d=5$, which we are going to follow as well. The brane is represented by the usual FLRW geometry, while bulk spacetime is Minkowski. The DGP model introduces a single additional free parameter, namely a crossover scale $r_c$ measured in Mpc. The whole theory can hence be well-defined by an Einstein-Hilbert action integral \citep{2000PhLB..485..208D}:
\begin{equation}
\begin{gathered}
    \mathcal{L}_{\rm DGP} = M_*^3\int_{\mathcal{M}_{\rm bulk}} d^4x dy\sqrt{-h}\mathcal{R}\\
    +\int_{\mathcal{M}_{\rm brane}} d^4x \sqrt{-g}(\mathcal{E}_4+M_p^2R+\mathcal{L}_{\rm M}).
\end{gathered}
\end{equation}
In this action, $M_*$ is a five-dimensional Planck's constant, which determines  the mass scale of the bulk dimension. Besides, $y$ is the coordinate of the bulk dimension itself, and $h=\det h_{\mu\nu}$ is the determinant of the 5d spacetime metric $h_{\mu\nu}$. Similarly, $R$ and $\mathcal{R}$ represent the Ricci scalar, based on the four and five dimensional manifolds respectively. Besides, $\mathcal{E}_4=M_p^2\Lambda$ accounts for the cosmological constant that lives on a brane. For the sake of generality, the action defined above includes the contribution from a Lagrangian density $\mathcal{L}_{\rm M}[g,\Gamma,\chi_i]$, which can contain various matter fields, both perfect fluid and not. Each arbitrary field is represented by $\chi_i,\;i\in\mathbb{N}$ (note that in our case, as previously mentioned, we consider matter to be a perfect fluid). The relationship of those fields with gravity and spacetime geometry is determined by $g=\mathrm{det}\, g_{\mu\nu}$ and $\Gamma$ (Levi-Cevita affine connection). The integration is performed over a four-dimensional Lorentzian manifold $(\mathcal{M}_{\rm brane} = \mathbb{R}^+\times \mathbb{S}^3)$ while dealing with the brane and over a five-dimensional manifold $(\mathcal{M}_{\rm bulk} = \mathbb{R}^+\times \mathbb{S}^3 \times \mathbb{S}^1)$ while working with the bulk.
Under the least-action principle, it is straightforward to derive a modified Friedmann equation, governing the evolution of a DGP universe:
\begin{equation}
    H^2 = \frac{8\pi G}{3}\rho_{\rm tot} +\epsilon\frac{H}{r_c}+\frac{\Lambda}{3}.
    \label{eq:31}
\end{equation}
This study focuses on the normal branch of the DGP model (nDGP), which corresponds to $\epsilon=-1$. Under the quasistatic approximation, commonly used by Einstein-Boltzmann solvers, the $\mu-\gamma$ parameterization reduces to \citep{2006JCAP...01..016K}:
\begin{equation}
\begin{gathered}
    \mu(a) = 1+\frac{1}{3\beta},\hfill \\
    \gamma(a) = \bigg(1-\frac{1}{3\beta}\bigg)\bigg/\bigg(1+\frac{1}{3\beta}\bigg), \; \beta = 1 + 2Hr_c\bigg(1+\frac{\dot{H}}{3H^2}\bigg).\hfill 
    \end{gathered}
\end{equation}
The only unknown here is the Hubble parameter, which can be obtained using the modified Friedmann equation (\ref{eq:31}) and rewritten in a more familiar form:
\begin{equation}
    H(a) = H_0\sqrt{\Omega_{m0}a^{-3}+\Omega_{r0}a^{-4}+\Omega_\Lambda+\Omega_{rc}}-H_0\sqrt{\Omega_{rc}}.
    \label{eq:34}
\end{equation}
where $\Omega_{rc}=1/(4H_0^2r_c^2)$ with normalization condition $\sqrt{\Omega_{m0}+\Omega_{r0}+\Omega_\Lambda+\Omega_{rc}}-\sqrt{\Omega_{rc}}=1$. The temporal derivative of the Hubble parameter is derived under the assumption that in nDGP, the conservation law for energy-momentum tensor is the same as in GR.
\subsection{$k$-mouflage}
The final model we consider, $k$-mouflage model, is based on $k$-essence, a model showing dynamical attractor behavior. Similarly to the Jordan-Brans-Dicke case, it incorporates a scalar field, but with an arbitrary kinetic term contribution \citep{2009IJMPD..18.2147B}:
\begin{equation}
    S=\int_\mathcal{M}d^4x\sqrt{-g}\bigg[\frac{M_p^2}{2}R+\mathscr{M}^4K(\chi)\bigg]+S_{\rm M}.
\end{equation}
Here, $\mathscr{M}$ represents the scalar field energy scale, and $K(\chi)$ is the non-standard kinetic term, whereas $\chi$ is the rescaled scalar field
\begin{equation}
    \chi = -\frac{g^{\mu\nu}\partial_\nu \phi \partial_\nu\phi}{2\mathscr{M}^4}.
\end{equation}
The action integral above is expressed in terms of the Jordan frame metric tensor $g_{\mu\nu}$. Einstein and Jordan frames are related via the Weyl conformal transformation $\widetilde{g}_{\mu\nu}=A^2(\phi) g_{\mu\nu}$. The field equations for $k$-mouflage model are rather complex:
\begin{equation}
    \begin{gathered}
        3M_p^2H^2 = \frac{A^2}{(1-\epsilon_2)^2}\bigg(\rho+\frac{\mathscr{M}^4}{A^4}(2\chi K'-K)\bigg), \hfill \\
        -2M_p^2\dot{H}=\frac{A^2}{(1-\epsilon_2)^2}\bigg(\rho+\frac{\mathscr{M}^4}{A^4}(2\chi K'-K)+\mathscr{M}^4K\bigg),\hfill 
    \end{gathered}
    \label{eq:37}
\end{equation}
where
\begin{equation}
\begin{gathered}
    \epsilon_1 = \frac{2}{K'}\bigg(\epsilon_2M_p\bigg(\frac{\phi}{\ln a}\bigg)^{-1}\bigg)^2, \hfill \\
    \epsilon_2 = \frac{d\ln A}{d\ln a}. \hfill
\end{gathered}
\label{eq:epsilon}
\end{equation}
From Eq. (\ref{eq:37}), the Hubble parameter is derived in terms of the dimensionless mass densities for each matter and energy component, and a Hubble constant $H_0$ 
\begin{equation}
    H(a) = AH_0(1-\epsilon_2)\sqrt{\Omega_{m0}a^{-3} + \Omega_{r0}a^{-4} + \Omega_{\phi0}\frac{\rho_{\phi}}{\rho_{\phi0}}}.
\end{equation}
The scalar field energy density is given by $\rho_{\phi}=\mathscr{M}^4/A^4(2\chi K'-K)$. In $k$-mouflage models, mass densities follow the following relation: $\Omega_{m}+\Omega_{r}+\Omega_\phi = (1-\epsilon_2)^2$. The model can be conveniently expressed as a set of functions that determine the deviation from $\Lambda$CDM \citep{2019JCAP...05..027B}:
\begin{equation}
\begin{gathered}
    \mu(a) = (1+\epsilon_1)A^2,\hfill \\
    \gamma(a)=\frac{1-\epsilon_1}{1+\epsilon_1}A^2. \hfill
\end{gathered}
\end{equation}
The only unknowns left are the choice of $K(\chi)$, $A(\phi)$, and $\phi$. \texttt{MG-CLASS} assumes that the kinetic term is non-linear, while $A(\phi)$ is linear \citep{Brax:2014wla,2022JCAP...05..030S}:
\begin{equation}
    K(\chi) = -1+\chi + K_0\chi^m, \quad A(\phi) = 1+\beta\phi.
    \label{eq:44}
\end{equation}
Here, $K_0$, and $\beta$ are free parameters of the model, with $m=2$, making $K(\chi)$ of quadratic order. Plugging Eq. (\ref{eq:44}) into Eq. (\ref{eq:epsilon}), we can simplify those quantities as follows:
\begin{equation}
    \epsilon_1 = \frac{2\beta^2}{1+mK_0\chi^{m-1}},\quad \epsilon_2 = \frac{a\beta}{1+a\beta}.
    \label{eq:45}
\end{equation}
Thus, the rescaled scalar field is $\chi = A^2\dot{\phi}^2/2\mathscr{M}^4$.
The value of the scalar field at present is determined using the continuity equation:
\begin{equation}
    \Omega_{\phi 0 } = (1-\epsilon_{2,0})^2-\Omega_{r0}-\Omega_{m0}.
    \label{eq:46}
\end{equation}
From Equation (\ref{eq:45}), at the present time, $\epsilon_{2,0}=0$, giving $\Omega_{\phi 0 } = 1 -\Omega_{r0}-\Omega_{m0}$, which is very similar to the $\Lambda$CDM case. We can also fix the mass scale by manipulating the definition of the scalar field. However, in that case one needs make use of a set of initial conditions, and following the work \cite{Brax:2014wla}, we adopt $\phi(1)=0$, from which it follows that:
\begin{equation}
    \Omega_{\phi0} = \frac{\rho_{\phi0}}{3M_p^2H_0^2} = \frac{\mathscr{M}^4 (\chi_0 (3 K_0 \chi_0+2)+1)}{3M_p^2H_0^2}.
\end{equation}
Equating this to (\ref{eq:46}) allows us to solve for $\mathscr{M}^4$:
\begin{equation}
    \mathscr{M}^4 = -\frac{3 H_0^2 M_p^2 (\Omega_{m0}+\Omega_{r0}-1)}{\chi_0 (3 K_0 \chi_0+2)+1}.
\end{equation}
Finally, the exact form of the scalar field $\phi$ should be specified. It is calculated from the Klein-Gordon equation \citep{Brax:2015pka}:
\begin{equation}
    \frac{d}{dt}\bigg[A^{-2}a^3\frac{d\phi}{dt}K'\bigg]=-a^3\rho_m\frac{d\ln A}{d\phi}.
    \label{eq:49}
\end{equation}
We have defined each modified gravity theory that is of special interest to us in the current work. Now, we can proceed and derive some observables, such as halo mass function with the use of \texttt{MG-CLASS}. Similarly to the phenomenological theories of gravitation, for our future analysis we consider a range $\log_{10}r_c\in[2.5,4]$ (10 steps) and $\beta\in[0.1,0.5]$ (3 steps), $K_0\in [0,1]$ (10 steps). 

\section{Extracting HMF from linear matter power spectra}\label{sec:3}
To constrain models beyond $\Lambda$CDM, we first need to obtain the Halo Mass Function (HMF). This cosmological observable can be derived using the extended Press-Schechter (EPS) approach, introduced in the pioneering work of \citep{1974ApJ...187..425P}:
\begin{equation}
\frac{dn}{d\log M_h} = -\frac{\overline{\rho}}{M_h}f(\nu)\frac{d\log\sigma}{d\log M_h}.
\label{eq:50}
\end{equation}
Here, $M_h=(4\pi/3)\overline{\rho}(cR)^3=(cR)^3\Omega_{m}H_0^2/2 G$ is the halo mass, $R$ is it's corresponding radius. Besides, $\overline{\rho}$ is the background density, and $\sigma(R)$ is the variance of a density field,  $\nu=\delta_c/\sigma(R)$. We take $c\approx3.3$, which seems to reasonably well approximate the HMF even beyond the $\Lambda$CDM \citep{2021JCAP...12..044P}. For the $\Lambda$CDM cosmology, the linear density threshold is $\delta_c(z)= 1.686/D(z)$. We assume for first-crossing distribution to have a Sheth-Tormen form \citep{10.1046/j.1365-8711.1999.02692.x}:
\begin{equation}
f(\nu) = A\sqrt{\frac{2\nu^2}{\pi}}(1+\nu^{-2p})e^{-\nu^2/2}.
\end{equation}
Here, $A=0.3222$ and $p=0.3$ are dimensionless constants, that were derived from the best-fits to the simulational data. The variance of matter fluctuations is calculated as follows:
\begin{equation}
\sigma^2(R,z) = \int \frac{k^2}{2\pi^2}P_l(k,z)W_F^2(k,R)dk.
\end{equation}
We define $P_l(k,z)$ as the linear matter-power spectrum, while $W_F(k,R)$ is the so-called window function:
\begin{equation}
W_F(k,R) = (1+(kR)^\beta)^{-1}.
\end{equation}
Recent results show that the value $\beta=4.8$ provides a good fit to the data, extracted from $N$-body simulations \citep{Leo:2018odn}. As mentioned earlier, the result $\delta_c(z)=1.686/D(z)$ applies only to a $\Lambda$CDM model and will vary for its modifications (e.g., see \cite{2010PhRvD..81f3005S,2013PhRvD..87l3511L} as an example). To derive the value of a density contrast for modified $\Lambda$CDM, we need to explore the physics of spherical collapse.
\subsection{Modelling spherical collapse}
Under the spherical collapse formalism, the top-hat distribution is often assumed for simplicity, which is defined in the following way:
\begin{equation}
    \rho_c(r) - \rho = \begin{cases}
        \delta \rho &  \mathrm{if} \, r\leq R \\
        0 & \mathrm{if} \, r > R
    \end{cases}
    \label{eq:48}
\end{equation}
This indicates that the sphere of some radius $R$ is filled homogeneously, having the density $\rho_c>0$. Since the density is non-zero, after some time, the sphere should eventually collapse due to its own gravitational pull in the absence of pressure, as we are considering matter to be dust-like. However, after the collapse, Birkhoff's theorem may no longer apply within modified gravity theory. As a result, Equation (\ref{eq:48}) will not hold, thus some non-vanishing density contrast can appear outside of $R$ (see \cite{2009MNRAS.395L..25M} for the implications of that violation). The applicability of this theorem will be discussed separately for each model in the next several sections.

To derive the value of the critical overdensity, we first introduce the set of non-linear continuity and Euler equations in real space. The first equation, namely continuity follows from the fact that the covariant gradient of the stress-energy tensor vanishes:
\begin{equation}
    \nabla_\mu T^\mu_{\nu}= \partial_\mu T^\mu_\nu +\Gamma^\alpha_{\;\alpha\mu}T^\mu_\nu - \Gamma^\alpha_{\;\mu\nu}T^{\mu}_\alpha=0.
\end{equation}
Here, $\Gamma^\alpha_{\;\mu\nu}$ is the Christoffel symbol of the second kind:
\begin{equation}
    \Gamma^\alpha_{\;\mu\nu} = \frac{1}{2}g^{\alpha\beta}(\partial_\nu g_{\beta\mu}+\partial_\nu g_{\beta\mu} - \partial_\beta g_{\mu\nu}).
\end{equation}
Assuming that $\nu=0$, we can derive the continuity equation in the non-linear regime:
\begin{equation}
    \dot{\rho} +3H(\rho+p) = 0.
\end{equation}
Inside and outside of the spherical region, we have different expansion rates, and therefore different scale factors:
\begin{equation}
    \dot{\rho}_c+3\mathscr{H}(\rho_c+p_c)=0.
\end{equation}
With $\mathscr{H}=\dot{r}/r$ being the local Hubble parameter within the sphere, and $r$ being the local scale factor respectively. Now, it is possible to define the non-linear overdensity as
\begin{equation}
    \delta_{\rm nl} + 1 = \frac{\rho_c}{\rho}.
    \label{eq:53}
\end{equation}
Following the work of \citep{2007JCAP...11..012A}, we apply the first-order temporal derivative to the equation above
\begin{equation}
    \dot{\delta}_{\rm nl} =  3\frac{(H-\mathscr{H})(\rho_c+p_c)}{\rho}.
    \label{eq:54}
\end{equation}
For a pressureless fluid (such that $p=p_c=0$, i.e. dust), Eqs. (\ref{eq:53}) and (\ref{eq:54}) reduce to:
\begin{equation}
    \dot{\delta}_{\rm nl} = 3(1+\delta_{\rm nl}) (H-\mathscr{H}) .
\end{equation}
Taking the temporal derivative of the relation above again, we get the equation that governs the evolution of the fluid in the non-linear regime within the FLRW universe
\begin{equation}
    \ddot{\delta}_{\rm nl} = 3(1+\delta_{\rm nl})(\dot{H}-\dot{\mathscr{H}})+\frac{\dot{\delta}_{\rm nl}^2}{1+\delta_{\rm nl}}.
    \label{eq:54}
\end{equation}
Recall that the second Friedmann equations within inside and outside of a sphere are:
\begin{equation}
    \dot{H} = -\frac{4\pi G}{3}\rho-H^2, \quad \dot{\mathscr{H}} = -\frac{4\pi G}{3}\rho_c - \mathscr{H}^2.
\end{equation}
From this, we deduce that
\begin{equation}
    \dot{H}-\dot{\mathscr{H}}=\frac{4\pi G}{3}(\rho_c-\rho)+\mathscr{H}^2-H^2.
\end{equation}
Substituting these expressions into Eq. (\ref{eq:54}), we obtain the final equation for the growth of $\delta_{\rm nl}$:
\begin{equation}
    \ddot{\delta}_{\rm nl}+2H\dot{\delta}_{\rm nl}-4\pi G\mu(a)\rho\delta_{\rm nl}(1+\delta_{\rm nl})=\frac{4}{3}\frac{\dot{\delta}_{\rm nl}^2}{1+\delta_{\rm nl}}.
\end{equation}
Here transform the gravitational constant $G\to G\mu(a)$. This deviation arises due to the modification of gravity. Since solving for cosmic time is challenging, we instead express the above equation in the terms of the scale factor. For that we use a change of variables $\ddot{\delta}_{\rm nl}\to a^2H^2\delta''_{\rm nl}+a(H^2+\dot{H})\delta'_{\rm nl}$, $\dot{\delta}_{\rm nl}\to aH\delta'_{\rm nl}$, and $\rho=3\Omega_{m0}H_0^2a^{-3}$. This works within $\Lambda$CDM and modified $\Lambda$CDM theories with no non-minimal couplings between different sectors. Now, we have
\begin{equation}
\begin{gathered}
    \delta''_{\rm nl}(a) + \bigg(\frac{3}{a}+\frac{H'}{H}\bigg)\delta'_{\rm nl}(a) - \frac{3}{2}\frac{\Omega_{m0}\mu(a)}{a^5E^2}\delta_{\rm nl}(a)(1+\delta_{\rm nl}(a))\\
    =\frac{4}{3}\frac{\delta'^2_{\rm nl}(a)}{1+\delta_{\rm nl}(a)}.
\end{gathered}
\label{eq:60}
\end{equation}
The linear regime overdensity can be derived by combining Equations (\ref{eq:6}) and (\ref{eq:7}), as shown in \cite{2001PhRvD..63f3504E}:
\begin{equation}
    \ddot{\delta}_{\rm l} +2H\dot{\delta}_{\rm l}-4\pi G\mu(a)\rho\delta_{\rm l} \simeq 0.
\end{equation}
By applying the same transformation as for Eq. (\ref{eq:60}), this can be expressed in the terms of the scale factor \citep{2007PhRvD..76b3514T,2015PhRvD..92b3013N,2019PhRvD..99d3516A}:
\begin{equation}
    \delta''_{\rm l}(a) + \bigg(\frac{3}{a}+\frac{H'}{H}\bigg) \delta'_{\rm l}(a)  -\frac{3}{2}\frac{\Omega_{m0}\mu(a)}{a^5E^2}\delta_{\rm l}(a) = 0.
    \label{eq:52}
\end{equation}
Thus, linear overdensity evolution is recovered in Equation (\ref{eq:60}) if all contributions from $\mathcal{O}(\delta_{\rm nl}^2)$ terms are ignored. To determine the exact value of the critical overdensity, we first find the scale factor $a_c$ at which $\delta_{\rm nl}$ diverges (i.e. spherical collapse occurs), and then solve Eq. (\ref{eq:52}) such that $\delta_{\rm l}(a_c)=\delta_c$. Initial conditions for the non-linear differential equation (\ref{eq:60}) are assumed to match with the Einstein-de Sitter (EdS) case, since at an initial scale factor $a_i=10^{-5}$, matter dominates ($\Omega_m(a_i)\sim 1$) and modified gravity does not play a significant role. Consequently, at $a_i$, the density perturbation takes the form $\delta_{\rm nl}(a_i)=c_1a_i^{-3/2}+c_2a_i$ \citep{Malekjani:2015pza,Malekjani:2016mtm}, and:
\begin{equation}
   \delta'_{\rm nl}(a_i) = \frac{\delta_{\rm nl}(a_i)}{a_i}.
    \label{eq:63}
\end{equation}
Those initial conditions are expected to lead to the collapse at the point when $\delta_{\rm nl}(a_c)\simeq 10^7$. Hence, we need to derive the value of $\delta_{\rm nl}(a_{i})$, to ensure that at the desired scale factor, $\delta_{\rm nl}(a_c)\simeq 10^7$. This can be achieved by adopting the algorithm presented in \cite{2017JCAP...10..040P}.

\begin{figure}[!htbp]
    \centering
    \includegraphics[width=0.95\linewidth]{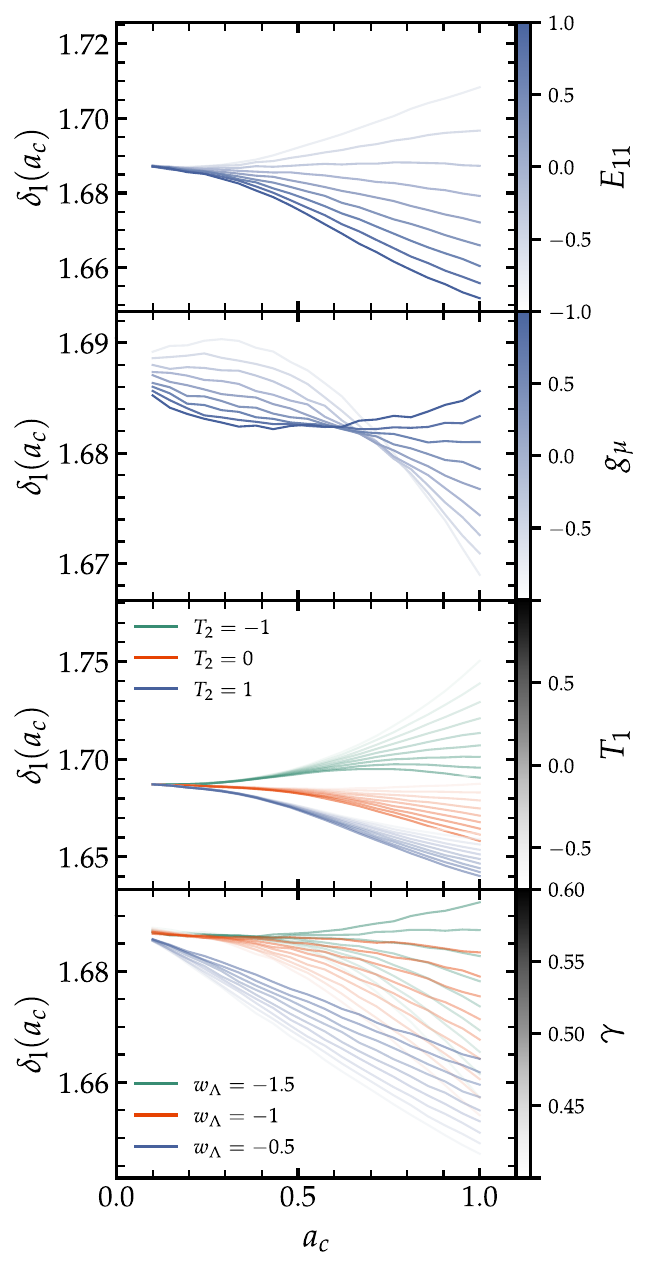}
    \caption{Linear density contrast for the spherical collapse occurring at the time $a_c$ within DES, z\_flex and Taylor expanded DES, varying $w_\Lambda$ and $\boldsymbol{\gamma}$ phenomenological theories of gravitation}
    \label{fig:3}
\end{figure}

Since the matter field shows a linear behavior at $a\ll1$, initial conditions from Eq. (\ref{eq:63}) apply for both non-linear and linear cases. We show the linear density threshold for each phenomenological modified gravity in Figure (\ref{fig:3}). 
As can be seen, the nDGP and $k$-mouflage models are not included in the plots. This is because both of those theories involve the use of Vainshtein screening, which suppresses the so-called "fifth" force at small scales, where GR should be recovered due to the Cosmological Principle. The derivation of Eq. (\ref{eq:60}) for screened models is different, and the methodology for that will be discussed in the next subsection.
\subsection{$\delta_c$ and screening of gravity}
As it was previously mentioned, the process of obtaining $\delta_c$ within the nDGP gravity differs from the phenomenological theories due to the unique nature of this theory. On subhorizon scales, it was shown that matter within nDGP gravity follows the evolution equation \citep{2010PhRvD..81f3005S}:
\begin{equation}
    \ddot{\delta}_{\rm nl} - \frac{4}{3}\frac{\dot{\delta}_{\rm nl}^2}{1+\delta_{\rm nl}}+2H\dot{\delta}_{\rm nl} = (1+\delta)\nabla^2\Psi.
    \label{eq:70}
\end{equation}
Bardeen's potential here deviates from the GR prescription by the so-called "brane-bending mode", a scalar field $\varphi$:
\begin{equation}
    \Psi = \Psi_N + \frac{1}{2}\varphi.
\end{equation}
The Newtonian potential obeys the standard Poisson equation, while the brane-bending mode is described by \cite{PhysRevD.75.084040}:
\begin{equation}
    \nabla^2\varphi + \frac{r_c^2}{3\beta}((\nabla^2\varphi)^2-\nabla_i \nabla_j \varphi\nabla^i \nabla^j \varphi)=\frac{8\pi G}{3\beta}\delta\rho.
\end{equation}
In \cite{2010PhRvD..81f3005S}, the full profile of the brane-bending mode was calculated, and it was demonstrated that the $\nabla^2\varphi$ is constant within the top-hat interior, similarly to the Newtonian potential $\Psi_N$. However, $\nabla^2\varphi$ retains a non-trivial dependence on the energy density perturbation \citep{2010PhRvD..81f3005S}:
\begin{equation}
    \nabla^2 \varphi = 8\pi \Delta G_{\rm eff} (R/R_*)\delta \rho.
\end{equation}
Here $\Delta G_{\rm eff}$ represents the difference between the modified gravitational constant and Newton's original gravitational constant. Additionally,
\begin{equation}
 \mu(x) = \frac{G_{\rm eff}(x)}{G} = \frac{2}{3\beta}\frac{\sqrt{1+x^{-3}}-1}{x^{-3}} + 1.
 \label{eq:74}
\end{equation}
and \citep{2014PhRvD..90b3508B}
\begin{equation}
    R_\star = \bigg(\frac{16G\delta M r_c^2}{9\beta^2}\bigg)^{1/3}, \quad R_\star = \bigg(\frac{\delta M}{4\pi M_p\mathscr{M}^2}\bigg)^{1/2}.
    \label{eq:75}
\end{equation}
These two expressions represent the Vainshtein radius, a characteristic scale for nDGP and $k$-mouflage respectively, at which the screening takes effect. By combining Poisson's equations for scalar field and Newtonian potential, and using Eq. (\ref{eq:70}), we get the evolution equation for non-linear overdensity in terms of the scale factor:
\begin{equation}
\begin{gathered}
    \delta_{\rm nl}''(a) + \bigg(\frac{3}{a}+\frac{H'}{H}\bigg)\delta'(a)-\frac{3}{2}\frac{\Omega_{m0}\mu(R/R_\star)}{a^5E^2}\delta_{\rm nl}(1+\delta_{\rm nl})\\
    =\frac{4}{3}\frac{\delta_{\rm nl}'^2(a)}{(1+\delta_{\rm nl})}.
\end{gathered}
\end{equation}
This closely resembles Eq. (\ref{eq:60}). However, in the case when gravity is screened, the effective gravitational constant not only depends on the scale factor but also on the screening radius. The condition $R\gg R_\star$ implies the fact that $G_{\rm eff}/G_N=1+1/3\beta$ holds for nDGP, as noted in the previous subsection.
Thus, in the case when the actual radius significantly exceeds the Vainshtein radius, the effects of screening can be ignored, and the equations from the previous subsection can be used to derive $\delta_c$. However, if one assumes that $\delta\rho/\rho\gg1$, then $G_{\rm eff}/G_N\propto (R/R_\star)^{3/2}$ \citep{2010PhRvD..81f3005S}. Linearizing gravity substantially simplifies the calculations, and as shown in the paper \cite{Schmidt:2009sv}, the matter power spectrum for both full and linearized nDGP is approximately the same. Besides, at high enough redshift $z\gg 0$, linearized and full nDGP gravity coincide, since $R/R_\star\to \infty$. But for consistency, we are going to keep the full nDGP physics and linearize it only for $\delta_{\rm l}$ calculations. Our results for $\delta_c$ may differ from the work of \cite{2010PhRvD..81f3005S}, as we do not adjust the effective dark energy equation of state in the nDGP case, because $H(a)$ would then correspond exactly to the $\Lambda$CDM prediction. Instead, we prefer to use the framework of Eq. (\ref{eq:34}).

Similarly to the nDGP scenario, Vainshtein-like screening also occurs within the $k$-mouflage cosmology, as mentioned in \citep{Brax:2014wla}. In this work, we use the fact that at small enough scales, as in $\Lambda$CDM, each spherically collapsing sphere can be treated separately. This approach was first noted by the Lemaitre and its detailed version for $k$-mouflage theory was presented in \citep{2014PhRvD..90b3508B}. Thus, at the scales $ctk/a\gg 1$, our approximation works well. These scales exceed those of typical galaxy clusters, and therefore one could compute such quantities as halo mass function under the assumption that the screening is absent. On the other hand, the derivation for $\delta_{\rm nl}$ gets complex beyond these scales, as screening should be taken into account as well.
\begin{figure}[!htbp]
    \centering
    \includegraphics[width=0.95\linewidth]{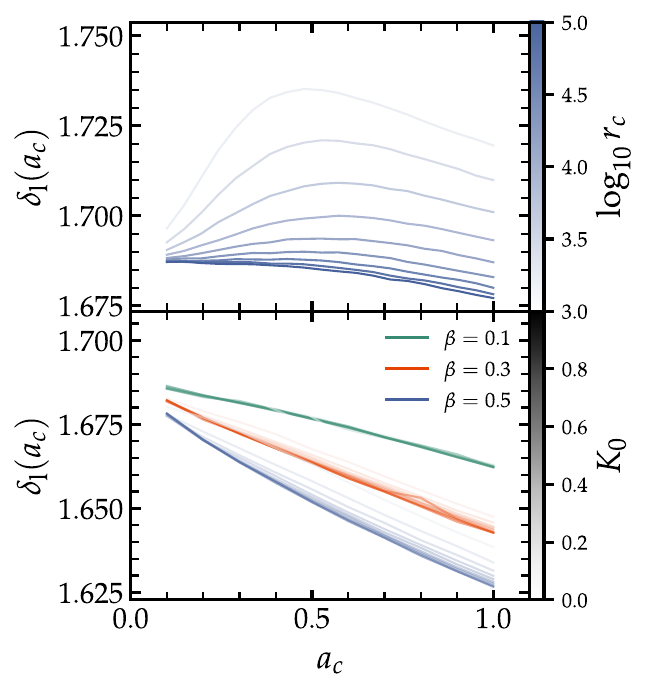}
    \caption{Linear density contrast for the spherical collapse occurring at the time $a_c$ within screened MOG models, namely nDGP and $k$-mouflage}
    \label{fig:4}
\end{figure}
With derivation of $d n/d\log M_h$ from the Eq. (\ref{eq:50}), we can now finally determine the observationally constrainable quantities, as outlined in the next subsection.
\section{Stellar mass function and density estimates}\label{sec:4}
The Halo Mass Function (HMF) cannot yet be constrained by observational data directly; instead, existing constraints are only being derived using cosmological simulations. One cannot state the robust conclusions on the validity of a theory using the cosmological simulations alone. However, several quantities can be derived from the HMF, that in turn are constrained by the observational data. These are the Stellar Mass Function (SMF) and the corresponding Stellar Mass Density (SMD). In this study, we are particularly interested in JWST observations of galaxies across the redshift range $z\sim 0-17$. To obtain SMF and SMD, we apply an analytical approach, similar to the one used in \citep{2023arXiv230314239D}. In this approach, we first assume that each halo of mass $M_h$ contains mass according to the relation $M_g=(\Omega_b/\Omega_m)M_h$. Some portion of baryonic mass within the halo converts to stars via star formation processes, such that $M_\star=\epsilon_\star M_g$. Here, $\epsilon_\star$ is the Star Formation Efficiency (SFE), which varies from one model of star formation to another. Typically, the value of star formation efficiency $\epsilon_\star$ ranges from $2\%$, up to $100\%$ in the extreme cases, as discussed in a greater detail by \citep{2021ApJ...911..128K}. Most theoretical studies at high redshifts $z\sim 5-10$ assume that $\epsilon_\star = 0.1-0.3$ \citep{2023arXiv230611993H,2023JCAP...10..072A,2023arXiv230805606G}. 

In our case, we can follow the paper \cite{2023arXiv230314239D}, where they assumed two models of star formation, but it appears that their motivation for the first model is vague, and thus it is more appropriate to assume a more complex fit for $\epsilon_\star$, derived from the computationally-expensive cosmological simulations or the observational data. For the latter, we have chosen to use an approximation, derived from several deep sky surveys in the redshift range $z\sim0-8$, as presented in the \cite{2013ApJ...770...57B, 2017MNRAS.470..651R}.

\textit{Model I}, Rodriguez-Puebla et al. approximation for $\epsilon_\star$: They introduce a parametrization of stellar mass to halo relation, which gives a median stellar mass for each halo mass bin, based upon the formalism laid out in \cite{2013ApJ...770...57B}, namely
    \begin{equation}
        \log_{10}M_\star = \log_{10}(\varepsilon M_1) + f\left(\log_{10}\left(\frac{M_h}{M_1}\right)\right) - f(0)
        \label{eq:64}
    \end{equation}
    with
    \begin{equation}
        f(x) = -\log_{10}(10^{\alpha x}+1)+\delta \frac{(\log_{10}(1+\exp(x)))^\gamma}{1+\exp(10^{-x})}
    \end{equation}
with $x=\log_{10}(M_h/M_1)$. Unlike the \cite{2013ApJ...770...57B} case, this parameterization takes into account more accurate predictions for scatter in the SMHR, that are closer to the observational data at lower redshifts (see \cite{2014ApJS..210....3M}). According to \citep{2017MNRAS.470..651R}, best-fit functions have the following form:
\begin{equation}
\begin{gathered}
		\log(\varepsilon(z)) = -1.758 +  
		\mathcal{P}(0.110,-0.061,z) 
		\nu(z)  + 
		\mathcal{P}(-0.023,0,z)\hfill\\
		\log(M_{1}(z)) = 11.548  +  
		 \mathcal{P}(-1.297,-0.026,z)\nu(z)  \hfill\\
		\alpha (z) = 1.975  + 
		 \mathcal{P}(0.714,0.042\pm0.017,z)\nu(z) \hfill\\
		\delta(z) =3.390  + 
		 \mathcal{P}(-0.472,-0.931,z)\nu(z)  \hfill\\
	\gamma(z) = 0.498 + \mathcal{P}(-0.157,0,z)\nu(z) \hfill
\end{gathered}
\end{equation}
Here, $\nu(z) = \exp(-4/(1+z)^{2})$ and $\mathcal{P}(x,y,z)=yz-xz/(1+z)$ \citep{2017MNRAS.470..651R}. We account for observational systematics and various biases in SMHR by adopting the formulation \citep{2013ApJ...770...57B}:
\begin{equation}
    \log_{10}\bigg(\frac{M_{\star,\rm meas}}{M_{\star,\rm true}}\bigg) = \mu.
    \label{eq:83}
\end{equation}
Clearly, measured and true stellar masses deviate by a factor $10^\mu$. The purpose of this offset is to account for possible uncertainties in the Initial Mass Function (IMF), the stellar population synthesis model, star formation history model, and redshift uncertainties for each galaxy in the sample. An additional parameter $\kappa$ has to be introduced into relation (\ref{eq:83}) to account for the mass-dependent evolution of active galaxies. Further details on these parameters and their best-fit values can be found in \cite{2013ApJ...770...57B}. For the fraction of passive galaxies, one can consider the formula given in \cite{2011ApJ...739...24B}.

\textit{Model II}: Another way to parameterize the Stellar Mass to Halo Relation (SMHR) is to use the well-known and fairly straightforward double power-law best fit, introduced in the pioneering work of \cite{2010ApJ...710..903M,2017MNRAS.464.1365M}:
\begin{equation}
    \epsilon_\star = \frac{\epsilon_{\star,0}}{\left(\frac{M_h}{M_{h,0}}\right)^{\gamma_{\rm lo}} + \left(\frac{M_h}{M_{h,0}}\right)^{\gamma_{\rm hi}}}.
    \label{eq:68}
\end{equation}
Here, $\epsilon_{\star,0}$ represents the star formation efficiency at the peak mass $M_{h,0}$, while $\gamma_{\rm lo}$ and $\gamma_{\rm hi}$ are power-law indices. The values of these indices either suppress or enhance the star formation rate at low and high halo masses. Based on \cite{2017MNRAS.464.1365M} and the subsequent paper \cite{2024arXiv240615548S}, we take the following values that fit the observational data up to redshifts of $z\sim 12$: $\log_{\rm 10}(M_{h,0}/M_\odot)=12$, $\gamma_{\rm lo}=-0.6$ and $\gamma_{\rm hi}=0.5$. We assume a fiducial value of $\epsilon_{\star,0}$, specifically $\epsilon_{\star,0}=0.21$. The double power-law function above is redshift independent, though the recent data in \cite{2009ApJ...696..620C} and other studies suggests that redshift dependence in SMHR should be present for the accurate representation of the observational data. The redshift-dependent baryon conversion ratio can be easily implement, following \cite{2017MNRAS.464.1365M}:
\begin{equation}
    \epsilon_{\star,p} (z) = \epsilon_{\star,0}\bigg(\frac{1+z}{7}\bigg)^{\gamma_\star}.
\end{equation}
The amplitude of star formation efficiency is fixed, but the peak halo mass can vary with redshift as follows:
\begin{equation}
    M_p (z) = M_0\bigg(\frac{1+z}{7}\bigg)^{\gamma_M}.
\end{equation}
The indices introduced above vary in the range $-1 \leq \gamma_{\star/M} \leq 1$. A redshift-independent SMHR is recovered when $\gamma_\star = \gamma_M = 0$.

\textit{Model III}: This is an extreme case using $\epsilon_\star=1$. This choice of constant baryon conversion ratio is likely not viable, since there is substantial evidence from the JWST CEERS survey that $\epsilon_\star$ should vary with redshift, particularly in the range $z\sim4-8$ \citep{2023arXiv231114804C}. Moreover, the $100\%$ conversion is not realistic and contradicts the observational data previously mentioned. Thus, this case will only be used for illustrative purposes.

In this work, we assume that the least massive halos are of mass $M_{h,\rm cut}=10^{6.5}M_\odot$, which sets the cut-off scale of the stellar mass via the SFE relation. We now can define the stellar mass function itself in terms of the halo mass function. However, while deriving the HMF we did not account for many baryonic effects that arise, for example, AGN feedback, supernova feedback, nucleosynthesis (see how those effects contribute to star formation and IMF in the analysis of \texttt{EAGLE} \cite{2015MNRAS.450.4486F}, and \texttt{STARFORGE} \cite{2022MNRAS.515.4929G} simulations). For a detailed explanation on how to include those effects into analytical models, refer to \cite{2020MNRAS.491.5083S}. Our double power-law SMHR already accounts for both stellar and AGN feedback, as mentioned in \cite{2010ApJ...710..903M}. Stellar feedback strength is encoded into the parameter $\gamma_{\rm lo}$, while AGN is encoded into $\gamma_{\rm hi}$. On the contrary, since the Behroozi et al. approach relies solely on observational data for best fits, i.e. is practically model-independent, it also implicitly accounts for any baryonic physics present.
\subsection{Stellar Mass Function}
We derive the stellar mass function using the methodology of \cite{2023arXiv230311368B}, which gives the following definition for SMF:
\begin{equation}
    \phi_{\rm fid}(M_\star) = \frac{dn}{d\log_{10} M_h}\frac{d\log_{10} M_h}{d\log_{10} M_\star}.
\end{equation}
Here we have used the halo abundance matching method, explained in \cite{2023arXiv230311368B}. The equation above would work in an ideal scenario where no scatter is present in the SMHR relation. However, observational data, especially at higher redshifts, have significant uncertainties in $M_\star$, and hence scatter. The implementation of scatter into the SMF can be done using the formula provided in \cite{2010ApJ...717..379B}: 
\begin{equation}
    \phi_{\rm true}(M_\star) = \int^\infty_{-M_\star}\phi_{\rm fid}(M_\star + \Delta M_\star)P(\Delta M_\star)d\Delta M_\star.
\end{equation}
In this equation, $P(\Delta M_\star)$ represents a log-normal Gaussian kernel of width $\sigma_{\rm sf}$. Its value is redshift-dependent and defined in \cite{2017MNRAS.470..651R}. In addition to accounting for scatter in the observational data, we must get rid of statistical biases that may be present in the measurements of stellar mass \citep{2013ApJ...770...57B} and the fraction of passive galaxies. We will not delve into details on the matter of derivation of $f_{\rm passive,obs}(M_\star,z)$, and instead refer to the Appendix C of \cite{2013ApJ...770...57B}. We can now introduce the measured SMF as follows:
\begin{equation}
    \begin{gathered}
        \phi_{\rm meas}(M_{\star}) = f_{\rm passive,obs}(M_{\star})\phi_{\rm true}(M_{\star}10^{-\mu}) \\
        + \phi_{\rm true}(M_{\star}10^{-\mu-\kappa})(1-f_{\rm passive,obs}(M_{\star}10^{-\kappa})).
    \end{gathered}
\end{equation}
The observed SMF is the measured one multiplied by an appropriate factor $c(z)$, namely the stellar mass completeness function:
\begin{equation}
    c(z) = \begin{cases}
      1&\mathrm{if}\; z<1\\
      c_i(z) + (1-c_i(1)) &\mathrm{if}\; z>1
    \end{cases}
\end{equation}
Incompleteness in the measurements of stellar masses arises mainly in the high-redshift surveys due to the non-ideal target selection criteria and many other factors. Recent studies have demonstrated that the stellar mass function completeness in CMASS BOSS \citep{2016MNRAS.457.4021L}, DESI \citep{2023arXiv231009329Y} and COSMOS \citep{2023A&A...677A.184W} surveys can be modeled through the exponential fit with two degrees of freedom: amplitude $A$ and onset redshift $z_c$ \citep{2013ApJ...770...57B}:
\begin{equation}
    c_i(z) = 1-\frac{A}{\exp(z_c-z)+1}.
    \label{eq:76}
\end{equation}
The statistical bias corrections $\kappa$, $\mu$ and constants $A$, $z_c$, are provided in  \cite{2013ApJ...770...57B}. 

Note that scatter corrections are only applied to the first two models described above, i.e. that associated with \cite{2017MNRAS.470..651R} (Eqs. \ref{eq:64}-{\ref{eq:83}}) and the double power-law SMHR models (Eq. \ref{eq:68}). Model III, representing the extreme case with $100\%$ SFE, is not directly connected to any observational data, and hence there is no scatter present. Additionally, we do not apply the survey completeness correction (\ref{eq:76}) to the double power-law SMHR, as it is calibrated on the observational UVLF data, which incorporates those corrections \textit{ab initio}.
\subsection{Stellar Mass Density}
Finally, we define another quantity of special interest: cumulative halo mass density, which can be derived using the following equation \citep{2023NatAs...7..731B}:
\begin{equation}
    \rho (>M_h,z) = \int^\infty_{M_h} dM_h \left(M_h \frac{dn(M_h,z)}{dM_h}\right).
    \label{eq:SMD}
\end{equation}
Here, $n(M_h,z)$ represents the number density of halos of mass $M_h$ per unit volume. It is not possible to integrate up to infinity, as the HMF is a numerical rather than analytical quantity. Hence, we adopt an upper limit of an integral $M_{h,\rm max}=10^{18}M_\odot$. Eq. (\ref{eq:SMD}) relates to the total mass concentrated in a halo. On the other hand, the cumulative stellar mass density only accounts for a stellar mass within a halo:
\begin{equation}
    \rho_\star(>M_\star,z) = f_\star\rho(>M_\star/f_\star,z).
\end{equation}
Here, we define $f_\star=\epsilon_\star (\Omega_b/\Omega_m)$. Note that even for our modified cosmologies, the fraction $\Omega_b/\Omega_c$ is assumed to be constant throughout the redshift range we consider, because both $\Omega_b\propto a^{-3}$ and $\Omega_c\propto a^{-3}$. This assumption will not hold if one supposes some non-standard interactions between dark and baryonic sectors, causing the loss of baryons through cosmic time. With this, we can now proceed to derive another probe.
\section{Ultra-Violet Luminosity Function}\label{sec:5}
Another useful observable which can be utilized is the Ultra-Violet Luminosity Function (UVLF). To derive the UVLF from the HMF, we use the relation similar to the one used for the stellar mass function, namely 
\begin{equation}
\phi(M_{\rm UV}) = \frac{dn}{dM_{\rm UV}} = \frac{dn}{d\log_{10}M_h}\bigg|\frac{d\log_{10}M_h}{d\rm M_{\rm UV}}\bigg|    .
\end{equation} 
Here, $\rm M_{\rm UV}$ represents the dust-attenuated magnitude in the UV range. In order to fully specify $\rm M_{\rm UV}$, several things must be done. We begin by mapping between the Star Formation Rate (SFR), characterized by the value of $\dot{M}_\star$, and UV galaxy luminosity $L_{\rm UV}$:
\begin{equation}
    \dot{M}_{\star} = \kappa_{\rm UV} L_{\rm UV}.
    \label{eq:90}
\end{equation}
Typically, SFR is expressed in units of $\rm M_\odot\rm\, yr^{-1}$, and UV luminosity in units of $\rm erg\,s^{-1}\,\rm Hz^{-1}$. Note that some previous studies retain the dependence on $h$, reduced Hubble constant. This has been avoided in the present paper. In Equation (\ref{eq:90}), $\kappa_{\rm UV}=1.15\times 10^{-28}$ is the conversion constant, introduced in the review article by \cite{Madau:2014bja}. This conversion factor is derived from the well-known Salpenter IMF (Salpeter 1955), and uses a UV wavelength of $1500\textup{\AA}$. For details on how the value of $\kappa_{\rm UV}$ changes with redshift, pivot wavelength $\lambda_{\rm UV}$, metallicity $Z/Z_\odot$ and IMF, refer to \cite{2013pss6.book..141B, 2022ApJ...938L..10I, Madau:2014bja}. Since the star formation rate is obtained from the dark matter halo mass, we need to calculate the halo Mass Accretion Rate (MAR) $\dot{M}_h$ first.
\subsection{Mass Accretion History }
$\dot{M}_h(z)$ is typically derived from cosmological simulations, for instance see \cite{2009ApJ...707..354Z, 2021MNRAS.508..852D} for cold and warm dark matter cosmologies respectively. However, we can also achieve that analytically using the approximation derived from the aforementioned simulations by averaging over merger trees \citep{2014MNRAS.445.1713V}, or from statistical data \citep{2009MNRAS.398.1858M}. In the current study, we assume the prescription for MAR derived in \cite{2006MNRAS.372..933N}, which uses the familiar Extended Press-Schechter formalism:
\begin{equation}
    \frac{dM_h}{dz} = \sqrt{\frac{2}{\pi}}\frac{M_h}{\sigma_{q}^2-\sigma^2}\frac{\delta_c}{D^2(z)}\frac{dD}{dz}.
    \label{eq:91}
\end{equation}
In Eq. (\ref{eq:91}), $\sigma_q = \sigma(M_h(z)/q)$, where $q$ is a free parameter that ranges from $2$ to some $q_{\rm max}$. The choice of $q_{\rm max}$ depends on the cosmological model adopted, but $q$ is a free parameter, that varies within the given bound and can be derived using data from cosmological simulations. The approach taken in the present paper is to adopt the analytic expression given by \cite{2015MNRAS.450.1514C}:
\begin{equation}
\begin{gathered}
    q = 4.137\widetilde{z}_f^{-0.9476} \hfill \\
    \widetilde{z}_f =  -0.0064(\log_{10}M_0)^2+0.0237\log_{10}M_0+1.8837 \hfill.
\end{gathered}
\end{equation}
The Eq. (\ref{eq:91}) can then be transformed into a Mass Accretion History (MAH) of the form:
\begin{equation}
    M(z) = M_0(1+z)^\alpha e^{\beta z},
    \label{eq:MAH}
\end{equation}
by integrating, while noting the fact that
\begin{equation}
\begin{gathered}
    \beta = -f(M_0),\hfill \\
    \alpha = \bigg[\delta_c \sqrt{\frac{2}{\pi}}\frac{dD}{dz}\bigg|_{z=0}+1\bigg]f(M_0)\hfill.
\end{gathered}
\label{eq:alphabeta}
\end{equation}
Here, $M_0$ is the halo mass at present epoch, and the function $f(M_0)=(\sigma_{q}^2-\sigma^2)^{-1/2}|_{z=0}$ \citep{2015MNRAS.450.1514C}. The form (\ref{eq:MAH}) is motivated by the study \cite{2002ApJ...568...52W}. This study suggests that CDM halos behave as $M(z)\sim e^\beta z$ at high redshifts, and as $M(z)\sim (1+z)^\alpha$ at lower redshifts. This transition arises due to the development of matter density, that is slowed down at late times, when dark energy component dominates. The MAR is obtained from Eq. (\ref{eq:MAH}) by differentiating with respect to redshift, and multiplying the resulting expression with $dz/dt=-(1+z)H(z)$. This leads to \citep{2015MNRAS.450.1514C}:
\begin{equation}
\begin{gathered}
    \dot{M}_h(z) = 71.6\mathrm{M}_\odot \mathrm{yr}^{-1}\bigg(\frac{M(z)}{10^{12}M_\odot}\bigg)\bigg(\frac{h}{0.7}\bigg)\\
    \times [-\alpha-\beta(1+z)]E(z).
\end{gathered}
\label{eq:103}
\end{equation}
Subsequent work \citep{2015MNRAS.450.1521C} indicates that taking the mean $\langle\alpha\rangle=0.24$ and $\langle\beta\rangle=-0.75$ works well over the halo mass range $10^8\leq M_h \leq 10^{14}$ in the context of standard LCDM. Nevertheless, some theories exhibit sufficient deviations in $P_{\rm m}(k)$ from the $\Lambda$CDM theory, which in turn will lead to the deviation in $\langle\alpha\rangle$ and $\langle\beta\rangle$ values, as power spectrum is directly incorporated into the value of $f(M_0)$. To address this issue, we compute these parameters explicitly with Eq. (\ref{eq:alphabeta}), taking $\delta_c$ from Section (\ref{sec:3}) and $D(a)$ from Section (\ref{sec:2}). Parameters are defined over a discrete grid of halo masses $10^{6.5}\leq M_h[\mathrm{M}_\odot]\leq 10^{18}$ and then linearly interpolated. This way, there are no inconsistencies, even when the departure from $\Lambda$CDM is significant. 

Given the relationship in Eq. (\ref{eq:90}), we can now derive the UV luminosity. But the derived luminosity does not take into account various effects which can affect it on the way to the telescope detector. Those effects, and how they can be taken into account when calculating UVLF are discussed in the next subsection.
\subsection{Dust attenuation and other effects in UVLFs}
From (\ref{eq:90}), we compute the intrinsic UVLF by integrating the following expression \citep{2021MNRAS.504.1555M}:
\begin{equation}
    d\phi(L_{\rm UV}) =  \frac{dn}{d\log_{10}M_h}\frac{d\log_{10}M_h}{dL_{\nu}} dL_{\rm UV} .
\end{equation}
However, this expression does not account for an absolute UV magnitude, namely $\rm M_{\rm UV}$. This is important, as most observations of the UVLF are expressed in terms of it. To relate the UV luminosity to its absolute magnitude, we factor in dust attenuation, which affects UVLF significantly at $M_{\rm UV}\lesssim -20$ (see \cite{2023MNRAS.525.3254S} for comparison of intrinsic and corrected UVLF). Effects of dust are included with the well-known IR excess $L_{\rm IR}/L_{\rm UV}$ (IRX)-$\beta$ relation \citep{1999ApJ...521...64M}
\begin{equation}
    \rm IRX = 4.43 + 1.99\beta.
\end{equation}
Here, $\beta$ is the UV spectral slope. Consequently, $\beta$ can be related to the absolute magnitude $M_{\rm UV}$ as follows \citep{2023MNRAS.520...14C}:
\begin{equation}
    \beta = -0.17 \rm M_{\rm UV} -5.4.
\end{equation}
This expression was calibrated using JWST data to work within the redshift range $z\sim 4-16$. Finally, we account for the scatter arising from variation in the HMF (if one uses Sheth-Tormen \citep{10.1046/j.1365-8711.1999.02692.x}, Tinker \citep{2008ApJ...688..709T}, or any other kind of HMF approach), AGN/stellar feedback models, and IMF assumptions. Scatter is modelled by smoothing the UVLF with a Gaussian kernel of width $\sigma_{\rm UV}$. For clarification and methods, see the study by \cite{2023MNRAS.525.3254S}. As noted in \cite{2014MNRAS.444.2960D, 2016ApJ...825....5S}, at higher redshifts ($z\geq 8$), $\sigma_{\rm UV}\sim 0.4 \rm\; dex$, down from $\sigma_{\rm UV}\sim 0.5-0.6\; \rm dex$ at $z\leq4$. Additional scatter is observed due to the uncertainty in the stellar mass estimates, as the UVLF is connected to the star formation rate via Eq. (\ref{eq:90}).  \cite{2023MNRAS.525.3254S} estimate that $\sigma_{\rm UV}\sim 2.5$ dex in total at $z\sim 16$ and higher redshifts. In this work, we take $\sigma_{\rm UV}$ to be both mass and redshift dependent, following \cite{2024arXiv240513108G,2024arXiv240615548S}:
\begin{equation}
    \sigma_{\rm UV}(M_{\rm halo}) = \max [\mathcal{A}-\mathcal{B}(\log _{10}(M_{\rm halo}/M_\odot)-10)+\mathcal{C},\sigma_{\rm min}].
    \label{eq:88}
\end{equation}
In Eq. (\ref{eq:88}), we use $\mathcal{A}=1.1$, $\mathcal{B}=0.34$ and $\sigma_{\rm min}=0.2$ dex. We also set $\mathcal{C}$ to be redshift-dependent, linearly interpolated from $\mathcal{C}(z\lesssim10)=0$, $\mathcal{C}(z\sim12)=0.3$, $\mathcal{C}(z\sim14)=0.5$ \citep{2024arXiv240615548S}. Those parameters were derived on the basis of $\Lambda$CDM cosmology. Clearly, this is an approximation, as we are dealing with non-standard cosmology in this paper, but there are no estimates of $\sigma_{\rm UV}$ for our models yet available in the literature.

To calculate the UVLFs numerically, we have made use of the code \texttt{highz-empirical-variability}
\footnote{\href{https://github.com/XuejianShen/highz-empirical-variability}{https://github.com/XuejianShen/highz-empirical-variability}}, documented and firstly used in the paper \cite{2023MNRAS.525.3254S}. The code was properly modified to accommodate for the modified theories of gravity.
\subsection{Star Formation Rate Density}
We also need to calculate the Star Formation Rate Density (SFRD). It is defined in a similar way to the stellar mass density. The only difference is that we change mass inside the integral (\ref{eq:SMD}) to be the corresponding mass accretion rate. This leads to the following definition of the SFRD:
\begin{equation}
    \rho_{\rm SFR}(>M_h, z) = \int_M^\infty \frac{dn(M_h)}{dM_h}\dot{M}_\star dM_h.
\end{equation}
The SFRD will be used jointly with other quantities to obtain constraints from JWST in the next sections.
\section{Epoch of Reionization}\label{sec:6}
The Epoch of Recombination, which occurred at $z\sim 1100$, converted most of the free electrons and protons into neutral hydrogen and helium. After this epoch, with few sources of ionising radiation, the \textit{Dark Ages} lasted until $z\sim 30$, when the formation of first stars and galaxies brought it to an end. Radiation from the first luminous objects ionized neutral atoms, initiating the rise of the EoR. However, the process of reionization was not homogeneous - clouds of ionized gas were first formed at around $z\sim 12$ in massive galaxy clusters, in which there are embedded powerful sources of ionizing radiation. Voids were ionized later, at $z\sim 6$ due to the gradual increase of an ambient UV radiation.

There are several physical quantities which can be derived during the EoR, but we focus on the ionized hydrogen fraction $x_{\rm HII}$ and optical depth $\tau_{\rm reion}$. We are not concerned here with the reionization of helium, which occurs much later. A parameter of particular interest, $Q_{\rm HII}$, measures the volume filling fraction of \HII{}. The end of reionization epoch is reached as $Q_{\rm HII}\sim 1$. Note that the time evolution of $Q_{\rm HII}$ is governed by the equation below \citep{2003ApJ...588L..69W,1999ApJ...514..648M}:
\begin{equation}
    \frac{dQ_{\rm HII}}{dt} = f_{\rm esc}\frac{\dot{n}_{\rm ion}}{\overline{n}_{\rm H}}-C_{\rm HII}\alpha_{B}(T_{\rm HII})\overline{n}_{\rm H}(1+z)^3x_e.
    \label{eq:105}
\end{equation}
Here, $f_{\rm esc}$ represents the fraction of ionized photons that escape galaxies and reach the Inter-Galactic Medium (IGM). Some galaxies are very dusty - and thus the escape fraction for such objects should be very small, as most photons will be absorbed. In order to correctly reproduce predictions for $x_{\rm HII}$, we need to account for both the populations of dusty and unobscured galaxies. A paper that explores this topic further is \cite{2021Natur.597..489F}. We take the value assumed by the \texttt{SIMBA} hydrodynamical simulation suite, namely $f_{\rm esc}=0.25$ \citep{2022ApJ...931...62H}. The exact values of the escape fraction vary in the literature, but generally lie in the range $f_{\rm esc}\in[0.1,0.3]$. Additionally, there is some evidence present for $f_{\rm esc}$ evolving with redshift \citep{2012MNRAS.423..862K,2016arXiv160503970P}. 

Another unknown in Eq. (\ref{eq:105}) is the mean number density of all hydrogen species at $z=0$, denoted by $\overline{n}_{\rm H}$. $C_{\rm HII}$ is the clumping factor of \HII{}. We take it to be $C_{\rm HII}=3.0$ following \cite{2015ApJ...810..154K}. This value has been derived using a series of cosmological simulations, avoiding a redshift dependence, since the data is too uncertain to determine that reliably.

Besides, $\alpha_{B}$ represents the case-B recombination coefficient, which sums over only excited states, i.e. all states excluding the ground one. case-A recombination sums over all hydrogen states. We can mention that the choice of case-B is appropriate, since the ionized hydrogen that is produced by the recombined neutral hydrogen in the ground state is usually absorbed by nearby thick clouds of gas, and hence does not reach the observer. This is also supported by the fact that at higher redshifts, an intergalactic medium can be considered to be highly opaque \citep{2007MNRAS.377.1175D}. The recombination coefficient mainly depends on the temperature of ionized hydrogen, namely $T_{\rm HII}$. Following the work \cite{2023ApJ...947...28G}, it is fitting to assume $T_{\rm HII}=2\times 10^4$ K \citep{2015ApJ...802L..19R}, resulting in a recombination coefficient $\alpha_B = 2.5\times 10^{-13}\;\rm cm^3\;s^{-1}$. For a detailed discussion on the choice of recombination coefficient, refer to \cite{2023RNAAS...7...90N} and references therein.

The only undefined quantity left in Eq. (\ref{eq:105}) is the production rate of ionizing photons, which is usually calculated from \citep{2023ApJ...947...28G, 2021MNRAS.504.1555M}: 
\begin{equation}
    \dot{n}_{\rm ion} = N_{\rm ion} \int_{M_{\rm min}}^\infty dM_h\left(\frac{dn(M_h)}{dM_h}\dot{M}_\star\right).
    \label{eq:107}
\end{equation}
The value of $N_{\rm ion}$, standing for the total flux of Lyman-continuum photons per solar mass per second, is usually derived from a Stellar Population Synthesis code (some of the widely used options include \texttt{STARBURST99} \cite{1999ApJS..123....3L}, or alternatively, \texttt{BPASS} \cite{2017PASA...34...58E}) or from the observational data. Additionally, sophisticated and computationally expensive hydrodynamical simulations of structure formation can also be used for that purpose. For instance, in one of the recent works, \texttt{THESAN-HR} simulations \citep{2024MNRAS.527.2835S} have been used to explore the degeneracy of $N_{\rm ion}$ between CDM and WDM/FDM/sDAO models. In this paper, following the prescription of \cite{2013ApJ...768...71R, 2015ApJ...802L..19R}, we adopt the value of $N_{\rm ion}$, derived from the Hubble Ultra Deep Field (HUDF) data, such that $\log_{10}N_{\rm ion} \approx 53.14$. 

An expression for $Q_{\rm HII}$ is given next. This equation must be evaluated numerically due to the complicated nature of Hubble parameter and ionizing photon production rate. In Eq. (\ref{eq:105}), $x_e$ represents the abundance of electrons and is determined by $x_e = n_e/(n_{\rm H} + n_{\rm He})$, assuming an absence of metals. Here, $n_e$ is the corresponding free electron number density. It is difficult to calculate the value of $n_e$ without making assumptions regarding the ionization of other elements. For simplicity we assume that during the period of $20\geq z \geq 6$, $^3\rm He$ is being ionized only into \HeII (with \HeIII forming after the EoR). We can write the fraction as:
\begin{equation}
    x_e = Q_{\rm HII}\bigg(1+\frac{Y_{\rm He}}{4}\bigg),
\end{equation}
where the abundance of helium $Y_{\rm He}$ was given in Section (\ref{sec:2}).  The other, non-analytical solution to this problem is to run a full-physics radiative transfer simulation, and calculate the electron abundance at each time step explicitly. Since the methodology derived here is analytic, we do not use this approach here. Now, one can express the ionized hydrogen filling fraction in the following way \citep{2016MNRAS.457.4051K}:
\begin{equation}
\begin{gathered}
    Q_{\rm HII}(z_0) = \frac{1}{\overline{n}_{\rm H}}\int^\infty_{z_0}\frac{f_{\rm esc}\dot{n}_{\rm ion}}{(1+z)H(z)}\exp\bigg[-\alpha_{\rm B}(T_{\rm HII})\overline{n}_{\rm H}C_{\rm HII}\\
    \times \left(1+\frac{Y_{\rm He}}{4}\right)\int^z_{z_0}dz'\frac{(1+z')^2}{H(z')}\bigg]dz.
\end{gathered}
\label{eq:93}
\end{equation}
In the expression above, we have opted to express $Q_{\rm HII}$ as a redshift dependent quantity, rather than using dependence on the scale factor. The reason for this is that we later compare calculations of Eq. (\ref{eq:93}) with empirical data, which is itself represented as a function of redshift.

As an initial condition for the differential equation, one can choose $Q_{\rm HII}$(\textit{z}=10$)=0.2$, which lies within the 68\% confidence interval of a joint Planck + HST constraint \citep{2015ApJ...802L..19R}. However, we choose to follow \cite{2019PhRvD..99b3518C}, and set the initial condition at higher redshift. This way, $Q_{\rm HII}$(\textit{z}=20$)=10^{-13}$. Such initial condition is set without the loss of generality, as even the most conservative models of reionization predict a negligibly small ionized hydrogen content at $z\gtrsim 20$. 

Finally, in addition to $Q_{\rm HII}$, a constraint on the total optical depth of the CMB can be derived using EoR quantities \citep{2020A&A...641A...6P}:
\begin{equation}
    \tau_{\rm reion} = \overline{n}_{\rm H} \sigma_T\int^{z_{\rm max}}_0 cx_e\frac{(1+z)^2}{H(z)}dz.
    \label{eq:109}
\end{equation}
We take the Thomson scattering cross section as $\sigma_T = 6.65\times 10^{-33}\;\rm cm^2$ and set $z_{\rm max} = 20$. Reionization will have a negligible effect on $\tau$ at this redshift, which is evident from \cite{2022LRCA....8....3G}).  \textit{Planck 2018} has put the following constraint on optical depth: $\tau=0.054 \pm 0.007$ \citep{2020A&A...641A...6P}. This value will be compared to the theoretical predictions for each theory of modified gravity in the next section.

The lower bound of integral (\ref{eq:107}), $M_{\rm min}$, is the minimal mass of CDM halos at virial temperature of $T_{\rm vir}\approx10^4\rm K$, assuming that baryons collapse at overdensity $\delta_b>100$ \citep{2001PhR...349..125B}. This quantity is typically calculated from the spherical collapse formalism.
\subsection{Determining $M_{\rm min}$}\label{sec:6.1}
Now we want to define the value of minimal CDM halo mass, which itself depends on the chosen cosmological background. To derive the minimal virial mass, one should first define the virial temperature as follows \citep{2014MNRAS.439.3798F, 2001PhR...349..125B}:
\begin{equation}
    T_{\rm vir} = \frac{\mu_{\rm mol} m_{\rm p}v_{\rm vir}^2}{2k_{\rm B}}.
    \label{eq:tvir}
\end{equation}
Here symbols $\mu_{\rm mol}$, $m_{\rm p}$, and $k_{\rm B}$ have their usual meaning, namely mean molecular weight, proton mass, and Boltzmann constant. We take mean molecular weight to be $\mu = (1-Y_{\rm He}+Y_{\rm He}m_{\rm H}/m_{\rm He})^{-1}\approx 1.22$ \citep{2004ApJ...600..508S}. This works well for our primordial gas, which mainly consists of neutral hydrogen and helium. This assumption can also help us to derive the mean hydrogen density \citep{2022GReGr..54..102C, 2013ApJ...768...71R}:
\begin{equation}
\overline{n}_{\rm H} = \frac{\rho_{\rm H}}{m_p} = (1-Y_{\rm He})\Omega_{b0} \rho_{\rm cr}.
\end{equation}
Therefore, the only unknown left is the virial velocity which is derived from the virial theorem in a familiar manner:
\begin{equation}
    v_{\rm vir} = \sqrt{\frac{GM}{R_{\rm vir}}},\quad R_{\rm vir} = \sqrt[3]{\frac{3GM}{4\pi \rho_{\rm vir}}}.
    \label{eq:rvir}
\end{equation}
The virial density is related to the average matter density through $\rho_{\rm vir} = \Delta_{\rm vir}\overline{\rho}$. In the $\Lambda$CDM model, virial overdensity can be well approximated as $\Delta_{\rm vir}\approx 18\pi^2$ at $z\gg0$. However, in modified gravity scenarios, this may not be true, requiring us to derive the virial quantities anew using the formalism for spherical collapse outlined Section (\ref{sec:3}). 
\subsubsection{Phenomenological MG theories}\label{sec:6.1.1}
We approach this problem by adopting the formalism presented in \cite{2009PhRvD..79h3518S} and references therein. First, we rearrange Eq. (\ref{eq:60}). Let the mass of a sphere be $M=4\pi/3R^3\overline{\rho}(1+\delta_{\rm nl})$. It remains constant through cosmic time due to the conservation of mass. By substituting the expression for $\delta_{\rm nl}$, derived from the mass-radius relation, into Eq. (\ref{eq:60}), we infer the evolution equation for the radius of the sphere:
\begin{equation}
    \frac{\ddot{R}}{R} = \dot{H} + H^2 - \frac{4\pi G\mu(a)\overline{\rho}\delta_{\rm nl}}{3}.
    \label{eq:111}
\end{equation}
For simplicity, we presume the absence of shear stress and dynamic friction \citep{2000MNRAS.314..279E,1998A&AT...16..127D}. We can now use Eq. (\ref{eq:111}) and the virial theorem to derive $R_{\rm vir}$. The virial theorem postulates that any virial structure is in an equilibrium state, meaning it is self-gravitating and completely stable. This equilibrium  state is usually described by the equivalence of total kinetic energy $\langle K \rangle$ and potential energy $\langle U \rangle$:
\begin{equation}
    \langle K \rangle +\frac{1}{2}\langle U \rangle = 0.
    \label{eq:112}
\end{equation}
For a spherically symmetric distribution of matter, kinetic energy is expressed as an integral \citep{2024PhRvD.109b3535A}
\begin{equation}
    \langle K \rangle = \frac{1}{2}\int \overline{\rho}\mathbf{v}^2d^3\mathbf{x} = \frac{3}{10}M\dot{R}^2.
    \label{eq:100}
\end{equation}
Subsequently, the potential energy is given by \citep{2012JCAP...06..019B}:
\begin{equation}
    \langle U \rangle = -\int \overline{\rho}\mathbf{x}\cdot \nabla\Psi d^3\mathbf{x} = -\frac{3M}{R^3}\sum_i \int^R_0r^3\frac{d\Psi_i(r)}{dr}dr.
    \label{eq:114}
\end{equation}
Here, $\Psi_i$ is a set of all terms contributing to the Bardeen potential, derived from the Poisson equation using Green's function:
\begin{equation}
    \nabla^2\Psi_{\rm N+MG} = 4\pi G\mu(a)\overline{\rho}\delta_{\rm nl}\Leftrightarrow \Psi _{\rm N+MG}= -\frac{G\mu(a)\delta M}{R}.
\end{equation}
Clearly, in our case, it is enough to divide the potential into two parts: Newtonian and Modified Gravity. An additional, effective contribution to the potential from dark energy is:
\begin{equation}
    \begin{gathered}
    \nabla^2\Psi_{\rm eff} = -3\bigg(\frac{\ddot{a}}{a}\bigg) = 4\pi G(\overline{\rho}+(1+3w_{\rm eff})\overline{\rho}_{\rm eff}), \\
    \Leftrightarrow \Psi_{\rm eff} = -\frac{1}{2}\bigg(\frac{\ddot{a}}{a}\bigg)r^2.
\end{gathered}
\end{equation}
This allows us to reformulate Eq. (\ref{eq:111}) into a simpler form, $\ddot{R}/R=-1/3\nabla^2(\Psi_{\rm N+MG}+\Psi_{\rm eff})$. For phenomenological theories of gravitation, this also allows for the derivation of potential energy from Eq. (\ref{eq:114}) explicitly \citep{2011JCAP...04..025K}:
\begin{equation}
    \langle U \rangle = -\frac{3}{5}\frac{G\mu(a)M\delta M}{R}+\frac{3}{5}\bigg(\frac{\ddot{a}}{a}\bigg)MR^2.
\end{equation}
The formalism outlined in this section should be used with care whilst assuming nDGP and $k$-mouflage cosmological backgrounds. While the Vainshtein screening is active, since the effective gravitational constant is radius dependent, the derivation of the Poisson equation becomes non-trivial. For the nDGP case, this problem was addressed in \cite{2010PhRvD..81f3005S}. Simply put, only the potential energy is modified by an additional term $U_\varphi$. Besides, it is possible to linearize gravity at some scales, simplifying the derivation. We return to this problem in Section (\ref{sec:6.1.3}).

\begin{figure}[!htbp]
    \centering
    \includegraphics[width=0.95\linewidth]{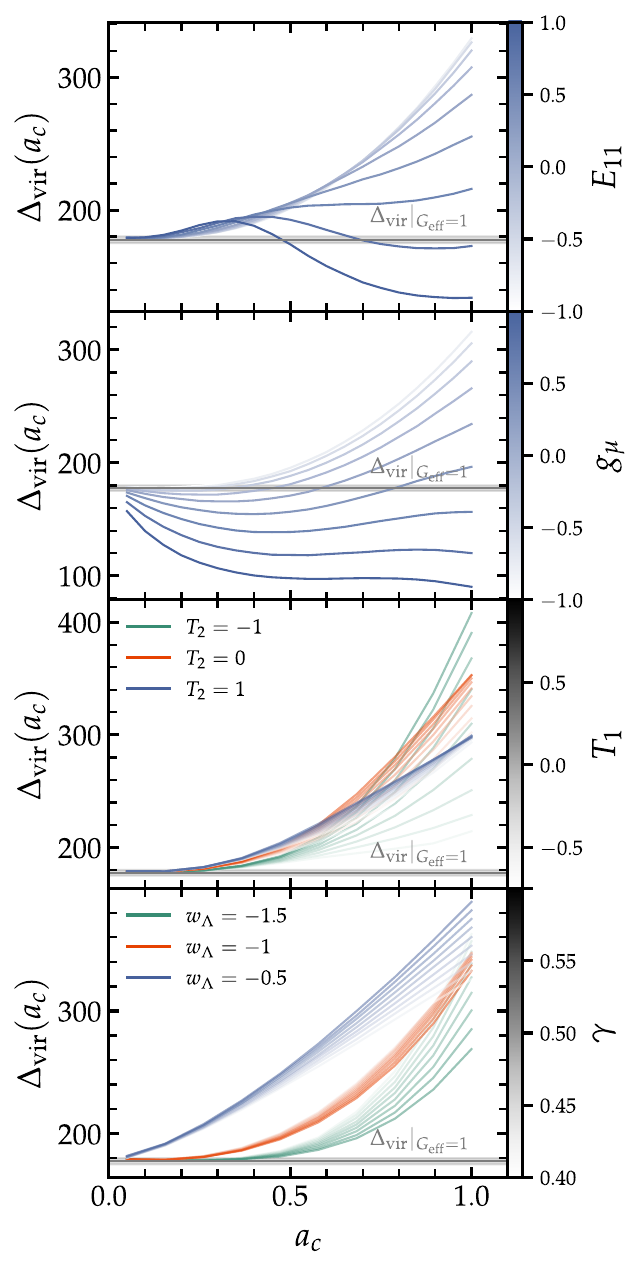}
    \caption{Virial overdensity for phenomenological models}
    \label{fig:5}
\end{figure}

For phenomenological gravity, the virial radius is found by solving Eq. (\ref{eq:111}) until $a=a_{\rm vir}$, where (\ref{eq:112}) is satisfied. Initial conditions are set at $a=a_c$ such that $R\approx 0$. A second initial condition is imposed by applying the mass conservation, i.e. we differentiate the mass-radius relation at the value $a=a_c$ and set the resulting expression to zero. Once done, the expression for $R_{\rm vir}$ is substituted into Eq. (\ref{eq:rvir}) and then into Eq. (\ref{eq:tvir}). The assumption that $T_{\rm vir} = 10^4$K (see discussion immediately preceding the beginning of Section (\ref{sec:6.1})) yields the value of $M_{\rm min}$. Alternatively, using the fact that $\langle K \rangle = \frac{1}{2}Mv_{\rm vir}^2$ along with the virial theorem (\ref{eq:112}) and Eq. (\ref{eq:tvir}), we deduce
\begin{equation}
\begin{gathered}
    \frac{v_{\rm vir}^2}{\rm km \;s^{-1}} = \frac{3}{5}\left(GMH_0\sqrt{\Omega_{m0}a_c^{-3}\Delta_{\rm vir}/2}\right)^{2/3}\\
    \times\left(\frac{1}{\Delta_{\rm vir}}\left[1+\frac{\overline{\rho}_{\rm eff}}{\overline{\rho}}(1+3w_{\rm eff})\right]+\mu \left[1-\frac{1}{\Delta_{\rm vir}}\right]\right) .
\end{gathered}
\end{equation}
This then determines the minimal mass
\begin{equation}
\begin{gathered}
    \frac{M_{\rm min}}{M_\star} = \left(\frac{10k_{\rm B}T_{\rm min}}{3\mu_{\rm mol}m_{\rm p}}\right)^{3/2}\left(GH_0\sqrt{\Omega_{m0}a_c^{-3}\Delta_{\rm vir}/2}\right)^{-1}\\
    \times \left(\frac{1}{\Delta_{\rm vir}}\left[1+\frac{\overline{\rho}_{\rm eff}}{\overline{\rho}}(1+3w_{\rm eff})\right]+\mu \left[1-\frac{1}{\Delta_{\rm vir}}\right]\right) ^{-3/2}.
\end{gathered}
\label{eq:120}
\end{equation}
In our code, we implement the latter approach.
Figure (\ref{fig:5}) shows the resulting virial overdensity for all three phenomenological theories of modified gravitation. As $z\to\infty$, all phenomenological theories converge to $\Lambda$CDM. This behaviour is expected, since $\Omega_{\Lambda}\to0$ in the early universe.
\subsubsection{Case with $w_\Lambda\neq -1$}
When we modify the nature of dark energy itself by assuming a kind of non-standard equation of state, we need to tweak previously introduced expressions. As noted in \cite{1998ApJ...508..483W}, for quintessence cosmology, the virial theorem has an additional term:
\begin{equation}
    \langle K \rangle+\frac{1}{2}\langle U \rangle = \langle U _\Lambda\rangle.
\end{equation}
The potential energy contribution for DE is given by $\langle U _\Lambda\rangle = -4/5\pi G\rho_\Lambda MR^2$. For the $w_\Lambda\rm CDM$ model, dark energy does not cluster, and thus Birkhoff's theorem remains valid. Hence, the radius evolution is governed by:
\begin{equation}
    \frac{\Ddot{R}}{R} = -\frac{4\pi G}{3} ((3w_\Lambda + 1)\overline{\rho}_\Lambda+\overline{\rho}+\overline{\rho}\delta_{\rm nl}).
    \label{eq:119}
\end{equation}
This is very similar to the Eq. (\ref{eq:111}), derived for phenomenological modified gravity. We assume that $w_\Lambda\neq -1$, with a constant equation of state parameter. The corresponding dark energy density is $\overline{\rho}_\Lambda=3H_0^2\Omega_\Lambda a^{-3(1+w_\Lambda)}$. This can be inferred from the usual conservation equation in the absence of interactions between sectors. Since dark energy contributes to spherical collapse only on the background level, its terms in Eq. (\ref{eq:119}) are absorbed into the Hubble parameter using the second Friedmann equation.

We derive $\Delta_{\rm vir}$ using the expression for radius from Eq. (\ref{eq:119}) and the same initial conditions that were assumed in the phenomenological scenario. The radius is substituted into the virial theorem based on Eqs. (\ref{eq:100}) and (\ref{eq:114}). The results are then plotted in Figure (\ref{fig:6}).
\subsubsection{Case with the present screening field}\label{sec:6.1.3}
In order to preserve the cosmological principle, for both nDGP and $k$-mouflage cosmologies an additional scalar field must be introduced. Due to the coupling between the scalar field with gravity, some additional terms will arise in the radius evolution equation. As shown in \cite{2010PhRvD..81f3005S}, we get
\begin{equation}
    \frac{\ddot{R}}{R} = \dot{H} + H^2 - \frac{4\pi G_{\rm eff}(R/R_\star)\overline{\rho}\delta_{\rm nl}}{3}.
\end{equation}
Interestingly, the virial condition itself remains unchanged. For screened gravity, kinetic energy is the same as in phenomenological models (see Eq. (\ref{eq:100})). On the contrary, potential energy is modified:
\begin{equation}
    \langle U \rangle =-\frac{3}{5}\frac{GM^2}{R} -\frac{3}{5}\frac{\Delta G_{\rm eff}M\delta M}{R}+\frac{3}{5}\bigg(\frac{\ddot{a}}{a}\bigg)MR^2.
\end{equation}
The virial radius is then defined using the approach laid out in Section (\ref{sec:6.1.1}). Minimal halo mass is expressed by Eq. (\ref{eq:120}), under the change of variables $\mu \to G_{\rm eff}(R/R_\star)/G_N$. This result is valid for nDGP. 
However, since the notion of screening for $k$-mouflage differs from nDGP (see Eq. (\ref{eq:75})), we have to make an appropriate change for $k$-mouflage, as discussed in \cite{2014PhRvD..90b3508B}. Note that for $k$-mouflage, at scales $ctk/a\gg1$, screening can be neglected. For small enough redshift, these scales correspond to $\sim $Mpc, at which galaxy clusters reside. For example, a Coma cluster with radius $R\sim 20$ Mpc (upper limit from \cite{2013A&A...553A.101C}) lies within the range of unscreened gravity. Thus, ignoring the effects of screening, we can introduce the equation for radius evolution within $k$-mouflage theory as follows \citep{2014PhRvD..90b3508B}:
\begin{equation}
    \ddot{R} + \frac{d\ln A}{dt}\dot{R}-\bigg(\frac{\Ddot{a}}{a}+\frac{\dot{a}}{a}\frac{d\ln A}{dt}\bigg)R=-\frac{1}{a}\nabla(\Psi_N+\ln \widetilde{A}).
    \label{eq:123}
\end{equation}
Here, $\widetilde{A}$ is a non-linear quantity, constructed from its background and perturbed components. The background contribution is calculated from Eq. (\ref{eq:44}) and the perturbed contribution arises due to $\delta\varphi$. We can then derive a Bardeen potential using the modified Poisson equation \citep{2014PhRvD..90b3508B}:
\begin{equation}
    \nabla \Psi_N = A\frac{G\delta M}{R^2}, \;\; \nabla\ln \widetilde{A} = -\frac{2\mathscr{B}^2}{K'}\nabla\Psi_N, \;\; \mathscr{B} = M_p\frac{d\ln A}{d\varphi}.
    \label{eq:124}
\end{equation}
From BBN constraints, we obtain $A\simeq1$ and $\mathscr{B}\simeq \beta$ \citep{2014PhRvD..90b3508B}. It is important to note that Eq. (\ref{eq:124}) is valid only for the case when $\delta_{\rm nl}\lesssim 200$. This is suitable for computing virial quantities. However, it will fail at smaller scales, such as those corresponding to the Milky Way core or the Solar System, where GR should be restored via the screening mechanism absent from the aforementioned equation. Therefore, we set the non-linear density threshold at the time of spherical collapse to $\delta_{\rm nl} = 200$, which has been shown to be an effective choice for the derivation of the halo mass function (for instance, see \cite{2015PhRvD..92d3519B}). Using the prescription outlined in \cite{2014PhRvD..90b3508B}, we rearrange Eq. (\ref{eq:123}), and define a new variable, $y(t)=R(t)/a(t)\overline{R}_0$. An obvious initial condition for the radius is to take $\overline{R}_0$. This variable is related to the non-linear density contrast via $\delta_{\rm nl}+1=y(t)^{-3}$. This expression for non-linear density contrast was then substituted into the Eq. (\ref{eq:123}) to get the following differential equation:
\begin{equation}
\begin{gathered}
    \delta''_{\rm nl}(a) + \bigg(\frac{3}{a}+\frac{H'}{H}+\frac{\epsilon_2}{a}\bigg)\delta'_{\rm nl}(a) - \frac{3}{2}\frac{\Omega_{m0}(1+\epsilon_1)}{a^5E^2}\\\times\delta_{\rm nl}(a)
    (1+\delta_{\rm nl}(a))
    =\frac{4}{3}\frac{\delta'^2_{\rm nl}(a)}{1+\delta_{\rm nl}(a)}.\end{gathered}
\end{equation}
In \cite{2015PhRvD..92d3519B}, it was demonstrated that the $k$-mouflage gravitational potential differs from $\Lambda$CDM by a factor of $\mu(a)=1+\epsilon_1(a)$. Since there is no explicit radius dependence in $\mu$ for $k$-mouflage, the computation of kinetic and potential energies is very similar to the phenomenological gravity discussed earlier. 

\begin{figure}[!htbp]
    \centering
    \includegraphics[width=0.95\linewidth]{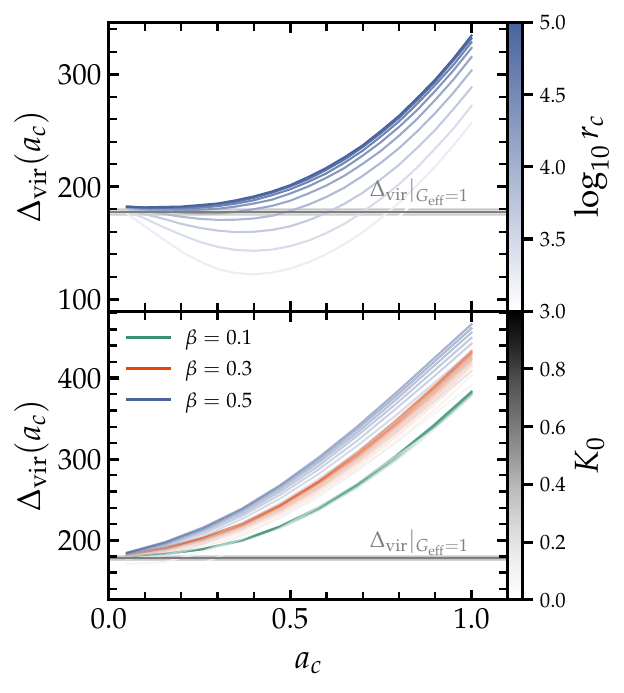}
    \caption{Virial overdensity for the theories of modified gravitation with the screening field present, namely nDGP and $k$-mouflage}
    \label{fig:6}
\end{figure}

We plot the numerical solutions for both nDGP and $k$-mouflage for $\Delta_{\rm vir}$ in Figure (\ref{fig:6}). Interestingly, similar to the phenomenological case, for screened theories, $\Delta_{\rm vir}(a_c)\to 18\pi^2$ as $a_c\to0$. This is not a trivial result, as screened theories modify not only the effective dark energy but also add some additional matter terms to the respective Lagrangians that can decay at a different rate compared to the dark energy component, hence possess a non-zero mass fraction at a higher redshift. The consequence of this is that in some cases deviations from $\Lambda$CDM can be observed at very high redshifts. 

Now, with the help of virial radius, we obtain the virial temperature using Eq. (\ref{eq:tvir}). This allows us to place a lower bound on halo mass, at which the gas within the halo can cool efficiently. $M_{\rm min}$ is then substituted into Eq. (\ref{eq:107}) to derive the hydrogen ionization fractions. Theoretical predictions for $Q_{\rm HII}(z)$ and $\tau_{\rm reion}(z)$ are subsequently compared to the observational data in the next section.

\begin{figure*}[!htbp]
    \centering
    \includegraphics[width=0.95\linewidth]{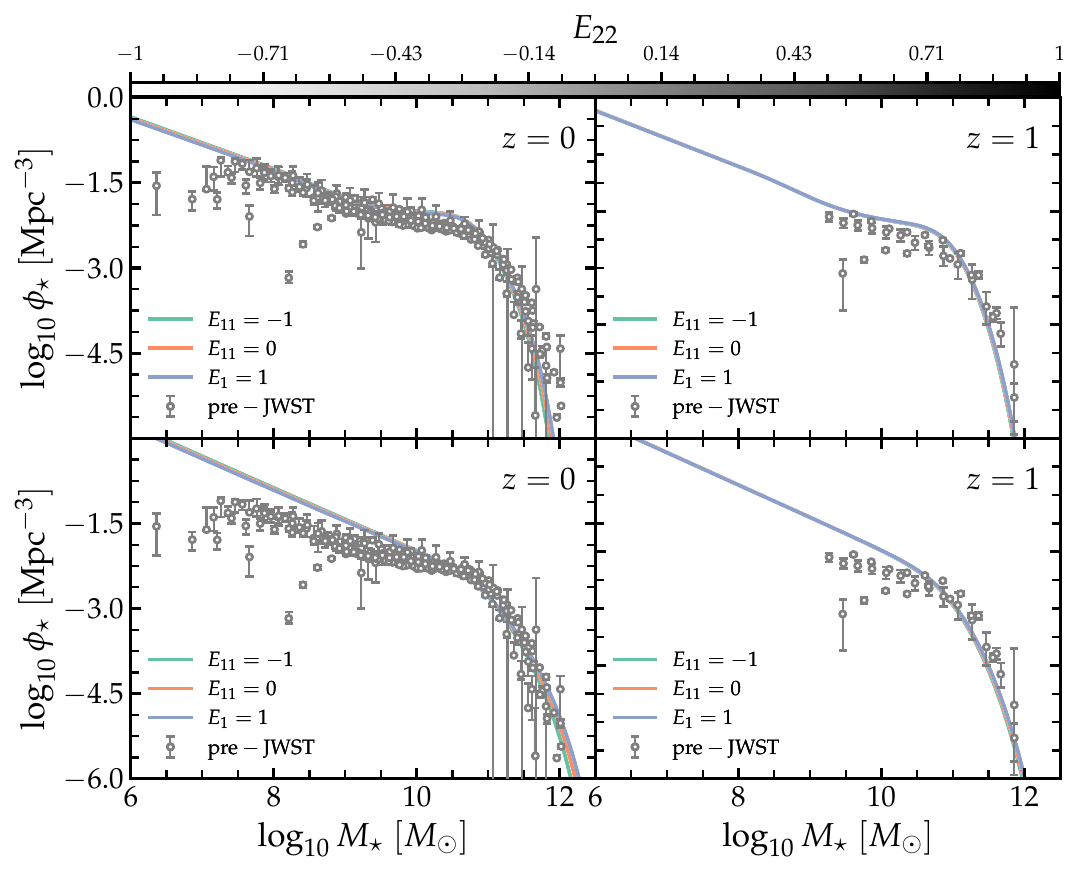}
    \caption{Stellar mass function for the Planck parameterization under both Rodriguez-Puebla (top row) and double power-law (bottom row) SMHR. Observational data is provided as a compiled pre-JWST constraints (gray circles) and JWST data (black triangles)}
    \label{fig:SMF_E11}
\end{figure*}

\begin{figure*}[!htbp]
    \centering
    \includegraphics[width=0.95\linewidth]{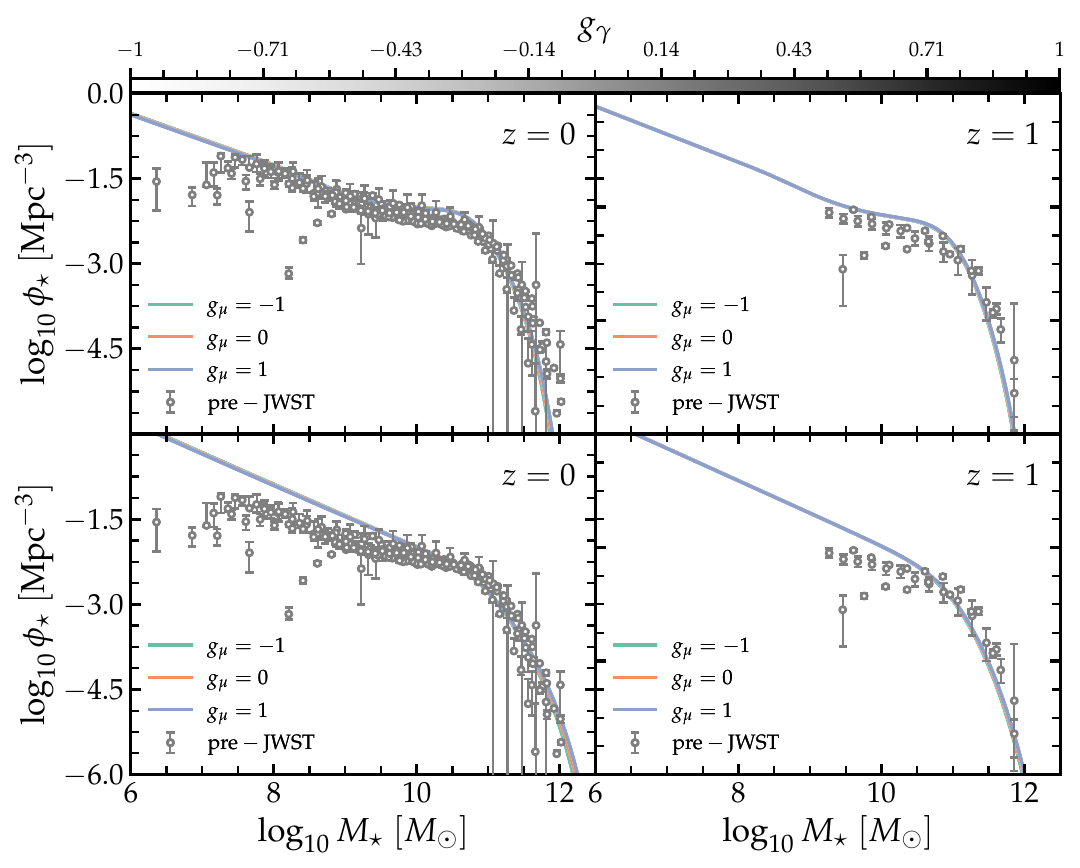}
    \caption{Figure (\ref{fig:SMF_E11}) continued, but for the z\_flex parameterization}
    \label{fig:SMF_gmu}
\end{figure*}

\begin{figure*}[!htbp]
    \centering
    \includegraphics[width=0.95\linewidth]{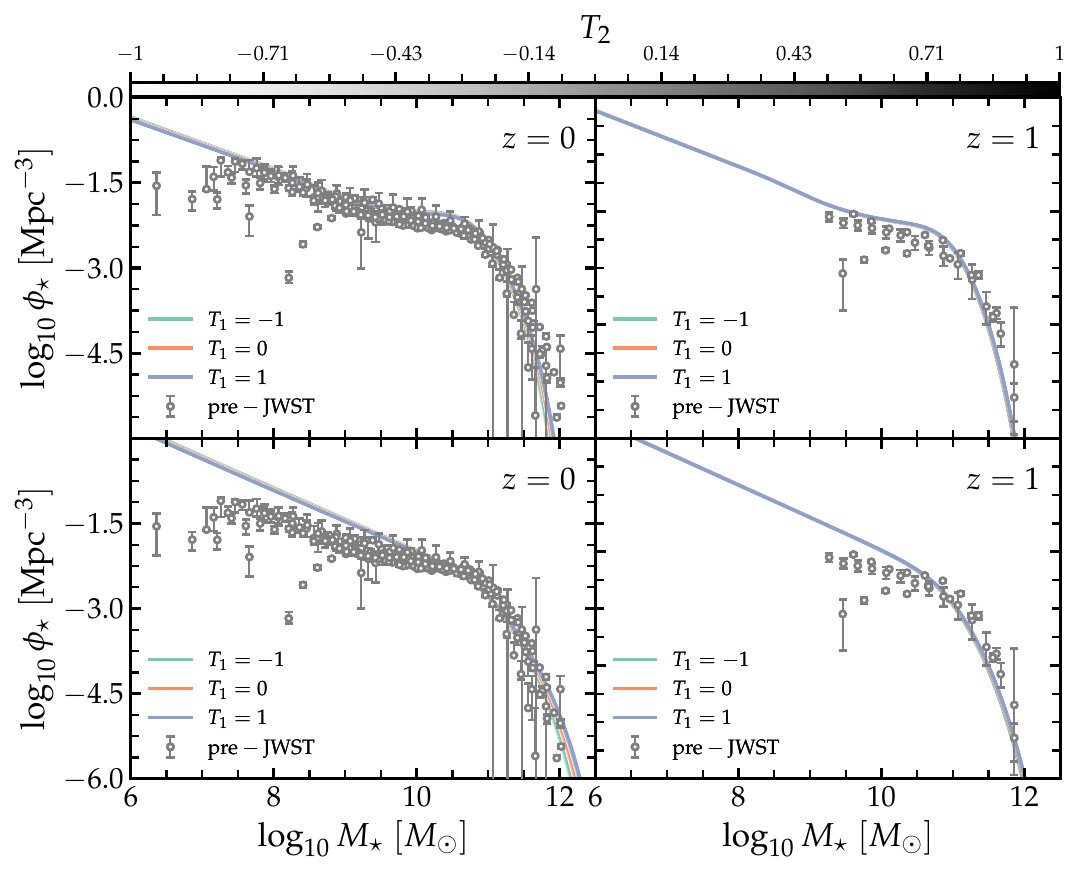}
    \caption{Figure (\ref{fig:SMF_E11}) continued, but for the DES parameterization}
    \label{fig:SMF_DES}
\end{figure*}

\begin{figure*}[!htbp]
    \centering
    \includegraphics[width=0.95\linewidth]{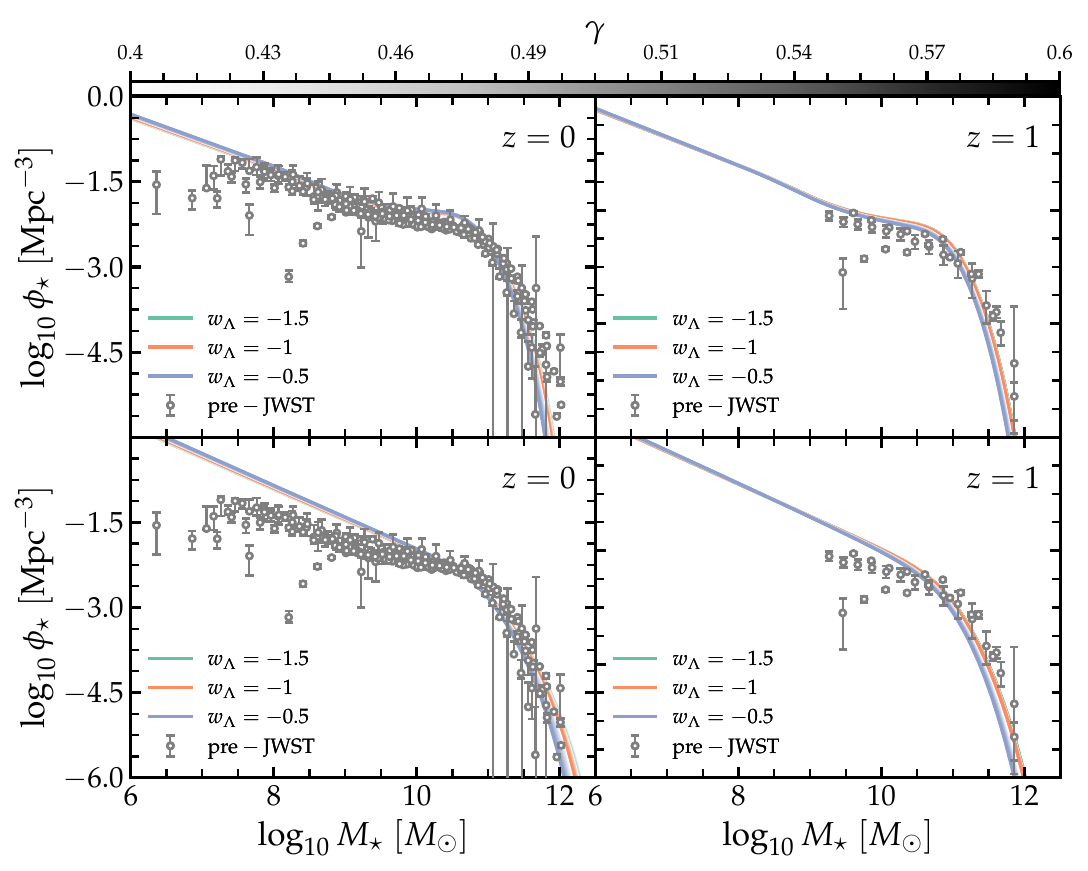}
    \caption{Figure (\ref{fig:SMF_E11}) continued, but for the $w\boldsymbol{\gamma}$CDM modified gravity}
    \label{fig:SMF_wCDM}
\end{figure*}

\begin{figure*}
    \centering
    \includegraphics[width=0.95\linewidth]{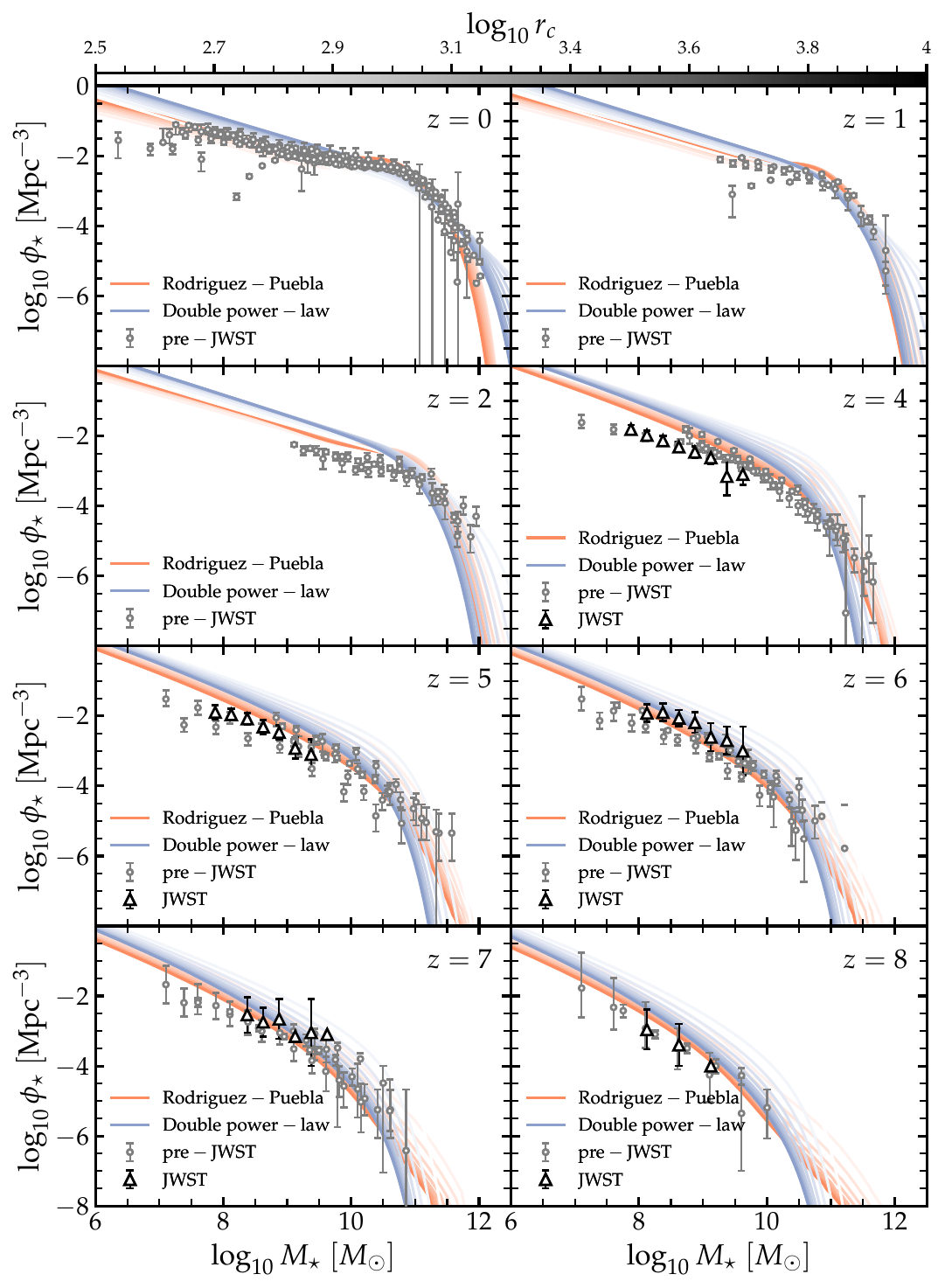}
    \caption{Stellar mass function for the nDGP gravity under both Rodriguez-Puebla (orange lines) and double power-law (blue lines) SMHR. Observational data is provided as a compiled pre-JWST constraints (gray circles) and JWST data (black triangles)}
    \label{fig:SMF_nDGP}
\end{figure*}
\begin{figure*}
    \centering
    \includegraphics[width=0.95\linewidth]{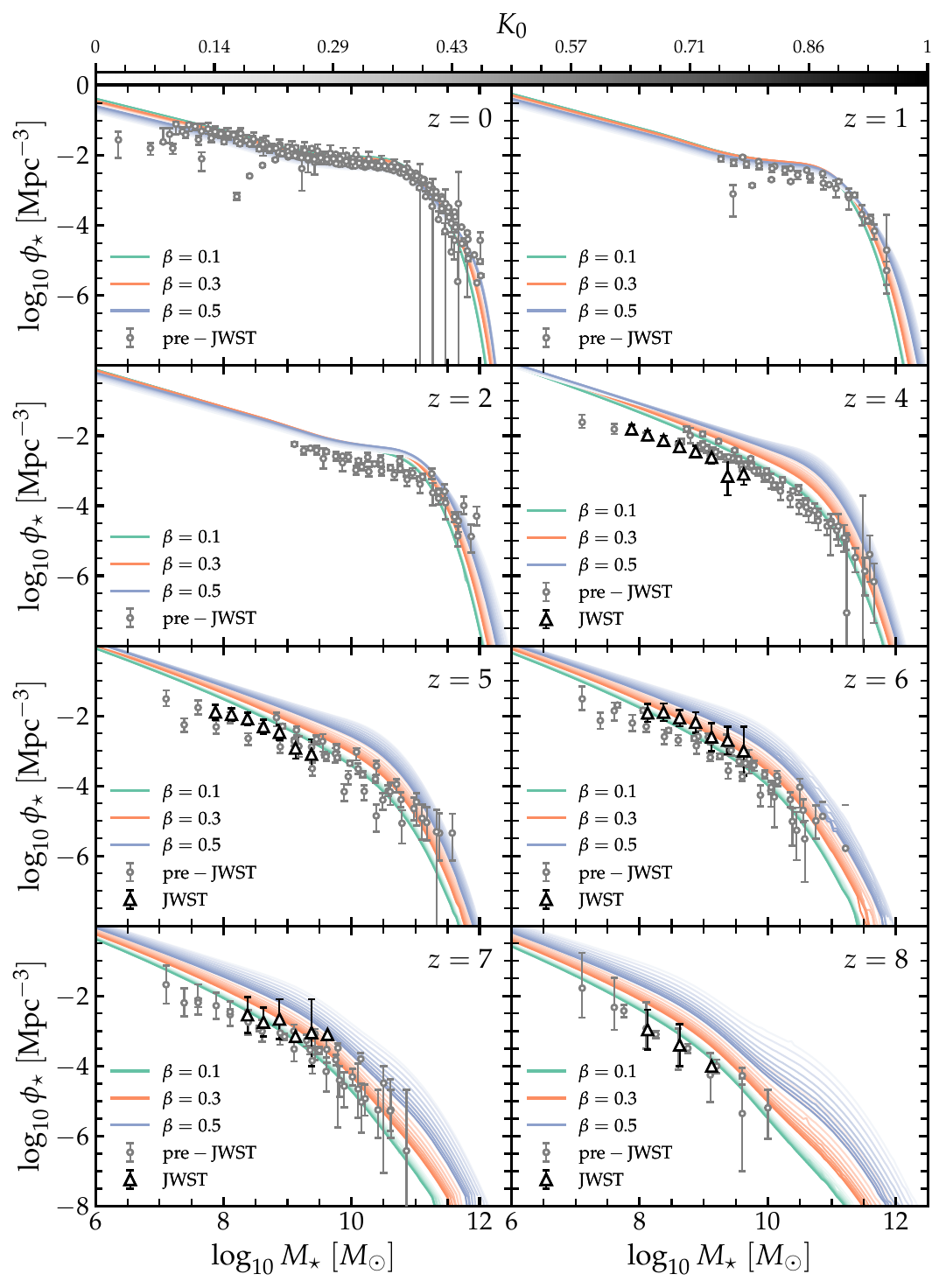}
    \caption{Figure (\ref{fig:SMF_nDGP}) continued, but for the $k$-mouflage gravity under the Rodriguez-Puebla SMHR}
    \label{fig:SMF_kmoufl_Puebla}
\end{figure*}
\begin{figure*}
    \centering
    \includegraphics[width=0.93\linewidth]{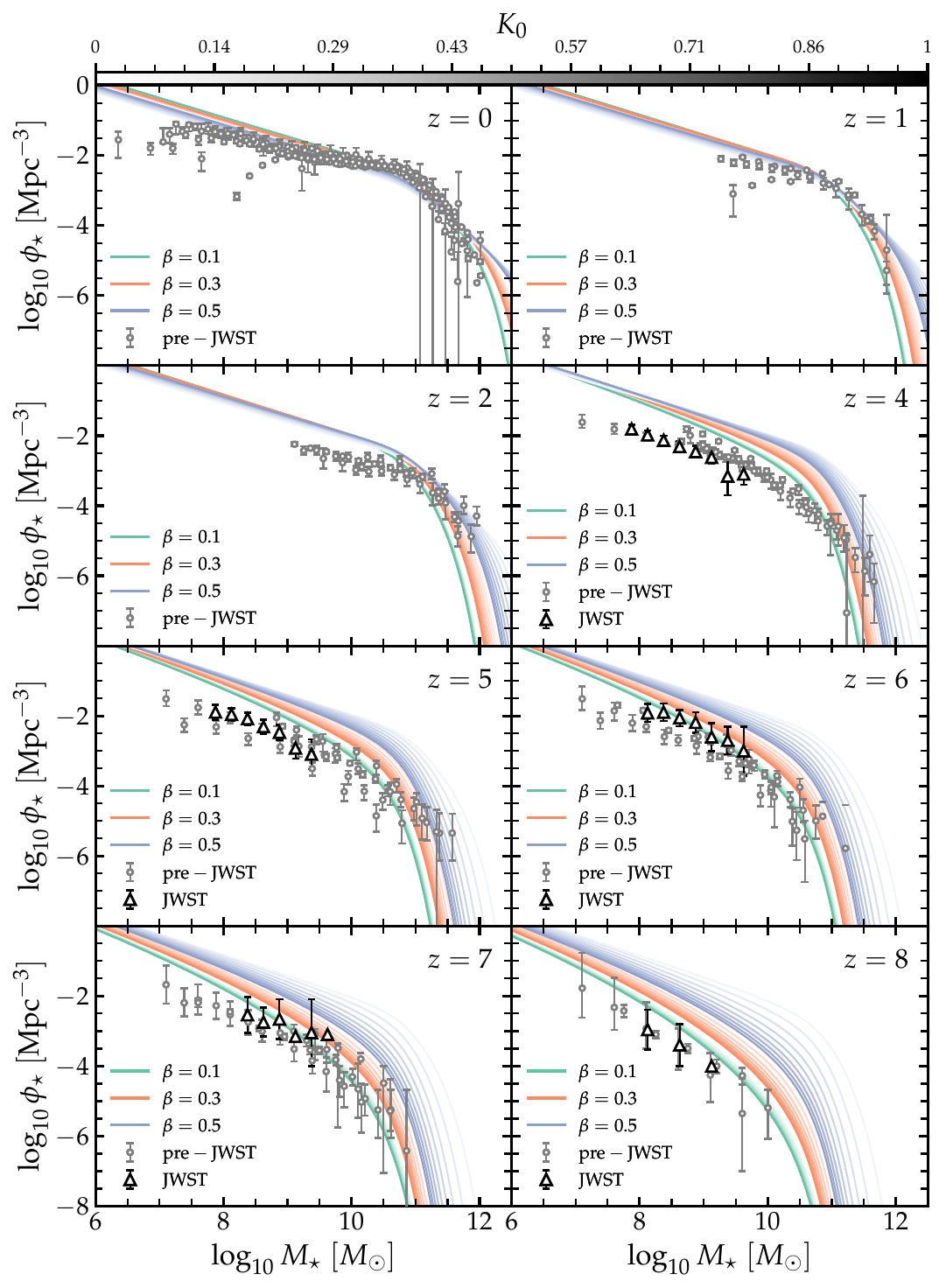}
    \caption{Figure (\ref{fig:SMF_nDGP}) continued, but for the $k$-mouflage gravity under the double power-law SMHR}
    \label{fig:SMF_kmoufl_double_power}
\end{figure*}
\begin{figure*}
    \centering
    \includegraphics[width=0.93\linewidth]{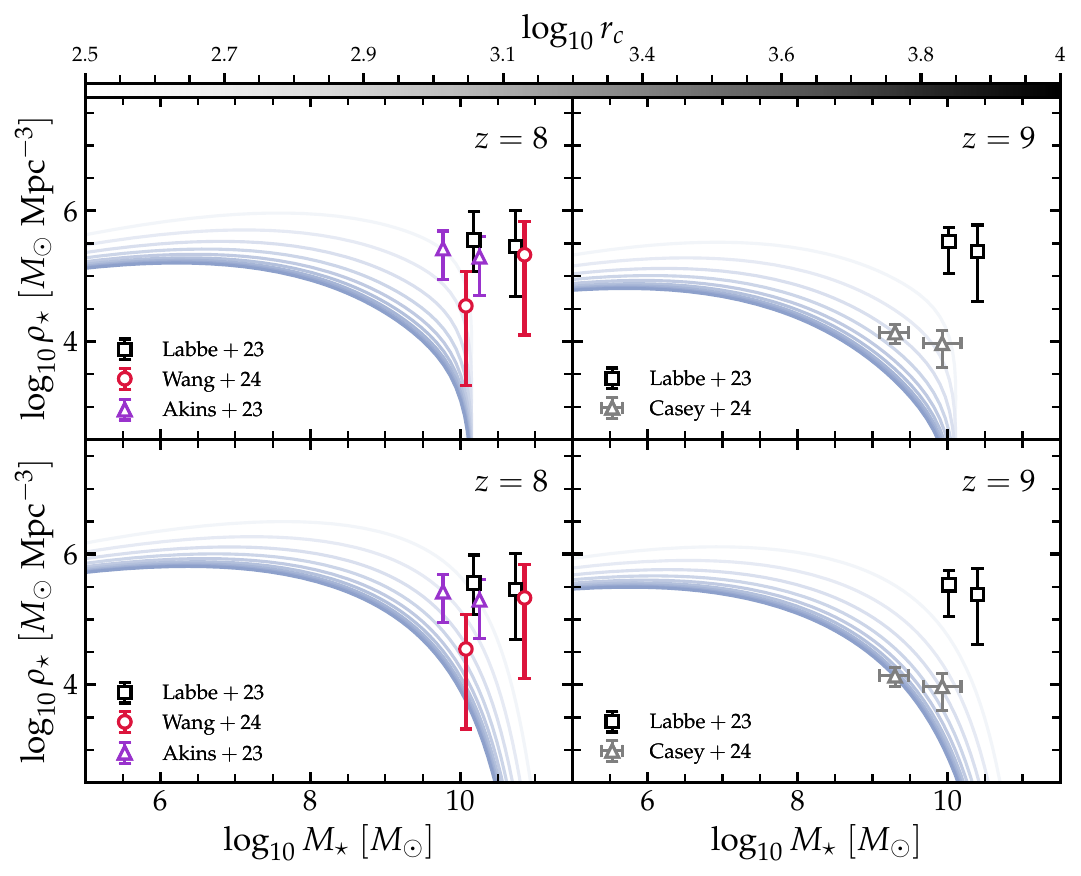}
    \caption{Stellar mass density for nDGP modified gravity theory versus recent observational data using Rodriguez-Puebla SMHR (upper two subplots) and double power-law SMHR (lower two subplots)}
    \label{fig:SMD_nDGP}
\end{figure*}
\begin{figure*}
    \centering
    \includegraphics[width=0.93\linewidth]{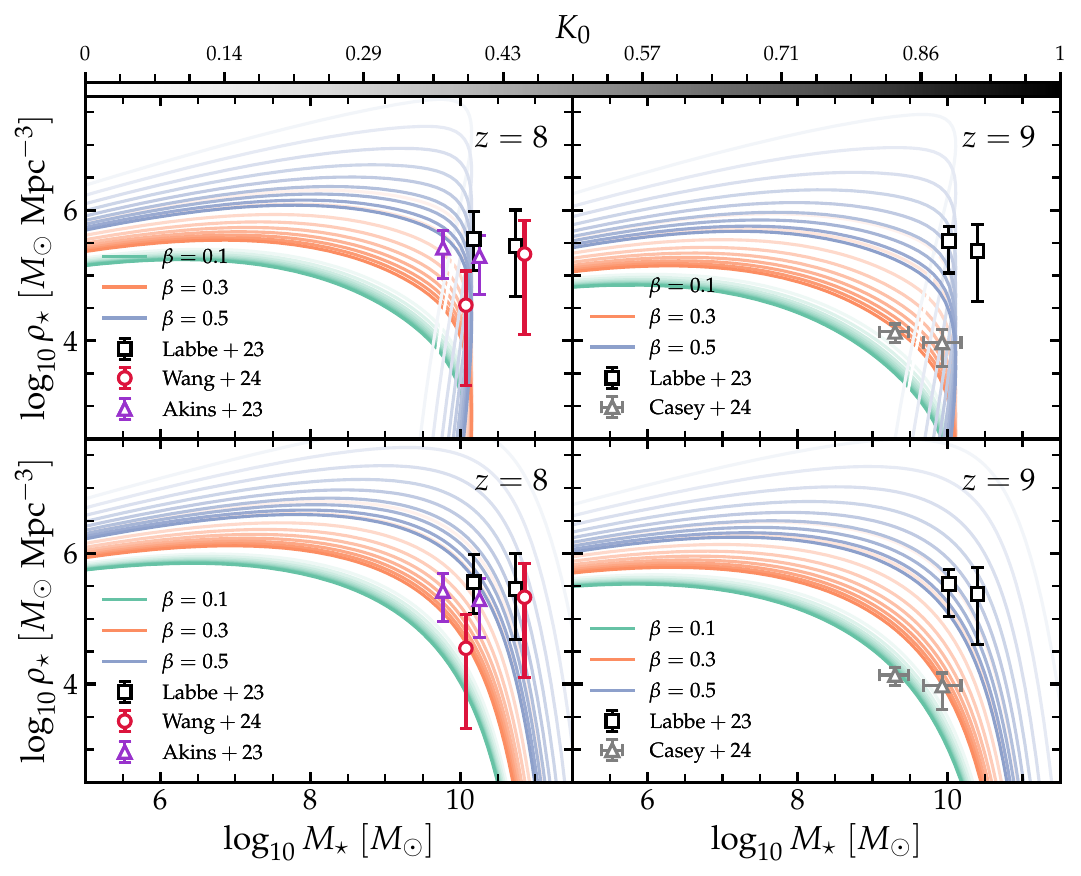}
    \caption{Figure (\ref{fig:SMD_nDGP}) continued, but for the $k$-mouflage gravity}
    \label{fig:SMD_kmoufl}
\end{figure*}
\begin{figure*}
    \centering
    \includegraphics[width=0.92\linewidth]{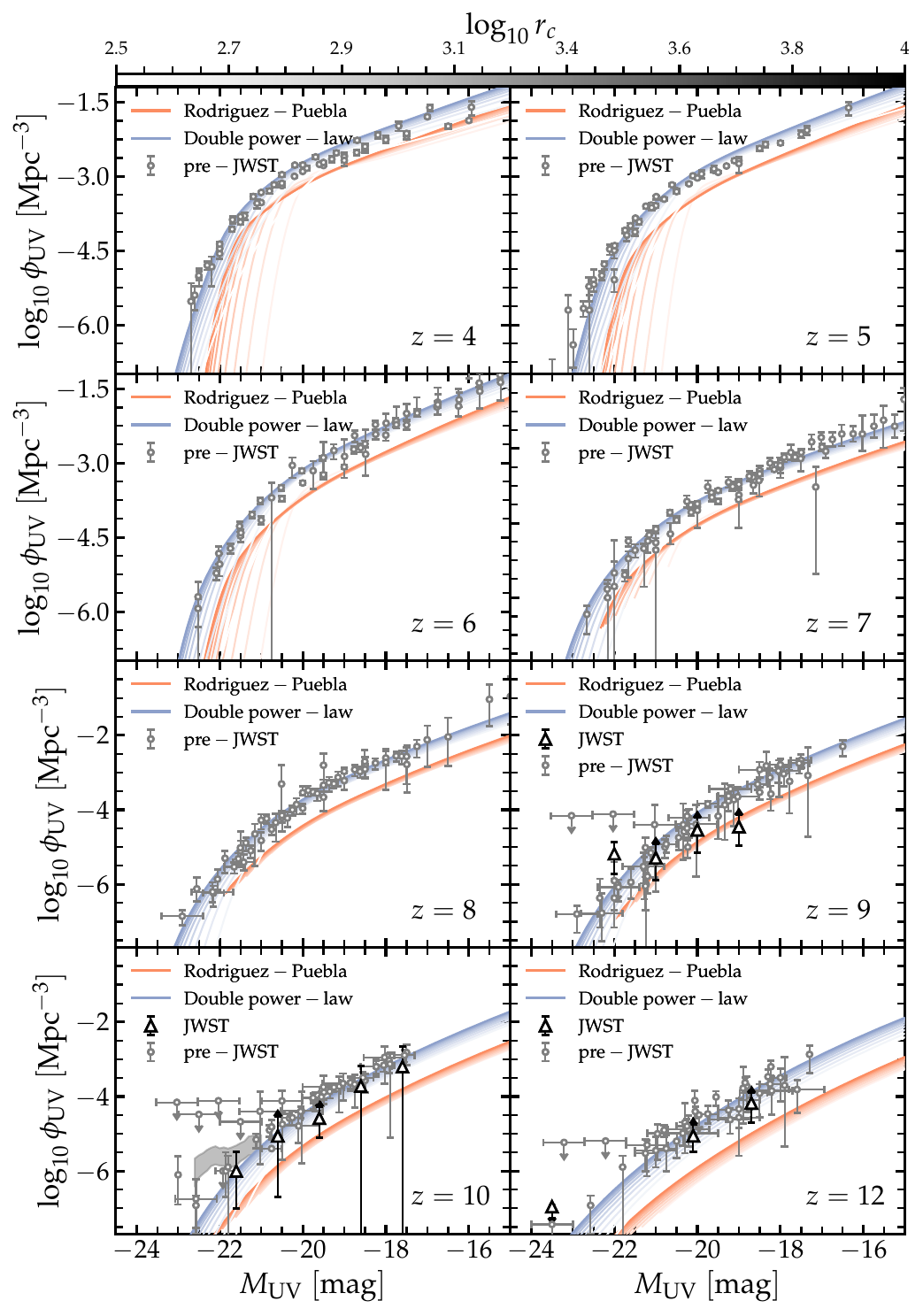}
    \caption{Ultra-violet luminosity function for the nDGP gravity under both Rodriguez-Puebla (orange lines) and double power-law (blue lines) SMHR. Observational data is provided as a compiled pre-JWST constraints (gray circles) and JWST data (black triangles). Arrows correspond to the upper/lower bounds}
    \label{fig:UVLF_nDGP}
\end{figure*}
\begin{figure*}
    \centering
    \includegraphics[width=0.93\linewidth]{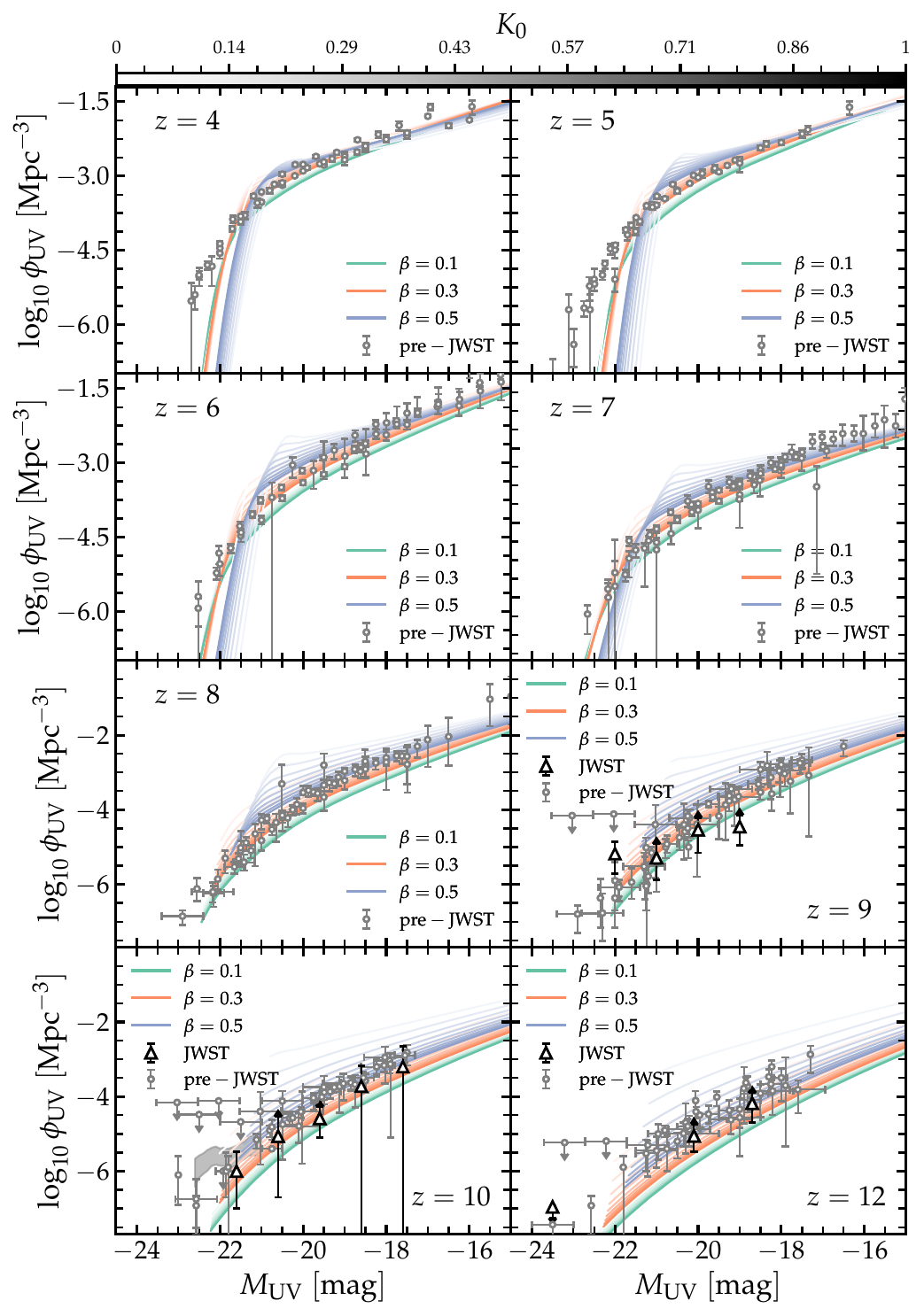}
    \caption{Figure (\ref{fig:UVLF_nDGP}) continued, but for the $k$-mouflage gravity under the Rodriguez-Puebla SMHR}
    \label{fig:UVLF_kmoufl_Puebla}
\end{figure*}
\begin{figure*}
    \centering
    \includegraphics[width=0.93\linewidth]{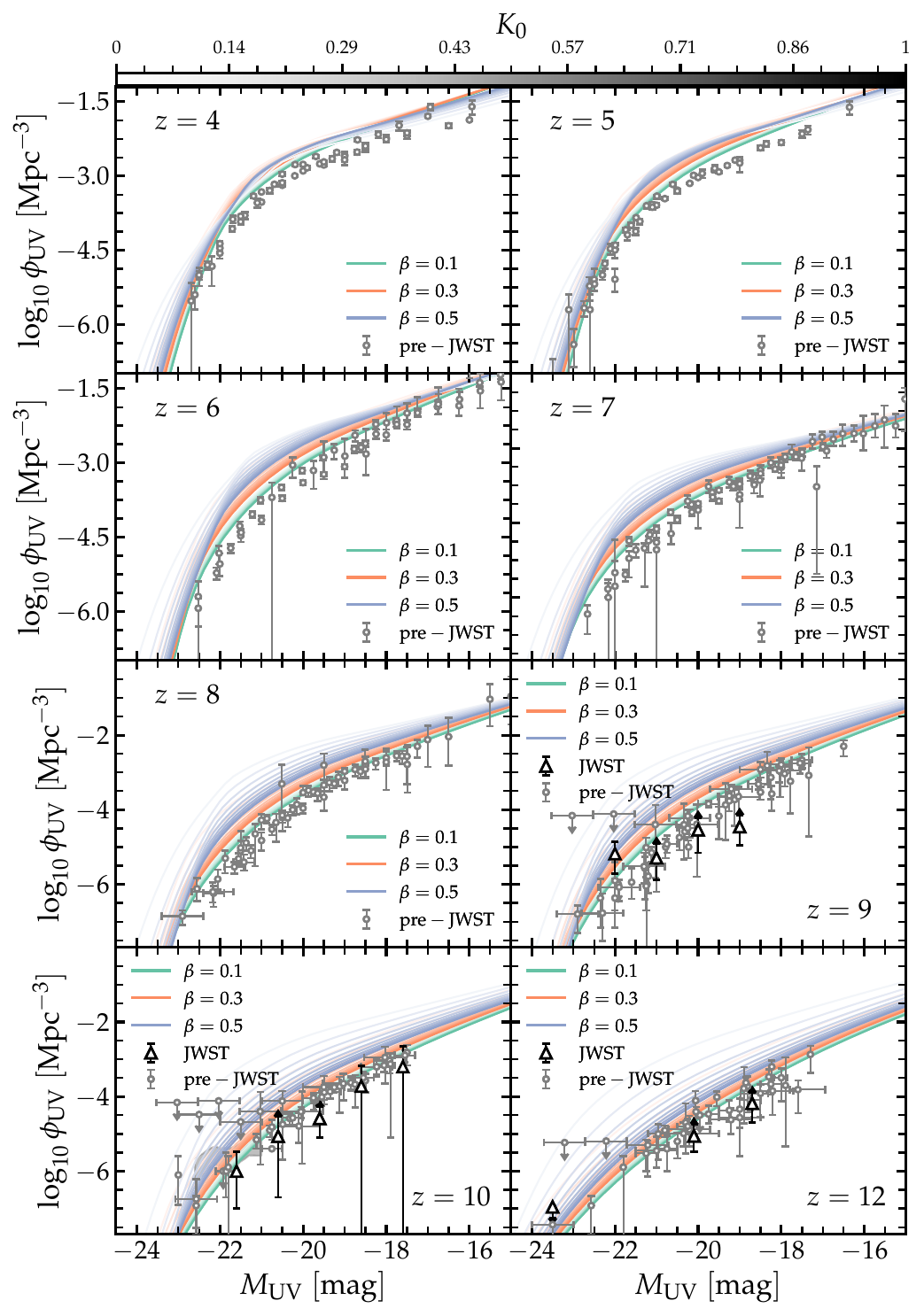}
    \caption{Figure (\ref{fig:UVLF_nDGP}) continued, but for the $k$-mouflage gravity under the double power-law SMHR}
    \label{fig:UVLF_kmoufl_double_power}
\end{figure*}

\begin{figure*}
    \centering
    \includegraphics[width=0.95\linewidth]{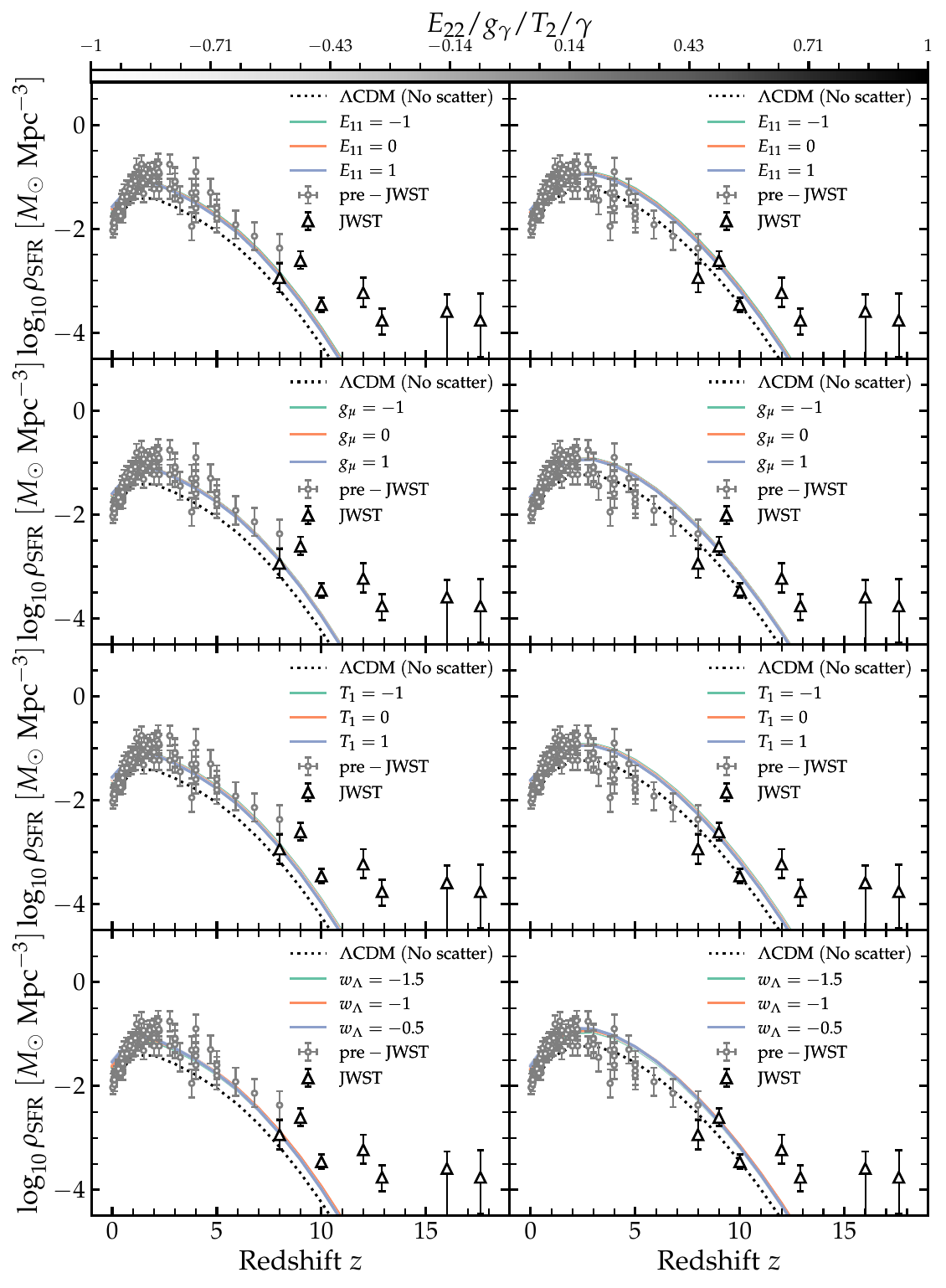}
    \caption{Star formation rate density for Planck (first row) and z\_flex (second row), DES (third row) and $w\boldsymbol{\gamma}$CDM modified gravity theories. We display results for the Rodriguez-Puebla SMHR in the first column and double power-law SMHR in the second column respectively. Observational data is provided as a compiled pre-JWST constraints (gray circles) and JWST data (black triangles)}
    \label{fig:SFRD_pheno}
\end{figure*}

\begin{figure*}
    \centering
    \includegraphics[width=0.95\linewidth]{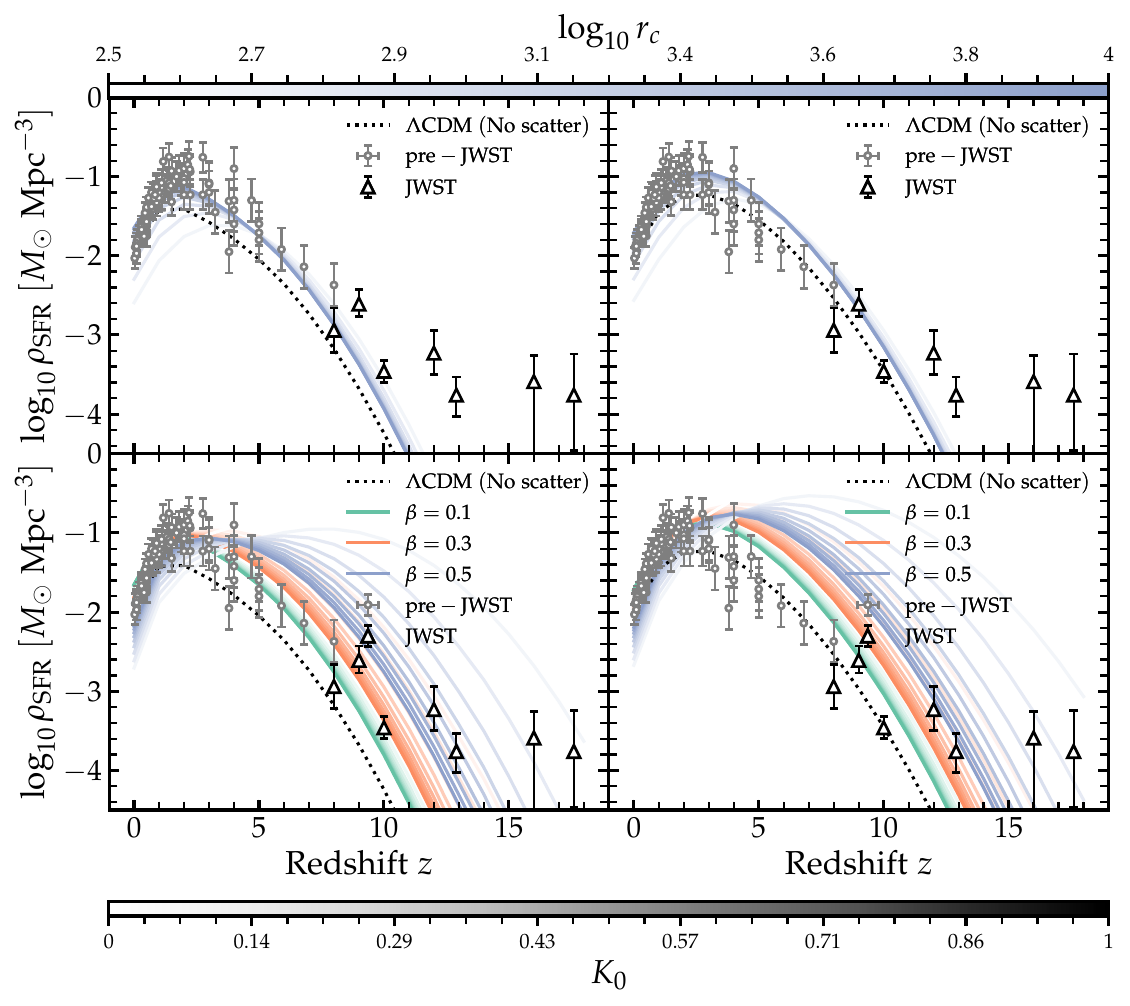}
    \caption{Figure (\ref{fig:SFRD_pheno}) continued, but for nDGP (first row) and $k$-mouflage (second row) modified gravity theories}
    \label{fig:SFRD_screen}
\end{figure*}

\begin{figure*}[!htbp]
    \centering
    \includegraphics[width=0.95\linewidth]{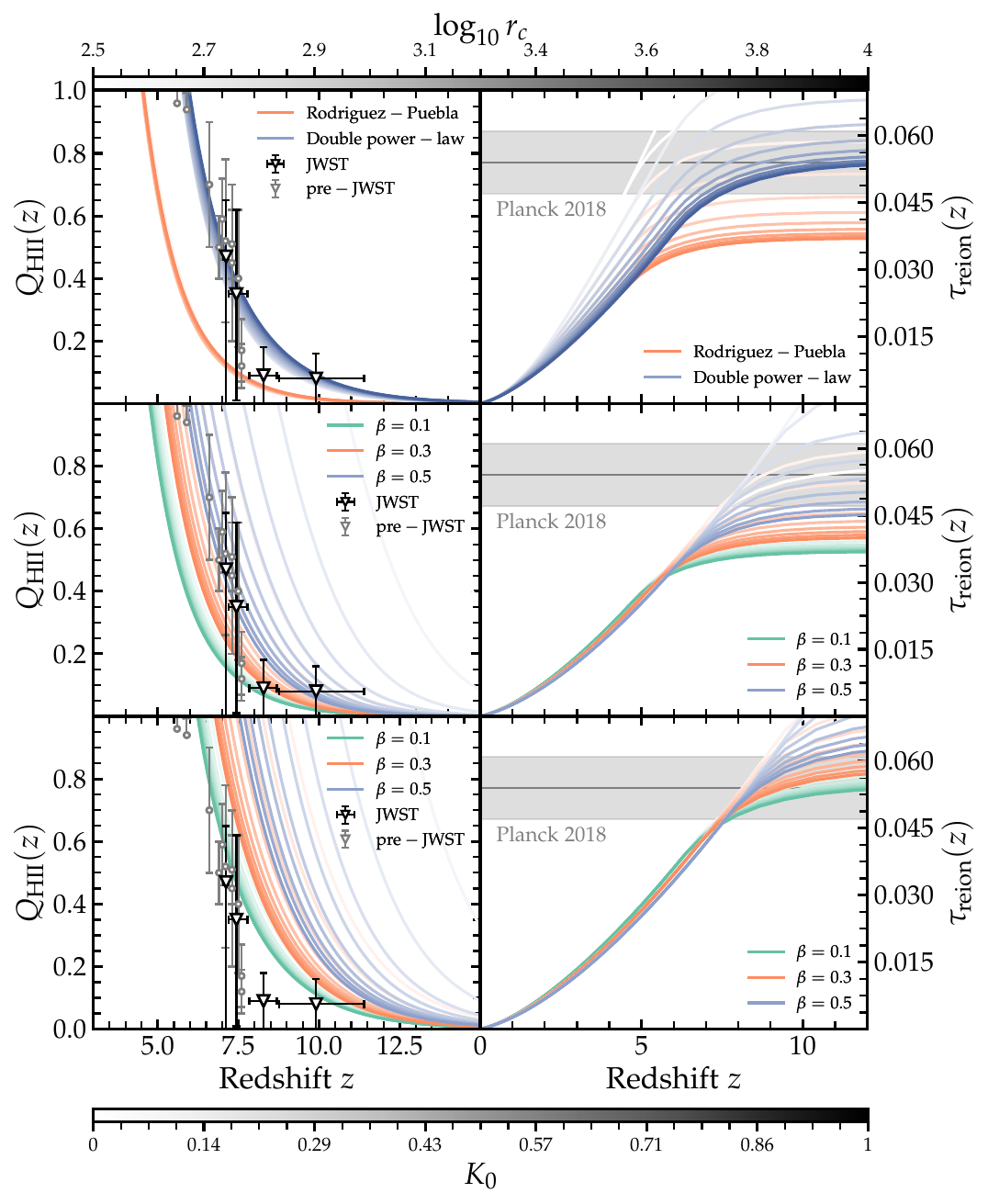}
    \caption{Volume filling fraction of HII (left column) and optical depth (right column) for nDGP (first row), $k$-mouflage Rodriguez-Puebla (second row) and double power-law SMHR (third row). Observational constraint on the optical depth by the \textit{Planck} probe is provided as a gray band}
    \label{fig:EoR_screen}
\end{figure*}
\section{JWST data and results}\label{sec:7}
Here, we introduce the data used to constrain the parameter space of our models. Note that we are going to utilize both JWST data and older measurements to ensure completeness.

\textit{Stellar Mass Function:} The first dataset includes SMF measurements in the range $4<z<7$ from the CANDELS-GOODS field by \cite{2014MNRAS.444.2960D}. A similar dataset, based upon both CANDELS/GOODS and HUDF fields, introduced in \cite{2016ApJ...825....5S}, covers galaxies with $4<z<8$. Additionally, the SMF derived from Hubble Frontier data by \cite{2019MNRAS.486.3805B} and \cite{2020ApJ...893...60K} spans the $6<z<9$ range. Results up to $z\sim 10$ were achieved through a continuous Spitzer/Infrared Array Camera (NIRC) observational campaign, as reported in \cite{2021ApJ...922...29S}. Finally, some preliminary SMF constraints for $4<z<8$ were derived using JWST observations \cite{2024ApJ...961..207N}. For lower redshifts, we use \cite{2001MNRAS.326..255C,2003ApJS..149..289B,2009ApJ...695..900Y,2009ApJ...707.1595D,2010A&A...523A..13P,2008ApJ...675..234P,2012MNRAS.421..621B} for the measurements of the SMF at $z\sim 0-1$ and \cite{2009ApJ...701.1765M,2011MNRAS.413.2845M,2014ApJ...783...85T} for the SMF at $z\sim 2$.

\textit{Stellar Mass Density:} The papers mentioned above not only include the SMF measurements, but also the SMD for the respective redshift ranges. We make use of one additional dataset, the early release of JWST-GLASS survey, which measured stellar masses for $z>7$ \citep{2023ApJ...942L..27S}. Previous studies have shown that $\Lambda$CDM can explain the observed SMD at smaller redshifts \citep{Madau:2014bja}, but newer JWST measurements at $z\sim8-10$ from \cite{2023Natur.616..266L} require star formation efficiency to be close $100\%$ when using $\Lambda$CDM. This is unrealistic, as most modern galaxy formation simulations and observational data infer that the SFE should be $\lesssim 10\%$ at high redshifts (e.g., \cite{2021Natur.597..485W, 2019A&A...626A..56P}). Additionally, the SFE should not be constant over the entire halo mass range, since star formation for massive galaxies is suppressed or quenched by various feedback mechanisms \citep{2012RAA....12..917S}. One way to address this problem is to invoke modified gravity. Because the most recent results from JWST provide a much more stringent test of $\Lambda$CDM, we choose to focus on that dataset.

\textit{Star Formation Rate Density:} At redshifts $z\lesssim 8$, we use the SFRD measurements from the compilation of \cite{Madau:2014bja, 2013ApJ...770...57B}. This dataset is based on both the Far Ultra-Violet and Infra-Red observations, and includes dust attenuation corrections. For higher redshifts reaching $z\sim 17$, we use JWST measurements from \cite{2023ApJS..265....5H} and \cite{2023MNRAS.523.1009B}.

\textit{Ultra-Violet Luminosity Function:} While some of the UVLF data is taken from the previously mentioned works, we also make use of high redshift data up to $z\sim 17$, derived using the JWST NIRCam instrument. This includes the following datasets: \cite{2023ApJ...951L...1P} for $8<z<13$, \cite{2022ApJ...940L..14N} for $10<z<12$, \cite{2023ApJ...946L..13F} for $z>10$, \cite{2023ApJ...954L..46L} for $z\sim9-12$, 
\cite{2023MNRAS.518.6011D} with $8<z<15$,
and \cite{2023ApJS..265....5H} with some galaxy candidates having redshift as high as $z\sim 16$. Finally, two works \cite{2023MNRAS.523.1036B} and \cite{2023MNRAS.523.1009B} managed to derive UVLFs at redshifts $8<z<15$ and $8<z<17$ respectively.

\textit{Epoch of Reionization:} For the neutral hydrogen fraction, we utilize JWST data from \cite{2023arXiv230600487U} covering the redshift span of $7<z<12$ and older datasets, derived from XShooter spectra \citep{2023MNRAS.525.4093G} at $z\sim5-6$ and QSO photometry \citep{2022MNRAS.512.5390G,2018ApJ...864..142D} at $z\gtrsim 7$. There are also some Ly-$\alpha$ and Ly-$\beta$ measurements present, namely \cite{2015MNRAS.447..499M,2022MNRAS.517.3263B,2018PASJ...70S..16K, 2014ApJ...797...16K,2019ApJ...878...12H} extending up to $z\sim 5$.

\subsection{Constraints from the Stellar Mass Function}

Our first probe of the modified gravity ansatz is the stellar mass function. It describes the change of the number density of galaxies with respect to the change in their stellar mass. This quantity has been actively studied across the redshift span of $0\leq z \leq 10$. Some of the works have been referenced in the previous subsection. We implement both Model I and Model II stellar mass to halo relations to probe the SMF. 

Firstly, let us discuss the phenomenological theories. Results for the Planck parameterization are placed onto the Figure (\ref{fig:SMF_E11}). In relation to screened models to be discussed in the next several paragraphs, for a phenomenological case we will be only considering $z=0$ and $z=1$ SMF, as anything further than that coincides with $\Lambda$CDM. But even at present, the HMF of phenomenological theories have a substantial deviation from the fiducial model at $M_{\rm h}\sim 10^{15}\rm M_\odot$ and above, which translates to $M_{\star}\gtrsim 10^{12}\rm M_\odot$ (this value strongly depends on the SMHR). There are only several data points that are located at such high stellar masses, each having large uncertainty, which makes constraining phenomenological theories challenging.

However, even taking these facts into account, one can notice that for Rodriguez-Puebla SMHR, there is a clear preference for $E_{11}\gtrsim1$, but the dependency of the SMF shape on $E_{22}$ is extremely weak and subsequently this parameter cannot be effectively constrained, which also applies to the double power-law SMHR. But, for the latter SMHR,$E_{11}\sim 0$ is a preferred choice, as $E_{11}\sim 1$ overestimates the SMF at higher stellar masses. Notably, the same behavior can be observed for the z\_flex parameterization (see Figure (\ref{fig:SMF_gmu})), where $g_\gamma$ cannot be properly constrained, but $g_\mu \gtrsim1$ is favored for Model I and $g_\mu\sim 0$ is favored for Model II SMHR. Note that $g_\mu \sim -1$ is still allowed allowed by the observational data. On the other hand, the DES parameterization shows more interesting results, allowing us to constrain both free parameters at low redshifts. Thus, as one can deduce from the Figure (\ref{fig:SMF_DES}), while working under the Rodriguez-Puebla stellar mass to halo relation, the case of $T_1\land T_2\gtrsim1$ is strongly preferred. Moreover, for the double power-law SMHR it is unexpected that exactly the opposite applies, i.e. $T_1\land T_2\lesssim-1$ is preferred but $T_2=0$ is not excluded.

Finally, as the last phenomenological theory we consider the well-known $w\boldsymbol{\gamma}$CDM with results being displayed in Figure (\ref{fig:SMF_wCDM}). Now one can notice the significant deviation at $z=1$ from $\Lambda$CDM unlike for the previous models. Additionally, Model I SMHR prefers $w\gtrsim-1.5$ while the Model II prefers the range $-1\lesssim w_\Lambda < \infty$ (note that the evolution of SMF beyond $w_\Lambda \sim -2$ is barely noticeable). The best-fit value of the growth index for the first SMHR appears to be $\boldsymbol{\gamma}\lesssim0.4$ while for the Model II it is somewhere in the range of $0.5\lesssim \boldsymbol{\gamma}\lesssim0.6$.

It is important to stress that since the first three phenomenological theories only marginally change the SMF, and therefore best fit and $\Lambda$CDM curves both reside within the uncertainties of the observational data, there is practically no improvement in the statistical sense under the assumption that for example $E_{11}\gtrsim1$ in relation to $E_{11}\sim0$. Subsequently, one can conclude that SMF is a poor probe of phenomenological modified gravity except for $w\boldsymbol{\gamma}$CDM which can show substantial deviation from $\Lambda$CDM at $M_\star \gtrsim 10^{10}M_\odot$.

The results for the first screened case, namely nDGP theory are located in the Figure (\ref{fig:SMF_nDGP}). It is interesting to note that at redshifts up to $z\sim 8$, Rodriguez-Puebla SMHR prefers $r_c\gtrsim10^3\rm \,Mpc$ with only lower redshift case allowing for $r_c\lesssim 10^{2.7}\rm \,Mpc$. On the other hand, double power-law SMHR shows strong evidence for $r_c\gtrsim10^{3.5}\rm \,Mpc$ at $0<z<4$ but can allow for $r_c\sim 10^3\rm \,Mpc$ at the higher redshift range, the beginning of Epoch of Reionization, i.e. $5<z<8$. However, for both kinds of stellar mass to halo relation, extreme cases of $r_c\lesssim10^{2.5}\rm \,Mpc$ are excluded.

On the other hand, let us investigate the more complex $k$-mouflage model, with the corresponding data being located in the Figures (\ref{fig:SMF_kmoufl_Puebla}) and (\ref{fig:SMF_kmoufl_double_power}). At low redshifts under Rodriguez-Puebla SMHR, interestingly the case of $\beta=0.5$ and $K_0\sim 1$ is preferred over smaller values of free parameters that lead to the $\Lambda$CDM like behaviour. However, at $z\geq 4$ the extreme case of $\beta=0.5$ is excluded and from $z\geq 8$ onwards even $\beta=0.3$ is excluded as well (vanishing $K_0$ may allow $\beta=0.3$ with tension being $<2\sigma$ as required, but it can be hypothesized that this will be excluded at even higher redshifts once such data will be made available in the near future). On the contrary, double power-law SMHR shows a strong preference for $\beta=0.1$ on the whole redshift span of our investigation. However, it is important to signify that with $\beta$ getting smaller, it becomes harder to constrain $K_0$, so in general, for this SMHR, the whole $0<K_0<1$ range is allowed (potentially, even bigger values can be used without the tension with respect to the observational data being present, but this is beyond the scope of the current study).

As a side note, it is essential to remark that the Rodriguez-Puebla SMHR seems to underestimate SMF at the higher stellar masses even for the $\Lambda$CDM, but the double power-law model does not have such a problem, so some parameter constraints that are aimed to fix this issue for the first SMHR may not coincide with the second model (i.e. first model suggests higher deviation from the fiducial cosmology than the second one). Most probably, this issue occurs due to the fact that Rodriguez-Puebla SMHR was calibrated to emulate the high-$z$ observational data rather than low-$z$. 

In addition, double power-law SMHR cannot reproduce the observed slope of the SMF at smaller stellar masses (starting at around $M_\star\sim 10^{10}M_\odot$). The correct slope is reproduced with a higher strength of SN Ia feedback (that suppresses star formation at the lower end of the SMF). Such inconsistency may hint to the fact that the SN Ia feedback calibrated with the help of the UVLF data rather than SMF/SMD may incorporate additional, yet unknown scatter.

\subsection{Constraints from the Stellar Mass Density}

The second set of constraints are imposed from the high-redshift observations of the stellar mass density made by JWST, with the results being displayed in the Figures (\ref{fig:SMD_nDGP}) and (\ref{fig:SMD_kmoufl}) respectively. We did not consider phenomenological modifications of gravity such as Planck/DES/$w\boldsymbol{\gamma}\rm CDM$, as they do not affect high-$z$ structure formation, and thus we only focus our attention on the screened theories.

For the first theory, namely nDGP, there is a significant preference of $r_{c}\leq 10^3$ Mpc, which can be relaxed in the double power-law SMHR case if one will consider higher than fiducial star formation efficiency at the peak mass $M_p$. However, even for small $r_c$ values, Rodriguez-Puebla SMHR could not satisfy SMD constraints completely, as at such high stellar masses $M_\star\sim 10^{12}M_\odot$, star formation is heavily suppressed due to the various feedback mechanisms.

On the other hand, for $k$-mouflage theory, $K_0< 1$ and $\beta\approx0.3$ was preferred as well as $K_0\geq 1 $ and $\beta\approx 0.5$, but the situation with Rodriguez-Puebla SMHR is similar, while double power-law case can satisfy all of the constraints.

\subsection{Constraints from the Ultra-Violet Luminosity Function}

It is important to denote that unlike the power law SMHR, for the Rodriguez-Puebla one we are going to limit the integration bounds of $M_{\rm UV}$ following the original work \citep{2017MNRAS.470..651R}, namely at $z=4$, the lower bound is $M_{\rm UV}=-22.6$, at $z=5$, $M_{\rm UV}=-23$, at $z=6$, $M_{\rm UV}=-22.5$, at $z=7$, $M_{\rm UV}=-22.75$ and finally at $z=8$, the lower limit is $M_{\rm UV}=-22$. For other redshifts, we linearly extrapolate the aforementioned limits. Also, in the current work, we are only going to derive the UVLF for two models, namely nDGP and $k$-mouflage, as any phenomenological model of our consideration will not show noticeable deviation from $\Lambda$CDM at $z\geq4$, and there is no UVLF observational data present at smaller redshifts.

We plot the theoretical predictions of UVLF on the Figures (\ref{fig:UVLF_nDGP}) for nDGP gravity and Figures (\ref{fig:UVLF_kmoufl_Puebla}), (\ref{fig:UVLF_kmoufl_double_power}) for $k$-mouflage. Constraints for each theory here are more uncertain than SMF/SMD as the ultra-violet variance is not as well constrained. However, it can be observed that scatter is around $\sigma_{\rm UV}\lesssim 1$ dex for $z\leq 8$ and then it goes up to $\sigma_{\rm UV}\sim 2$ dex for $z\sim12$. With the variation of the nDGP free parameter, namely $r_c$, it is possible to adjust the UVLF at brighter magnitudes. Specifically, for both SMHR cases, it can be observed that even if we apply proper $\sigma_{\rm UV}$ kernel, $\Lambda$CDM can diverge from the observational data at $M_{\rm UV}\lesssim-20$, but this can be fixed by setting nDGP parameter to $r_c\leq 10^3$ Mpc.

Interestingly enough, $k$-mouflage theory can satisfy all of the constraints up to $z\sim 12$ even with the Gaussian kernel of $\sigma_{\rm UV}=0.4$ dex, which is provided by the stellar mass scatter only. Case with $\beta=0.5$, $K_0\sim 0.1$ has practically perfect correspondence to the observational data for the Rodriguez-Puebla SMHR, but that is only true with $z\geq 8$ for the choice of the double power-law.

In general, we have noticed that UVLF is a worse probe of the screened modified gravity in relation to SMF or SMD, as the deviation from $\Lambda$CDM caused by the variation of free parameters is a lot less noticeable (especially at the brighter end of the UVLF) while comparing to the SMF. This has also been reported to be the case for other theories in \citep{2016ApJ...818...89S}.

Finally, it is important to signify that double power-law SMHR was not calibrated to work with UVLF in the $\Lambda$CDM, and therefore even with proper scatter setup, the results show up to tens of percent of deviation from the observational datasets at lower redshifts.

\subsection{Constraints from Star Formation Rate Density}
The other interesting probe of modified gravity may be the Star Formation Rate Density (SFRD), which is in our case derived from the mass accretion rate paired up with the star formation efficiency. Results for phenomenological theories of modified gravity are shown on the Figure (\ref{fig:SFRD_pheno}) and for screened theories on the Figure (\ref{fig:SFRD_screen}).

As usual, phenomenological theories show only a very small, in some cases sub-percent deviation from $\Lambda$CDM, and some free parameters are impossible to constrain because the variation of the letter is smaller than the uncertainties in the SFRD measurements. However, we can still see that $E_{11}\gtrsim 1$, $g_\mu \gtrsim1$ and $T_1\land T_2\gtrsim 1$ are preferred choices. On the other hand, $w\boldsymbol{\gamma}\rm CDM$ gives more significant deviation from the fiducial cosmology, and for low-$z$ SFRD constraints to be satisfied, $w_\Lambda\leq1$, which makes quintessence cosmology a viable choice.

Screened theories are far easier to constrain. As an example, at $r_c\leq 10^3$ Mpc, nDGP theory shows underwhelming SFRD prediction at lower redshifts but in turn, satisfies high-$z$ JWST data and vice-versa (which is true for both SMHR choices). In contrast, $k$-mouflage can satisfy both high and low redshift constraints up to $z\sim 12$ if one will take $\beta\sim 0.3$ and $K_0\sim 0.5$ (which seems to be the only possible choice, as $\beta=0.1$ has smaller than expected SFRD at high redshifts and $\beta=0.5$ at low redshifts respectively), while working under the Rodriguez-Puebla SMHR and $\beta \lesssim 0.1$, $K_0\sim 1$ under double power-law SMHR. It is evident that for both nDGP and $k$-mouflage models, for any choice of free parameters, there is still a minor inconsistency between theoretical and observational data at small redshifts. This phenomenon was previously found in the literature, for example in \texttt{EAGLE} simulation output \citep{2015MNRAS.450.4486F}, \texttt{Horizon Run 5} simulation output \citep{2021ApJ...908...11L} and \texttt{Illustris-TNG} simulation output \citep{2023MNRAS.524.2539P} respectively. The solution was to introduce an additional scatter of $0.2$ dex, which presumably arises due to some, yet unknown star formation mechanisms. Such scatter can also help to satisfy higher redshift constraints imposed by JWST and allow the case of $\beta=0.1$, decreasing the departure from $\Lambda$CDM for $k$-mouflage theory. Interestingly, a scatter of only $0.05$ dex or even less (depending on the value of a free parameter) is required for the double power law model. For any model and any scatter value, it is still not possible to produce the nearly constant SFRD, required by JWST at $z\gtrsim12$. In \citep{2023ApJS..265....5H}, it was also noted that SFRD at such redshifts cannot be reproduced with the constant SFE models. Unlike the lower redshift measurements of SFRD, JWST measurements rely on the photometric redshifts. Also, the sample size is small and it's robustness may change the observed SFRD by up to an order of magnitude \citep{2023MNRAS.523.1009B}.

Thus, Rodriguez-Puebla SMHR can very precisely reconstruct the SMF and SMD, but cannot reconstruct SFRD without the introduction of additional scatter. The opposite is true for the double power law. Such inconsistencies between SMF/SMD and SFRD have been noticed before on the whole $0<z<8$ range and are explored further in the work of \cite{2016ApJ...820..114Y}. The main issue is that Rodriguez-Puebla SMHR is calibrated on the basis of SMF and SMD observational data, but double power law SMHR rather uses UVLF. Since the connection between UV magnitude and halo mass is subject to a scatter that is yet to be properly determined, each model cannot satisfy all of the constraints imposed.

\subsection{Constraints from Epoch of Reionization}
The final set of constraints comes from the epoch of reionization, namely ionized hydrogen volume filling fraction $Q_{\rm HII}$ and optical depth $\tau$. Similarly to the UVLF and SMD cases, we do not consider phenomenological modifications of gravity in this section, as reionization happens when dark energy mass density is practically negligible and thus will have no effect on both $Q_{\rm HII}$ and $\tau$ (optical depth is integrated from the present time until the end of reionization, so it does capture the dark energy dominated epoch, but during that epoch, $Q_{\rm HII}\sim 1$ and thus the effect of the phenomenological modified gravity on the optical depth is also negligible).

\begin{figure*}[!htbp]
    \centering
    \includegraphics[width=0.95\linewidth]{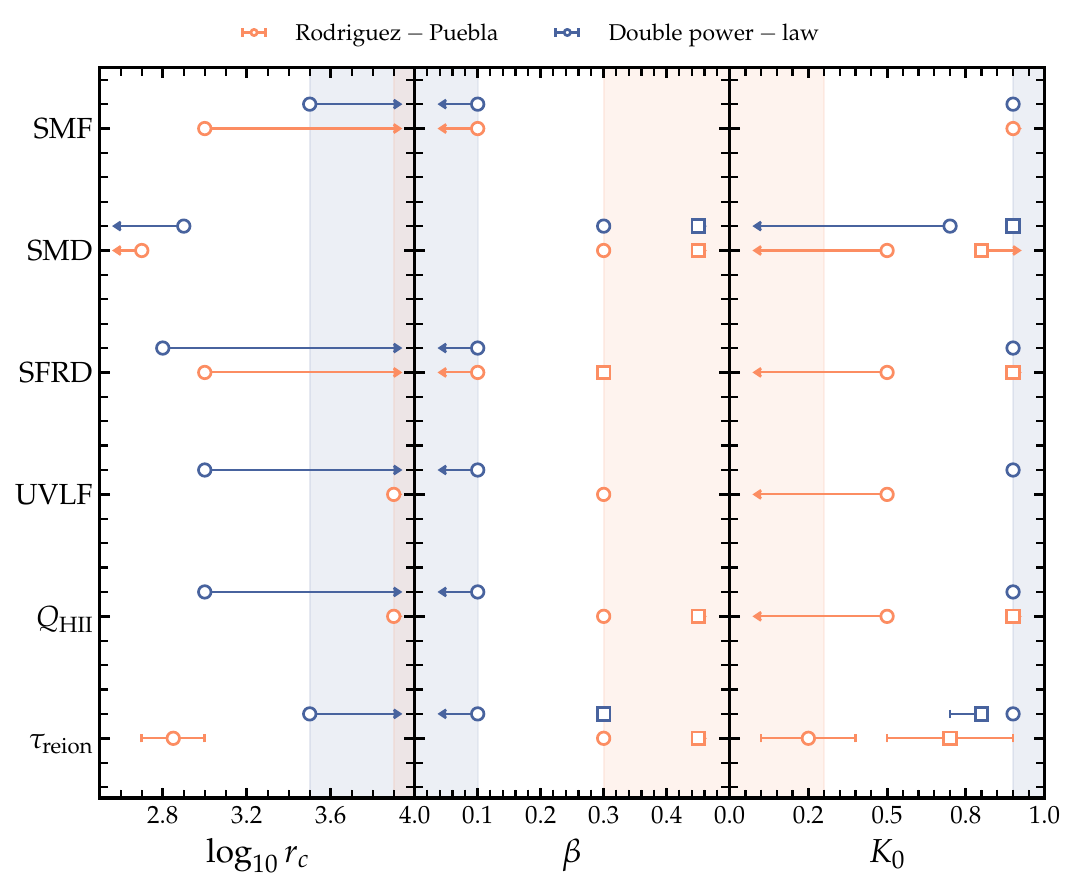}
    \caption{Compilation of the best-fit parameters determined by the visual inspection of various observables for nDGP (first column) and $k$-mouflage (second and third columns). Arrows signify an upper limit for a parameter, while shaded areas represent the parameter space preferred by the vast majority of observational datasets.}
    \label{fig:compilation}
\end{figure*}

Both volume filling fraction and the optical depth for screened theories are shown on the Figure (\ref{fig:EoR_screen}). Let us first consider the nDGP gravity. We note that under the Rodriguez-Puebla SMHR, the volume filling fraction of the ionized hydrogen predicted by nDGP cannot satisfy observational constraints for any value of $r_c$ and in general, the variation of $Q_{\rm HII}(z)$ with $r_c$ is only marginal. On the other hand, observational constraints for the double power-law SMHR are satisfied for the whole range $r_c\gtrsim10^3$ Mpc.

Under the condition that $r_c\lesssim10^{2.7}$ Mpc, the predicted optical depth for both SMHRs is higher than the one of \textit{Planck2018} measurement, but consistent with the reionization history based on the JWST-calibrated $N_{\rm ion}$ value \citep{2024MNRAS.tmpL..79M}. On the other hand, \textit{Planck} data implies that $10^{2.7}\,\mathrm{Mpc}\lesssim r_c\lesssim 10^{3}$ Mpc for Rodriguez-Puebla and that $r_c\gtrsim10^{3}$ Mpc for double power-law SMHR respectively.

The other model of our consideration, namely $k$-mouflage was able to satisfy reionization constraints for both SMHRs. Under the assumptions of a Rodriguez-Puebla SMHR, the preferred values are $\beta\sim0.3$ and $K_0\gtrsim0.5$ or $\beta \sim 0.5$ and $K_0\sim 1$. Besides, given double power-law SMHR, the only plausible choice is that $\beta\lesssim0.1$ and $K_0\sim 1$. 

One thing that we should denote is that we are using the star formation rate density already with the suitable scatter introduced (0.2 dex in the case of Rodriguez-Puebla and 0.05 dex in the case of double power-law SMHR respectively). Without this scatter, the reionization should occur later.

\section{Discussion and concluding remarks}\label{sec:8}
Most of the constraints and best-fit parameters for SMF, SMD, SFRD and UVLF, Epoch of Reionization have already been thoroughly discussed in previous subsections. We show the compilation of all of those results for screened theories of modified gravity in Figure (\ref{fig:compilation}). Note that in some cases a second solution is found i.e. the solution is degenerate. This has been indicated in Figure (\ref{fig:compilation}) using circle and square markers. Upper limits are indicated where appropriate. Such a degeneracy appears due to the fact that the variation of $\beta$ can increase the abundance of galaxies and $K_0$ can decrease it, thus it is possible to find several similar solutions for different choices of a pair $\{\beta, K_0\}$. The fact that we have only found two degenerate solutions at most is because we have considered a coarse grid for $\beta$, hence there will be more degenerate solutions for a grid of higher resolution.

On the Figure (\ref{fig:compilation}), we can see that the choice of $r_c\sim 10^{3.9}$ Mpc can provide the best fitting predictions for a wide range of observables and both stellar mass to halo relations. A similar situation may be observed in the case of $k$-mouflage, where $\beta\sim 0.1$, $K_0\gtrsim 0.9$ is highly preferred for double power-law SMHR and $\beta\sim 0.3$, $K_0\lesssim 0.3$ for Rodriguez-Puebla. Figure (\ref{fig:compilation}) does not illustrate the results for phenomenological theories, specifically for $w\boldsymbol{\gamma}$CDM due the incomplete set of constraints, but it is worth to mention that this model prefers $w_\Lambda\gtrsim -1$ instead of quintessence while using both SMF and SFRD. The only quantity that suggests the need for a substantial deviation from $\Lambda$CDM is the stellar mass density. It is interesting to note that the double power-law best-fit values are noticeably closer to the $\Lambda$CDM than Rodriguez-Puebla. This is an important result because it highlights the sensitivity of the adopted star formation history model as a test for new physics.

Another important result from the calculations described in this work is that whilst all other models fail, $k$-mouflage theory with the double-power law SMHR can satisfy both reionization and high redshift SMD constraints imposed by the JWST up to $M_\star \sim 10^{11}M_\odot$. We note that the Warm Dark Matter prescription of \cite{2023ApJ...947...28G} also achieves this. The reason that both phenomenological and nDGP theories fail (in this context) is that changes in $Q_{\rm HII}(z)$ within those theories are too small to provide good fits to the observational data. Even if it was possible to satisfy reionization constraints in the nDGP case, Figure (\ref{fig:SMD_nDGP}) shows that the observed SMD can only be reproduced for $r_c\lesssim10^{2.5}$ Mpc, which is way outside the observational bound, making this choice not viable.

The theories we have considered are only able to satisfactorily fit the observational SMF and UVLF if the UVLF to stellar mass relation has additional scatter, over and above that depicted in Equation 88. As it was previously discussed, this is most probably due to the fact that SMHRs considered in this work has been calibrated on different datasets. This problem has been addressed in a more recent work \cite{2019MNRAS.488.3143B}, where the authors combined SFR, UVLF and SMF measurements to deduce SMHR. Figure (\ref{fig:compilation}) shows several discrepancies between the model parameter estimates obtained using the Rodriguez-Puebla and double power law SMHR e.g. the \textit{Planck2018} $\tau_{\rm reion}$ parameter measurement. In that example, the discrepancy is a consequence of the Rodriguez-Puebla SMHR being calibrated using SFR/SMD data, whilst the double-power law was calibrated using UVLF data. 
The estimated $\tau_{\rm reion}$ parameter will depend on both the additional scatter between the UVLF and the SMF, and also the form used for $\dot{n}_{\rm ion}$; as noted by \cite{2019ApJ...879...36F}, the rate of production of Ly-C photons can also be defined in terms of the Ultra-Violet Luminosity Density (UVLD).
Compared to $\Lambda$CDM, the nDGP and $k$-mouflage theories always lead to an overabundance of UV sources and stellar mass overestimates. However, JWST data suggests that regardless of whether we use Rodriguez-Puebla or double power-law, we need fewer sources than predicted by $\Lambda$CDM at high redshifts (see Figure (\ref{fig:SMF_nDGP}) for example).

Even though we have taken into account several sources of scatter, it is possible that additional uncertainties are present, contributing to the scatter seen in the results presented in Figure (\ref{fig:compilation}). One example of a potential additional error source is the escape fraction of photons, namely $f_{\rm esc}$. Throughout the literature, various values for $f_{\rm esc}$ has been assumed and there is no verdict on what assumption is the correct one, as it is not something that can be directly observed. Hence, the variation of $f_{\rm esc}$ will lead to the variation of reionization quantities which in turn will change the constrained parameter space. 

Other parameters, such as $\epsilon_{\star,0}$, are not bounded at all. We had to assume the $\Lambda$CDM value in that case. This is motivated by the fact that $\Lambda$CDM prediction of the SMF \textit{should} be close to the observational data, which has been confirmed by the cosmological observations, but one still can keep it as a free parameter for the sake of generality. 

Another possible issue is that we have assumed \textit{Planck2018} cosmology that was established with $\Lambda$CDM as a background. In a perfect scenario, one will include cosmological parameters, such as $H_0$ or $\Omega_{m0}$ into the analysis as free variables. However, doing this is impractical in terms of computing time required.

Currently, there are several constraints imposed on the nDGP gravity in literature. For example, correlation function monopole and quadrupole, obtained from the SDSS DR7 sample, suggest that $r_c> 10^{2.53}$ Mpc at $2\sigma$ confidence \citep{2013MNRAS.436...89R}. However, analysis of the DR11 sample has a stricter constraint of $r_c>10^{3.65}$ Mpc \citep{2016PhRvD..94h4022B}, which is very close to what we have derived.  
We are not aware of constraints for the pair $\{\beta,K_0\}$ within the $k$-mouflage theory using the parameterization that we have assumed. However, a recent study using mock observations of Stage IV cosmic shear surveys \cite{2023arXiv230612368F} may lead to interesting constraints on nDGP and $k$-mouflage theories. 

The main purpose of the work described in this paper is to put forward new ideas as to how one might use new astronomical observations to constrain interesting theories of gravity beyond General Relativity. Clearly, a great deal of further work is required to identify the most effective ways of doing so. For example, after further optimization of our code, one could run more detailed Markov Chain Monte Carlo (MCMC) calculations, using thousands of steps in parameter space. This would then enable us to determine the redshift evolution of the preferred value of $\sigma_{\rm UV}$ (and its uncertainty), as a function of the relevant modified gravity free parameter. It may even be possible to make parameters such as the photon escape fraction $f_{\rm esc}$ or the double power law SMHR slopes $\gamma_{\rm lo}$, $\gamma_{\rm hi}$ as free variables, and derive constraints for those. Whilst this has been done within the framework of $\Lambda$CDM \citep{2019ApJ...878..114R}, the redshift evolution of those parameters will change significantly with the introduction of screened gravity.

\section*{Acknowledgements}
I am grateful to Prof. John Webb for a fruitful discussion and insightful comments on this work.

Sokoliuk O. performed the work in frame of the "Mathematical modeling in interdisciplinary research of processes and systems based on intelligent supercomputer, grid and cloud technologies" program of the NAS of Ukraine. This article/publication is based upon work from COST Action CA21136 "Addressing observational tensions in cosmology with systematics and fundamental physics (CosmoVerse)" supported by COST (European Cooperation in Science and Technology). In addition, this work has also benefited from the use of open-source \texttt{Python} packages: \texttt{SciPy} \citep{Virtanen:2019joe}, \texttt{Matplotlib} \citep{matplotlib}, \texttt{NumPy} \citep{numpy} and \texttt{emcee} \citep{Foreman-Mackey_2013}. Parts of the \texttt{Pylians3} code \citep{Pylians} were used to implement HMF derivation.
\section*{Data Availability}
We have made the codes available via the link \href{https://github.com/singlefrequency/scripts_JWST_MG}{\textsc{GitHub}}. Be aware that the code was made for internal use and therefore is poorly commented, for any questions regard to the corresponding author. All $\Lambda$CDM results produced by the code have been carefully checked for consistency with the data available in the literature to make sure that the code is working properly. For SMF/SMD comparison, we have used the data from \cite{2023arXiv230314239D,2023ApJ...947...28G}, for UVLF we have used \cite{2023MNRAS.525.3254S} and for spherical collapse, \cite{2010MNRAS.406.1865P}. Only small deviations have been observed for some quantities, and we believe that they arise due to the numerical precision limitations (for example, the size of the interpolation grid for the scale factor used to find spherical collapse initial conditions) and different cosmology assumed (for example, \textit{WMAP9} instead of \textit{Planck2018}).

\bibliographystyle{yahapj}


\end{document}